\documentclass[a4paper,fleqn,usenatbib]{mnras}

\usepackage{newtxtext,newtxmath}

\usepackage[T1]{fontenc}
\usepackage{ae,aecompl}

\usepackage{graphicx}	
\usepackage{amsmath}	
\usepackage{amssymb}	
\usepackage{gensymb}
\usepackage{threeparttable}
\usepackage{pdflscape}
\usepackage{CJK}

\newcommand{\angstrom}{\textup{\AA}}

\def\lsim{\mathrel{\rlap{\lower4pt\hbox{\hskip1pt$\sim$}}
    \raise1pt\hbox{$<$}}}                
\def\gsim{\mathrel{\rlap{\lower4pt\hbox{\hskip1pt$\sim$}}
    \raise1pt\hbox{$>$}}}                

\title[X-ray Enhancements of HRLQs at $z>4$]{Investigating the X-ray enhancements of highly radio-loud quasars at $z>4$}

\author[S. F. Zhu et al.]{S. F. Zhu,$^{1,2}$\thanks{E-mail: SFZAstro@gmail.com (PSU)}
W. N. Brandt,$^{1,2,3}$
Jianfeng Wu,$^4$
G. P. Garmire,$^5$
\newauthor
and B. P. Miller$^6$
\\
$^1$Department of Astronomy \& Astrophysics, The Pennsylvania State University, University Park, PA 16802, USA\\
$^2$Institute for Gravitation and the Cosmos, The Pennsylvania State University, University Park, PA 16802, USA\\
$^3$Department of Physics, 104 Davey Lab, The Pennsylvania State University, University Park, PA 16802, USA\\
$^4$Department of Astronomy, Xiamen University, Xiamen, Fujian 361005, China\\
$^5$Huntingdon Institute for X-ray Astronomy, LLC, 10677 Franks Road, Huntingdon, PA 16652, USA\\
$^6$Department of Chemistry and Physical Sciences, The College of St. Scholastica, Duluth, MN 55811, USA}

\date{Accepted XXX. Received YYY; in original form ZZZ}

\pubyear{2018}

\begin{document}
\label{firstpage}
\pagerange{\pageref{firstpage}--\pageref{lastpage}}
\maketitle
\begin{abstract}
We have investigated the jet-linked \mbox{X-ray} emission
from highly radio-loud quasars (HRLQs; $\log R>2.5$) at high redshift.
We studied the X-ray properties of 15 HRLQs at $z>4$,
using new {\it Chandra} observations for six objects
and archival {\it XMM-Newton} and {\it Swift}
observations for the other nine.
We focused on testing the apparent enhancement of jet-linked \mbox{X-ray} emission from HRLQs at $z>4$.
Utilizing an enlarged (24 objects) optically flux-limited sample with complete X-ray coverage,
we confirmed that HRLQs at $z>4$ have enhanced X-ray emission relative to
that of HRLQs at $z\approx$ 1--2 with matched UV/optical
and radio luminosity, at a \mbox{4.0--4.6}~$\sigma$ level; the X-ray enhancements are confirmed considering
both two-point spectral indices and inspection of broad-band spectral energy distributions.
The typical factor of enhancement is revised to $1.9^{+0.5}_{-0.4}$, which is smaller than but consistent with previous results.
A fractional IC/CMB model can still explain our results at high redshift,
which puts tighter constraints on the fraction of IC/CMB X-rays at
lower redshifts, assuming the physical properties of quasar jets do not have a strong redshift dependence.
A dominant IC/CMB model is inconsistent with our data.
\end{abstract}

\begin{keywords}
    quasars: general -- X-rays: galaxies -- galaxies: high-redshift
\end{keywords}

\section{Introduction}
\label{sec:intro}
Quasars (and their parent population of active galactic nuclei, AGNs) are ultimately powered by the accretion process,
where gravitational binding energy is released
as matter falls into the deep gravitational potential well of the supermassive
black hole (SMBH) located in the central region of the host galaxy.
The released energy is mainly in the form of quasi-thermal optical/UV photons,
likely radiated from an optically thick accretion disk, with
a mass-to-radiation conversion efficiency of $\sim0.1$.
Accompanying the accretion process,
a pair of highly collimated relativistic jets can sometimes launch from the vicinity
of the SMBH, perhaps by tapping the spin energy of the SMBH,
and extend to galactic and intergalactic scales \citep[e.g.][]{Begelman1984}.
These quasar jets can radiate across the whole electromagnetic spectrum
and are most easily detected in the radio band.
According to the flux ratio at rest-frame 5 GHz vs. 4400 $\angstrom$,
i.e. the radio-loudness parameter $R$ \citep[$\equiv f_{\rm 5\;GHz}/f_{\rm 4400\;\angstrom}$;][]{Kellermann1989},
the quasar population is divided into radio-quiet quasars (RQQs; $R<10$) and
radio-loud quasars (RLQs; $R>10$).\footnote{Use of the terms ``radio-loud AGN'' and ``radio-quiet AGN'' is sometimes
inappropriate; e.g. when the optical AGN continuum is obscured or when the
radio continuum has a strong contribution from non-jet emission \citep[e.g.][]{Padovani2017}.
However, for the powerful type~1 quasars with strong
jets under study here, use of this terminology is appropriate
(P. Padovani 2018, private communication).}
RLQs are found to be the minority, making up $\sim$10\% of the quasar population \citep[e.g.][]{Ivezic2004}.

X-ray emission is nearly universal from accreting SMBHs \citep[][and references therein]{Brandt2015}.
For RQQs, the primary power-law emission in X-rays (\mbox{$\sim$1--100 keV})
is thought to be created by UV photons from the accretion disk \mbox{inverse-Compton} (IC) scattering off
electrons in an optically thin and hot (\mbox{$\approx10^9$~K}) plasma above the disk, the so-called ``\mbox{accretion-disk} corona''.
RLQs have an additional jet-linked X-ray component \hbox{\citep[e.g.][]{Wilkes1987, Worrall1987}}, which
can outshine the coronal X-ray emission by a factor of \mbox{$\approx$ 3--30} in cases of large radio loudness \citep[e.g.][Miller11 hereafter]{Miller2011}.
This jet-linked X-ray emission is mainly attributed to IC emission of relativistic (non-thermal) electrons
that are accelerated by shocks/magnetic reconnection in the jet.

\begin{figure*}
\centering
\includegraphics[width=0.6\textwidth, clip]{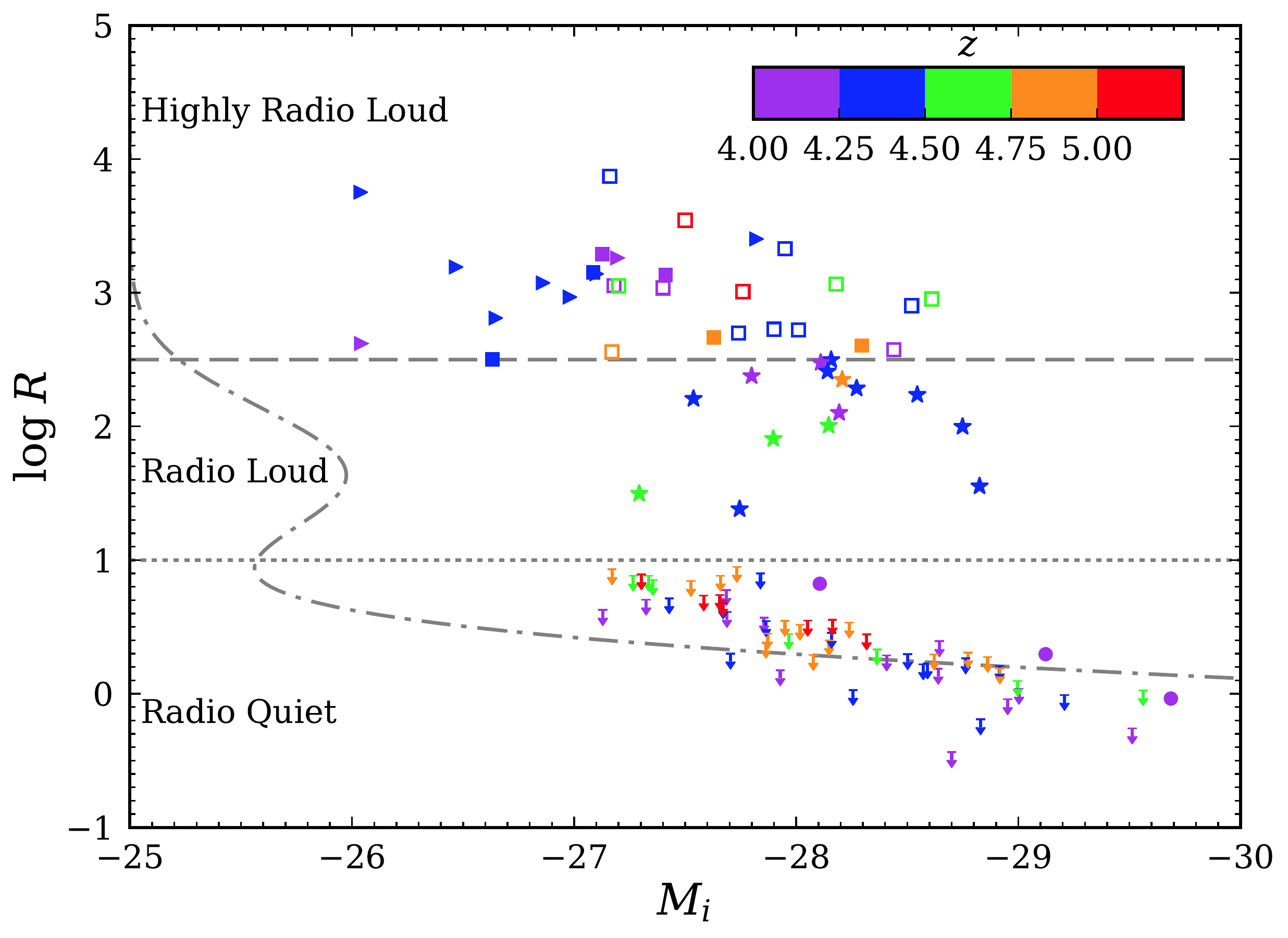}
\caption{The distribution of $z>4$ HRLQs in the $\log R$--$M_i$ plane,
compared to moderately radio-loud quasars and RQQs at $z>4$.
The filled squares and triangles are the {\it Chandra}
Cycle 17 objects and archival data objects, respectively.
The open squares are the HRLQs of Wu13.
The filled stars are moderately radio-loud quasars at $z>4$ (\citealt{Bassett2004, Lopez2006}; Miller11).
The filled circles and downward arrows
represent the radio-quiet SDSS quasars at $z>4$ that have sensitive X-ray coverage.
All the symbols are color-coded based on their redshifts using the 
color bar at the top right of the figure.
The dotted and dashed lines indicate our criteria for RLQs and HRLQs.
The dash-dotted curve (with an arbitrary linear scale) shows 
the radio-loudness distribution of SDSS quasars \citep{Ivezic2004}, which shows that
HRLQs reside in the tail of high radio-loudness.}
\label{fig:RMi}
\end{figure*}

The quasar population has long been known to
show strong cosmological evolution in number
density \citep[e.g.][]{Schmidt1968, McGreer2013, Yang2016a},
with RLQs likely evolving differently from RQQs \citep[e.g.][]{Ajello2009}.
However, the spectral energy distributions (SEDs) of quasars
generally show little evolution to $z>6$.
In X-rays, RQQs at $z>4$ have similar spectral
\citep[e.g.][]{Brandt2002, Vignali2003a, Shemmer2006, Nanni2017}
and variability \citep[e.g.][]{Shemmer2017} properties as those of appropriately matched RQQs at lower redshift,
in line with quasar properties in other bands \citep[e.g.][]{Jiang2006, Fan2012}.
Moderately \mbox{radio-loud} quasars ($1<\log R<2.5$) at $z>4$ also have similar X-ray properties
to their low-redshift counterparts \citep[e.g.][]{Bassett2004, Lopez2006, Saez2011},
while the highly \mbox{radio-loud} quasars (HRLQs; $\log R>2.5$)
show an apparent enhancement in the X-ray band at high redshift \citep[][Wu13 hereafter]{Wu2013}.

These X-ray studies of high-$z$ RLQs are inconsistent with one of the leading models
for quasar jets based on X-ray photometric imaging of low-$z$ objects,
where the IC process involving cosmic microwave background (CMB) photons is thought to play an important role.
CMB photons have long been proposed to be seeds for the IC process that
can effectively produce X-rays \citep[e.g.][]{Felten1966, Harris1979, Feigelson1995}.
One relevant case is quasar lobes \citep[e.g.][]{Brunetti1999}, where
relativistic electrons are coupled with CMB photons and produce X-ray emission.
After the discovery of the X-ray jet of PKS 0637$-$752 \citep[][]{Chartas2000, Schwartz2000} by {\it Chandra},
the application of the IC/CMB mechanism to kpc-scale quasar jets become popular.
The X-ray jet of PKS 0637$-$752 is so luminous (relative to the optical) that
it cannot be readily produced by other mechanisms \citep[e.g.][]{Schwartz2000, Harris2002},
while a modern version of the IC/CMB model
can explain \mbox{radio-to-X-ray} SEDs of individual jet knots and maintains the assumption of equipartition.
This modern version of the IC/CMB model has two essential requirements: that the \mbox{kpc-scale} quasar jets are
relativistic with bulk Lorentz factor $\sim$10 and are observed at
small angles to our line of sight \citep{Tavecchio2000, Celotti2001}.
These two ingredients naturally explain the \mbox{one-sidedness} of many X-ray jets
that are commonly detected in surveys of \mbox{low-$z$} quasar jets \citep[e.g.][]{Sambruna2004, Marshall2005, Kataoka2005}.

In spite of the apparent initial success of the (beamed) IC/CMB model in low-$z$ objects,
X-ray studies of high-$z$ RLQs provide a critical piece of evidence against
using this model to explain the dominant majority of the \mbox{X-ray} emission from quasar jets.
The CMB energy density has a strong cosmological evolution ($U_{\rm CMB}\propto(1+z)^4$),
which is not reproduced in the jet-linked X-rays from RLQs.
The \mbox{X-ray} luminosities of the few resolved jets at high redshift are usually only
a few percent that of the cores \citep[similar to large-scale jets at low redshift; e.g.][]{Siemiginowska2003, Yuan2003, Saez2011, Cheung2012, McKeough2016},
and useful X-ray upper limits on extended jet emission exist for many more RLQs (e.g. \citealt{Bassett2004, Lopez2006}; Wu13).
Additionally, the jet-linked core emission (which could include X-ray emission from foreshortened kpc-scale jets in some systems) at high redshift does not show
the dramatic enhancement predicted by the IC/CMB model (e.g. \citealt{Bassett2004, Lopez2006}; Miller11; Wu13).
Furthermore, there are other multiple lines of evidence against the \mbox{most-straightforward} IC/CMB model:
the tension between the observed and predicted relative brightness distribution in the \mbox{X-ray} and radio bands,
the excessive requirement for the jet power,
the need for extremely small viewing angles,
the high polarisation of the optical emission from some jet knots suggestive of synchrotron emission,
and the non-detections of $\gamma$-ray emission from quasar jets \citep[e.g.][]{Harris2006, Uchiyama2006, Meyer2014}.
Alternative models for the luminous low-$z$ \mbox{X-ray} jets often involve an {\it ad hoc} high-energy synchrotron component \citep[e.g.][]{Atoyan2004}.

Even if the IC/CMB process does not play a dominant role,
we should expect some IC/CMB X-ray emission from AGN jets; the question is the level of contribution from this
process \citep[e.g.][]{Harris2006}, which could be revealed by studying \mbox{high-redshift} \mbox{radio-luminous} quasars in X-rays.
HRLQs rank as the top 5\% of the RLQ population in radio loudness (see Fig.~1),
and HRLQs at $z>4$ harbour the most-powerful relativistic jets from the first SMBHs in
the early universe, when the CMB photon field is $>625$ times more intense than now.
Wu13 compared the X-ray emission of a sample of HRLQs at $z>4$ (median $z=4.4$) with that of another sample of HRLQs at $z<4$ (median $z=1.3$)
with matched UV/optical and radio luminosity,
and found an X-ray enhancement for the HRLQs at $z>4$ at a 3--4$\sigma$ level.
HRLQs at $z>4$ have stronger \mbox{X-ray} emission than their counterparts at $z<4$,
by a factor of $\approx$ 3 on average.
There is also evidence for a $5\sigma$ X-ray enhancement in another independent sample of HRLQs at $z=$3--4 that is drawn from Miller11.

To explain the redshift dependence of the relative X-ray enhancement of HRLQs, Wu13 proposed a {\it fractional} IC/CMB model,
in which CMB photons are relevant only on the scale of $\sim$1--5 kpc, with photons from the central engine dominating at smaller
distance \citep[e.g.][]{Ghisellini2009}.
At scales beyond a few kpc, the jet has already decelerated so that CMB photons in the rest frame of the
jet are not intense enough for the IC/CMB mechanism to be significant \citep[e.g.][]{Mullin2009, Meyer2016, Marshall2018}.
The cosmologically evolving IC/CMB \mbox{X-ray} emission only contributes a fraction of the overall X-ray emission from HRLQs with
the rest coming from (redshift-independent) IC processes on small scales that involve seed photons from the central engine.
The fraction was estimated to be $\approx$6\% at $z\approx1.3$ by Wu13 and rises with redshift.
Alternatively, the results of Wu13 can also be explained by a scenario where the star-forming activity
of the hosts provides infrared/optical photons that are IC scattered into the X-ray band. This scenario requires the host galaxies
of high-redshift quasars to have enhanced star-formation activity \citep[e.g.][]{Wang2011, Mor2012, Netzer2014}.
In this case, the IC/CMB process becomes even less relevant.

The sample of 17 HRLQs at $z>4$ used in Wu13 suffers from heterogeneity and limited size,
which renders their $\approx$4$\sigma$ results only suggestive.
Here, we aim at confirming the X-ray enhancement of HRLQs using a larger and more uniformly selected sample.
We obtained new {\it Chandra} observations for 6 HRLQs at $z>4$ and present their X-ray properties in the paper.
We also present X-ray properties of another nine HRLQs at $z>4$ that have archival {\it Swift} or {\it XMM-Newton} data.
We describe our sample selection in Section~\ref{sec:sample}, and X-ray data analyses in Section~\ref{sec:reduction}.
In the following sections, we adopt a flat $\Lambda$CDM cosmology, with $H_0=70.0$ km s$^{-1}$ Mpc$^{-1}$ and $\Omega_{\rm m}=0.3$ \citep[e.g.][]{Planck2016}.

\section{SAMPLE SELECTION}
\label{sec:sample}
\begin{table*}
\centering
\caption{X-ray observation log.}
\label{tab:obsLog}
\begin{threeparttable}[b]
\begin{tabular}{lccccccccc}
\hline
\hline
    Object Name & R.A. & Dec. & Instr. & $z$\tnote{a} & Obs. Date & Obs. ID\tnote{b} & Exp. Time\tnote{c} & FL\tnote{d}& Ref.\tnote{e} \\
    & (deg) & (deg) & &  & & & (ks) & & \\
\hline
\multicolumn{8}{c}{{\it Chandra} Cycle 17 Objects} \\
\hline
    SDSS J003126.79+150739.5 & 7.8617 & 15.1277 & ACIS-S & 4.296 & 2016/06/09 & 18442 & 5.4 &N&-\\
    B3 0254+434 & 44.4962 & 43.6438 & ACIS-S & 4.067 & 2015/12/12& 18449 & 5.5 & Y&1\\
    SDSS J030437.21+004653.5 & 46.1551 & 0.7816 & ACIS-S & 4.266 & 2015/11/28& 18443 & 5.9 & Y&-\\
    SDSS J081333.32+350810.8 & 123.3889 & 35.1363 & ACIS-S & 4.929 & 2015/12/19 & 18444 & 6.0 &Y&-\\
    SDSS J123142.17+381658.9 & 187.9257 & 38.2830 & ACIS-S & 4.115 & 2016/02/13 & 18445 & 6.0 &N&-\\
    SDSS J123726.26+651724.4 & 189.3594 & 65.2901 & ACIS-S & 4.301 & 2016/08/21 & 18446 & 7.9 &N&-\\
    SDSS J124230.58+542257.3 & 190.6274 & 54.3826 & ACIS-S & 4.750 & 2016/05/16 & 18447 & 4.9 &Y&-\\
    PMN J2314+0201 & 348.7030 & 2.0309 & ACIS-S & 4.110 & 2016/01/15 & 18448 & 5.9 &Y&2\\
\hline
\multicolumn{8}{c}{Archival Data Objects} \\
\hline
    SDSS J083549.42+182520.0 & 128.9559 & 18.4222&  XRT & 4.412 & 2017/01/10 -- 2017/05/25 & 00087221001 & 45.8 & N&-\\
    SDSS J102107.57+220921.4 & 155.2816 & 22.1560 & EPIC-pn & 4.262 & 2008/05/30 & 0406540401 & 8.1 &N&-\\
    SDSS J111323.35+464524.3  &  168.3473 &  46.7568 & XRT & 4.468 & 2016/07/05 -- 2016/07/20 & 00703176000 & 52.9 &N&-\\
    SDSS J134811.25+193523.6 & 207.0469 & 19.5899&  XRT & 4.404 & 2017/11/29 -- 2018/01/15 & 00087542001 & 46.6 &Y&-\\
    SDSS J153533.88+025423.3 & 233.8912 & 2.9065 &  XRT & 4.388 & 2017/01/06 -- 2017/01/26 & 00087222001 & 26.4 &Y&-\\
    SDSS J160528.21+272854.4 & 241.3675 & 27.4818 & EPIC-pn & 4.024 & 2011/05/01 & 0655571401 & 11.0 &N&-\\
    SDSS J161216.75+470253.6 & 243.0698 & 47.0482 &XRT & 4.350 & 2017/11/08 -- 2017/12/13 & 00088204001 & 48.7&N&-\\
    PMN J2134$-$0419&323.5501& $-4.3194$ & XRT & 4.346 & 2013/06/16 -- 2013/06/20 & 00032624001 & 25.1 & Y&2\\
    SDSS J222032.50+002537.5& 335.1354 & 0.4271 & XRT & 4.220 & 2013/07/01 -- 2013/08/29 & 00032626001 & 43.5 &Y&-\\
\hline
\end{tabular}
\begin{tablenotes}
\item[a] Redshifts for objects in the SDSS DR7 quasar catalog and the SDSS DR14 quasar catalog
    are from \citet{Hewett2010} and \citet{Paris2017}, respectively. Redshifts for other objects are from NED.
\item[b]
    We merged multiple observations of the same target for archival {\it Swift}/XRT data,
    while only the first observation is listed in the table.
    The full observation IDs are
    00087221001--00087221023 for SDSS J083549.42+182520.0,
    00703176000--00703176011 for SDSS J111323.35+464524.3,
    00087542001--00087542016 for SDSS J134811.25+193523.6,
    00087222001--00087222007 for SDSS J153533.88+025423.3,
    00032624001--00032624003 for PMN J2134$-$0419,
    00087543001--00087543018 (excluding 00087543013 {because it lacks PC-mode exposures}) for SDSS J161216.75+470253.6,
    and 00032626001--00032626005 for SDSS J222032.50+002537.5.
\item[c] For archival XRT data, this column refers to the LIVETIME from the merged event lists.
    For archival EPIC data, this column refers to the LIVETIME of the EPIC-pn CCD on which the source is detected, after filtering background flares.
\item[d] {This column indicates whether the quasar is included in the flux-limited (FL) sample or not.}
\item[e] References. (1) \cite{Amirkhanyan2006}; (2) \cite{Hook2002}.
\end{tablenotes}
\end{threeparttable}
\end{table*}

\begin{figure}
\centering
\includegraphics[width=0.45\textwidth, clip]{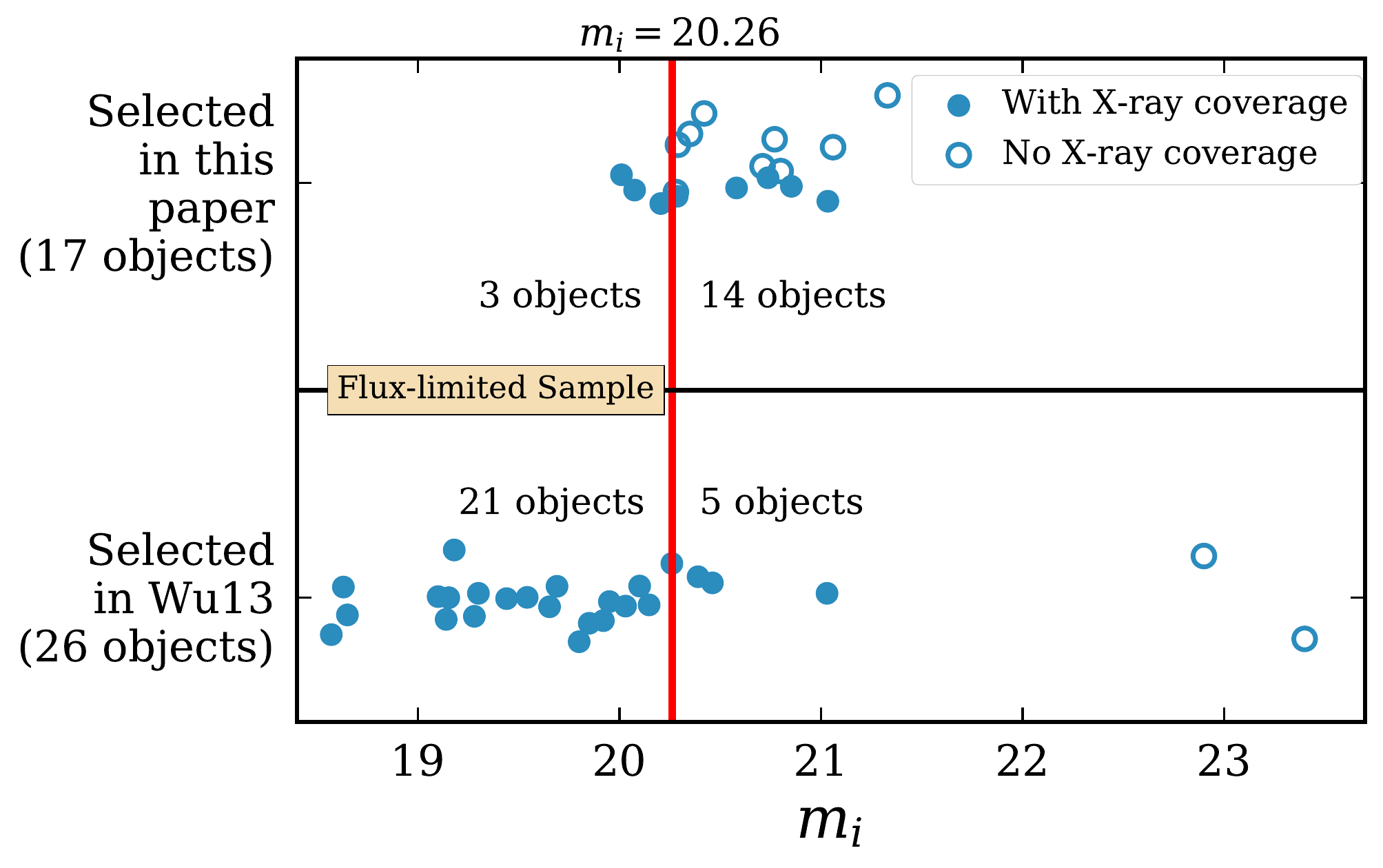}
\caption{The $m_i$ of the 43 HRLQs that were selected in this paper
(17 objects) and in Wu13 (26 objects).
The blue solid circle are the objects with sensitive \mbox{X-ray} coverage, while blue open
circles are the objects without sensitive X-ray coverage.
The vertical red line marks the magnitude cut for the flux-limited sample.
Object locations along the vertical axis are only used to distinguish between the objects selected by Wu13 and in this paper.
Additionally, each data point is also randomly perturbed in the vertical direction to avoid overlapping.}
\label{fig:sample}
\end{figure}

\begin{figure*}
\centering
\includegraphics[width=0.9\textwidth, clip]{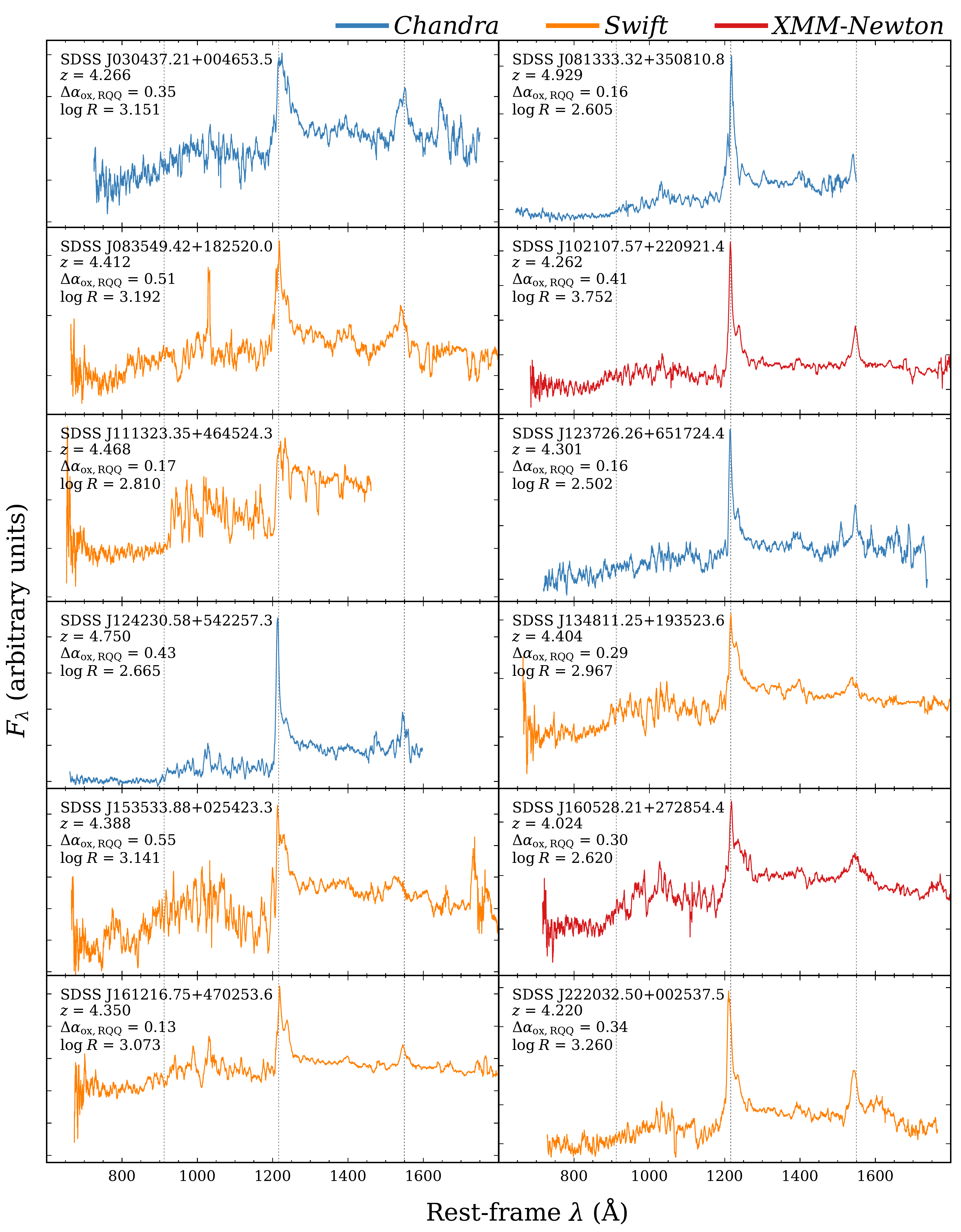}
\caption{The rest-frame UV spectra of the HRLQs that are in the SDSS quasar catalogs, ordered by RA.
The object name, redshift ($z$), $\Delta\alpha_{\rm ox, RQQ}$ (the difference
between the measured value of $\alpha_{\rm ox}$ and the expected $\alpha_{\rm ox,RQQ}$, see the description of
Column~16 in Section~\ref{sec:longTable}), and radio-loudness parameter ($\log R$) are shown in the top-left corner in each panel.
The spectra do not show strong dependence on $\Delta\alpha_{\rm ox,RQQ}$, $z$, or $\log R$.
We have plotted the spectra with different colors according to their X-ray data as labeled,
where the {\it Chandra} Cycle 17 objects are blue.
The $y$-axis is in linear scale with arbitrary units.
Each spectrum has been smoothed using a 21-pixel boxcar filter. 
Two emission lines (Ly$\alpha$ $\lambda$1216 and C {\sc iv} $\lambda$1549) and the Lyman limit have been labeled with
the dotted vertical lines.
Similar spectra can be found in Fig.~3 of Wu13 for the Wu13 objects.}
\label{fig:spec}
\end{figure*}

\begin{figure*}
\centering
\includegraphics[width=0.6\textwidth, clip]{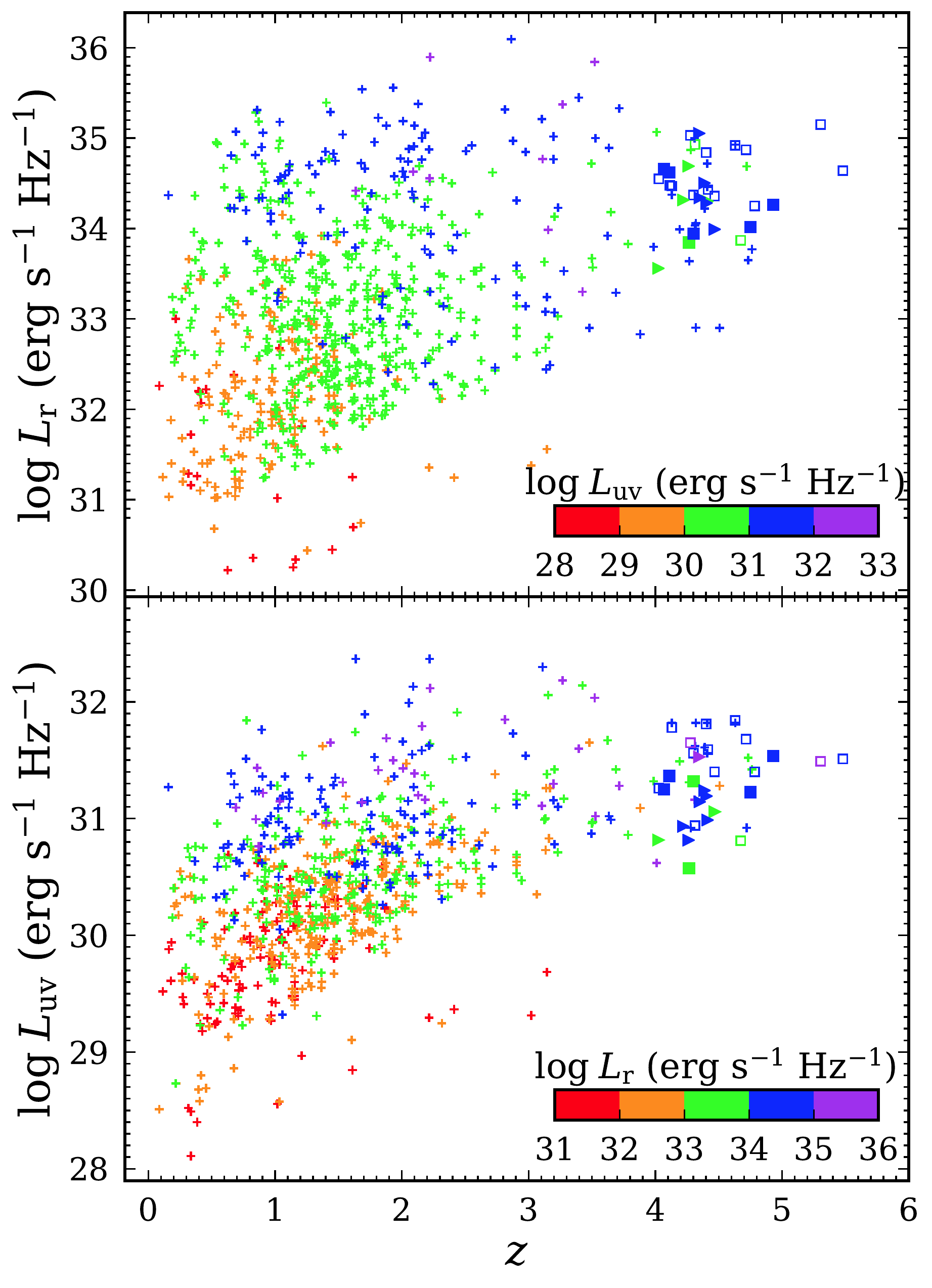}
\caption{The radio (rest-frame 5 GHz; upper panel) and UV (rest-frame 2500 \angstrom; lower panel) 
luminosities, plotted against redshift.
The filled squares and triangles are the {\it Chandra} Cycle 17 objects and archival-data objects, respectively.
The open squares are the high-redshift HRLQs of Wu13.
The plus signs are the radio-loud and radio-intermediate objects in the full sample of Miller11.
The upper and lower panels are color-coded
based on UV and radio luminosity, respectively.
Due to selection on $R$ and $m_i$, our sample and the sample
of Wu13 are composed of among the most-luminous objects in both the radio and UV bands.}
\label{fig:LuvLr-z}
\end{figure*}

\begin{table*}
\centering
    \caption{HRLQs at $z>4$ without vailable sensitive archival X-ray data.}
\label{tab:suplist}
\begin{threeparttable}[b]
\begin{tabular}{lccccccc}
\hline
\hline
Object Name & R.A. (J2000) & Dec. (J2000) & $z$ & $m_i$ & $M_i$ & $f_{1.4\;\rm GHz}$ & $\log R$\\
& (deg) & (deg) & &  & & (mJy)&  \\
\hline
SDSS J082511.60+123417.2  &  126.2984 &  12.5715 & 4.378 & 20.71 & $-$26.46 &  16.7 & 2.66\\
SDSS J094004.80+052630.9\tnote{a}  &  145.0200 &   5.4419 & 4.503 & 20.80 & $-$26.44 &  55.7 & 3.22\\
SDSS J104742.57+094744.9  &  161.9274 &   9.7958 & 4.252 & 20.29 & $-$26.77 &  18.9 & 2.58\\
SDSS J115605.44+444356.5  &  179.0227 &  44.7324 & 4.310 & 21.06 & $-$26.08 &  66.2 & 3.41\\
SDSS J125300.15+524803.3  &  193.2506 &  52.8009 & 4.115 & 21.33 & $-$25.66 &  55.9 & 3.47\\
SDSS J140025.40+314910.6\tnote{a}  &  210.1059 &  31.8196 & 4.640 & 20.28 & $-$26.89 &  20.2 & 2.61\\
SDSS J153830.71+424405.6  &  234.6280 &  42.7349 & 4.099 & 20.77 & $-$26.18 &  11.7 & 2.58\\
SDSS J154824.01+333500.1\tnote{a}  &  237.1001 &  33.5834 & 4.678 & 20.35 & $-$26.80 &  37.6 & 2.93\\
SDSS J165539.74+283406.7  &  253.9156 &  28.5685 & 4.048 & 20.42 & $-$26.51 &  23.0 & 2.73\\
\hline
\end{tabular}
\begin{tablenotes}
\item[a] {\it Chandra}/ACIS observations have been conducted or scheduled for SDSS J094004.80+052630.9, SDSS J140025.40+314910.6, and SDSS J154824.01+333500.1.
    Their X-ray data will become public after their proprietary periods.
\end{tablenotes}
\end{threeparttable}
\end{table*}

We started with a primary sample that was selected by Wu13 from the SDSS quasar
catalog Data Release 7 \citep[DR7; covering 9380 deg$^2$ of sky area;][]{Schneider2010} and NED.\footnote{\url{https://ned.ipac.caltech.edu/}}
They have utilized the 1.4~GHz NRAO VLA Sky Survey \citep[NVSS;][]{Condon1998},
which has provided homogeneous radio coverage for the full sky area of $\approx$~34,000~$\deg^2$ north of $\delta=-40\degree$.
For high-$z$ RLQs identified in current wide-field optical/UV surveys (i.e. $m_i\lesssim21$),
if an object satisfies the HRLQ criterion of $\log R>2.5$,
it should have been detected by the NVSS given its sensitivity
($\approx$~2.5~mJy).\footnote{We hereafter refer to objects at $z>4$ as high-redshift/high-$z$ objects and objects at $z<4$ as low-redshift/low-$z$ objects.}
Among the resulting sample of 26 HRLQs,\footnote{SDSS J003126.79+150739.5 and SDSS J123142.17+381658.9
are in Table~2 of Wu13. However, they do not satisfy the criterion of $\log R>2.5$
if we take their rest-frame 2500 \angstrom\ luminosities from spectral fitting \citep{Shen2011} instead of $m_i$.
We have thus revised the Wu13 sample from 28 objects to 26 objects.}
17 with sensitive X-ray coverage
(typically reaching $F_{\rm X}\approx10^{-14}$ erg cm$^{-2}$ s$^{-1}$ or better in the observed-frame 0.5--2 keV band)
have been studied in Wu13 while another two were studied by \cite{Sbarrato2015}.
The other five objects (see Table~2 of Wu13) with $m_i<21$
and lacking sensitive X-ray coverage were awarded {\it Chandra} time in Cycle~17.
The remaining two objects are fainter than $m_i=21$.
See Table~\ref{tab:obsLog} for the {\it Chandra} Cycle~17
observation log.\footnote{SDSS J003126.79+150739.5 and SDSS J123142.17+381658.9 were also awarded {\it Chandra} time in Cycle 17.
We analyze their X-ray data and report the results in Table~\ref{tab:photometry},
but will not show them in the figures or include them in the statistical tests.}

We furthermore searched in the SDSS quasar catalog Data Release 14 \citep[DR14;][]{Paris2017} for HRLQs at $z=$ 4.0--5.5,
and found another 16 HRLQs that were matched to the Faint Images of the
Radio Sky at Twenty-centimeters survey \citep[FIRST;][]{Becker1995}, which is designed to coincide with the primary region of sky covered by the SDSS.
Since the FIRST survey has a detection limit of $\approx$1~mJy,
all additional HRLQs in the SDSS quasar catalog DR14 with $m_i<21$ can be detected by FIRST if they satisfy
the criterion of $\log R>2.5$.\footnote{The typical radio fluxes of the high-$z$ HRLQs are $f_{\rm 1.4GHz}\ge20$ mJy.}
Another high-$z$ HRLQ, B3 0254+434 \citep{Amirkhanyan2006}, was selected in NED using the same method as Wu13.
See Section~\ref{sec:longTable} for details on the calculations of optical and radio luminosities and
radio-loudness parameters using optical and radio fluxes.
We retrieved the available sensitive archival X-ray
observations from HEASARC \footnote{\url{https://heasarc.gsfc.nasa.gov/cgi-bin/W3Browse/w3browse.pl}} of these new objects.
Five high-$z$ HRLQs (SDSS J083549.42+182520.0, SDSS J111323.35+464524.3, SDSS J134811.25+193523.6,
SDSS J153533.88+025423.3, and SDSS J161216.75+470253.6) have
useful deep ($\gtrsim25$ ks) {\it Swift} \mbox{X-ray} observations.
Two more (SDSS J102107.57+220921.4 and SDSS J160528.21+272854.4) are matched
with the {\it XMM-Newton} serendipitous-source catalog 3XMM-DR8 \citep{Rosen2016}.
See Table~\ref{tab:obsLog} for the observation log of the relevant {\it Swift} and {\it XMM-Newton}
archival data.\footnote{PMN~J2134$-$0419 and SDSS~J222032.50+002537.5 were selected by Wu13,
and their archival X-ray data have been analyzed by \cite{Sbarrato2015};
we analyzed the same data here to maintain consistency with the other objects.}
B3 0254+434 was awarded {\it Chandra} time in Cycle~17.
The rest of the objects that
lack publicly released sensitive archival X-ray observations are listed in Table~\ref{tab:suplist}.

We plotted the apparent $i$-band magnitudes and the
X-ray coverage of all the HRLQs that were selected in this
paper and in Wu13 in Fig.~\ref{fig:sample}, where the objects
with sensitive X-ray coverage were plotted as blue dots,
while objects without sensitive X-ray coverage were plotted as blue circles.
We here define our flux-limited high-$z$ sample by applying
an optical flux limit of $m_i\le20.26$.
The 24 HRLQs that satisfy this flux cut all have sensitive X-ray coverage,
among which 21 were selected in Wu13 and
three (B3 0254+434, SDSS J134811.25+193523.6, and SDSS J153533.88+025423.3)
were selected in this paper.
In comparison with Wu13, our flux-limited sample is not only
larger (twice as large) but also complete in its X-ray coverage (24/24 vs. 12/15),
thus suffering less from selection biases. For comparison, the optical flux limit of the Wu13 flux-limited sample
was $m_i=20$.

We have plotted the rest-frame UV spectra for the members of
our sample of HRLQs that are in the SDSS quasar catalogs in Fig.~\ref{fig:spec}.
It is apparent from their spectra that all of them are broad-line quasars, instead of BL~Lac objects;
the observed emission from the accretion disk and broad emission-line region
in the optical/UV is free from strong contamination by boosted jet emission.
The rest-frame UV spectra of the objects that were not in
the SDSS quasar catalogs can be found in \cite{Hook2002} for PMN J2134$-$0419 and PMN J2314+0201
and \cite{Amirkhanyan2006} for B3~0254+434.
The spectra of these 3 object show features of \mbox{broad-line} quasars as well.
We also plotted the radio and optical/UV luminosities of high-$z$ HRLQs against general RLQs in Fig.~\ref{fig:LuvLr-z};
their monochromatic luminosities are among the highest in both the radio
and optical/UV bands, with our sample extending to a slightly fainter range than Wu13 in the optical/UV.

RLQs with extended radio morphologies have systematically larger radio-loudness parameters
and also more powerful radio cores than quasars with compact radio morphologies \hbox{\citep[e.g.][]{Lu2007}}.
Our selection of HRLQs based on high $R$ thus should not cause a bias toward including quasars with
\mbox{core-only} morphologies or low intrinsic jet/core radio flux ratios, unless the cores
dominate the radio fluxes for quasars with extended morphologies or RLQs jets evolve with redshift.
High-redshift RLQs usually show compact radio morphology with few having
apparently extended structures, which is probably due to the steeper radio slope ($\alpha_{\rm r}<-0.5$)
and the cosmological surface brightness dimming of diffuse radio emission, i.e., $(1+z)^{-4}$.
Fifteen out of the 17 objects (except for B3 0254+434 and SDSS J1237+6517) in
table~\ref{tab:obsLog} are within the footprint of the FIRST survey,
and thirteen of them only show unresolved radio cores ($<5''$, or $<35$ kpc).
The remaining two (SDSS J0813+3508 and SDSS J2220+0025) are resolved
into multiple components \hbox{\citep[][]{Becker1995, Hodge2011}}
and have a linear extent of $\approx10''$ ($\approx70$ kpc).
The radio flux of the extended component of SDSS J0813+3508 is about half that of the core,
while the extended radio component is brighter than the core for SDSS J2220+0025.
Several quasars in Table~\ref{tab:obsLog} have been observed using Very Long Baseline Interferometry (VLBI).
Specifically, SDSS J$0813+3508$ and SDSS J$1242+5422$ were observed by \cite{Frey2010} at 1.6 GHz and 5 GHz,
and PMN J$2134-0419$ and SDSS J$2220+0025$ were observed by \mbox{\cite{Cao2017}} at 1.7 GHz and 5 GHz.
These observations can resolve structures on the scale of 1.2--25 milliarcseconds ($\approx$8--160 pc).
At such small scales, these quasars are often mildly resolved and show a compact core with a one-sided jet (SDSS J$0813+3508$ and
PMN J$2134-0419$) or an unresolved core (SDSS J$1242+5422$ and SDSS J$2220+0025$).
Note that VLBI observation of PMN J$2314-0419$ shows evidence of strong Doppler boosting \citep[][]{Cao2017}.

\section{X-ray Data analyses and multi-wavelength properties}
\label{sec:reduction}
In the below, we define the soft band, hard band, and full band to be 0.5--2 keV, 2.0--8.0 keV, and 0.5--8 keV in the observed frame, respectively.
\subsection{{\it Chandra} data analyses}
Eight RLQs were targeted with the Advanced CCD Imaging Spectrometer \citep[ACIS;][]{Garmire2003} onboard {\it Chandra}, using the back-illuminated S3 chip.
The {\it Chandra} data (see Table~\ref{tab:obsLog}) were first reprocessed using the standard CIAO (v4.9) routine {\sc chandra\_repro} and the latest
CALDB (v4.7.3). X-ray images and exposure maps were then generated using {\sc fluximage} in the three observed bands,
where the effective energy that was used to calculate the exposure map was chosen to be the geometric mean of the limits of each band.
All of the sources were detected by {\sc wavdetect} \citep{Freeman2002} in at least two bands
with a detection threshold $10^{-6}$ and wavelet scales of 1, $\sqrt{2}$, 2, $2\sqrt{2}$, and 4 pixels.
We performed statistical tests on the X-ray images and found no extended structure or large-scale jets.
Furthermore, we constrained any extended X-ray jets to be $\gtrsim$ 3--25 times fainter than the cores.
(see details in Appendix~\ref{sec:imgTest}).
Raw source and background counts were extracted using {\sc dmextract}.
The source region was a circle with a radius of $2.0''$, centred at the X-ray position from {\sc wavdetect},
and the background region was a concentric annulus with an inner radius of $5.0''$ and an outer radius of $20.0''$.
The offset between the X-ray position and optical position of each source ranges from $0.2''$ to $0.7''$.
All the background regions are free of X-ray sources except for that of SDSS J123142.17+381658.9,
in which we have excluded a source detected by {\sc wavdetect}.
The circular source region encloses $\approx95.9\%$ of the total energy at 1 keV
and $\approx90.6\%$ of the total energy at 4 keV.\footnote{\url{http://cxc.harvard.edu/ciao/ahelp/src\_psffrac.html}}
We also extracted source and background spectra using
{\sc specextract},\footnote{We used a larger background region to enclose more background events in extracting background spectra.}
which simultaneously produces response matrix files (RMFs) and ancillary response files
(ARFs).\footnote{Including both aperture-corrected and uncorrected ARFs.}

\begin{table*}
\centering
\caption{X-ray net counts, hardness ratio, and effective photon index.}
\label{tab:photometry}
\begin{threeparttable}[b]
\begin{tabular}{lccccc}\hline\hline
    Object Name & \multicolumn{3}{c}{Net X-ray Counts}& Band Ratio\tnote{a} & $\Gamma_{\rm X}$ \\
\hline
    &Full Band & Soft Band & Hard Band & & \\
    & (0.5--8 keV)& (0.5--2 keV)& (2--8 keV)& & \\
\hline
\multicolumn{6}{c}{{\it Chandra} Cycle 17 Objects} \\
\hline
SDSS J003126.79+150739.5&$14.8^{+4.4}_{-3.6}$&$12.5^{+4.0}_{-3.3}$&$2.1^{+2.0}_{-1.3}$ & $0.17^{+0.09}_{-0.14}$ & $2.43^{+1.20}_{-0.33}$ \\
    B3 0254+434 &$110.6_{-10.5}^{+11.2}$ & $60.4_{-7.6}^{+8.3}$ & $50.7_{-7.1}^{+7.8}$ & $0.84_{-0.20}^{+0.13}$ & $1.44_{-0.11}^{+0.21}$ \\
SDSS J030437.21+004653.5&$10.5^{+3.8}_{-3.0}$&$9.4^{+3.5}_{-2.8}$&$<4.1$ & $<0.44$ & $>1.79$ \\
SDSS J081333.32+350810.8& $28.7_{-5.2}^{+5.9}$ & $16.6_{-3.8}^{+4.6}$ & $12.1_{-3.3}^{+4.0}$ & $0.73_{-0.32}^{+0.19}$ & $1.35_{-0.17}^{+0.44}$ \\
SDSS J123142.17+381658.9&$25.3^{+5.5}_{-4.9}$&$15.6^{+4.4}_{-3.7}$&$9.8^{+3.7}_{-3.0}$ & $0.63^{+0.21}_{-0.29}$ & $1.37^{+0.44}_{-0.21}$ \\
SDSS J123726.26+651724.4&$13.6^{+4.2}_{-3.5}$&$10.4^{+3.7}_{-3.0}$&$3.1^{+2.4}_{-1.6}$ & $0.30^{+0.16}_{-0.21}$ & $1.92^{+0.90}_{-0.30}$ \\
SDSS J124230.58+542257.3&$15.8^{+4.5}_{-3.8}$&$13.5^{+4.1}_{-3.5}$&$2.1^{+2.0}_{-1.3}$ & $0.16^{+0.09}_{-0.12}$ & $2.38^{+1.13}_{-0.33}$ \\
PMN J2314+0201&$43.5^{+7.2}_{-6.5}$&$25.0^{+5.5}_{-4.8}$&$18.6^{+4.9}_{-4.2}$ & $0.74^{+0.20}_{-0.25}$ & $1.33^{+0.30}_{-0.17}$ \\
\hline
\multicolumn{6}{c}{Archival Data Objects} \\
\hline
SDSS J083549.42+182520.0 & 205.2$^{+17.1}_{-16.9}$ & 135.3$^{+13.9}_{-13.3}$& 69.8$^{+10.5}_{-10.1}$& 0.52$^{+0.08}_{-0.09}$& 1.56$^{+0.15}_{-0.10}$ \\
SDSS J102107.57+220921.4 & 49.8$^{+13.9}_{-14.2}$ & 37.7$^{+9.6}_{-9.4}$ & $<26.6$ & $<0.71$ &  $>1.30$ \\
SDSS J111323.35+464524.3 & $33.3_{-7.5}^{+8.2}$ & $21.9_{-5.9}^{+6.6}$ & $11.4_{-4.4}^{+5.1}$ & $0.52_{-0.29}^{+0.17}$ & $1.51_{-0.21}^{+0.59}$ \\ 
SDSS J134811.25+193523.6 & 78.4$^{+12.7}_{-12.0}$ & 52.5$^{+10.5}_{-9.8}$ &  26.0$^{+7.4}_{-6.8}$ & 0.49$^{+0.14}_{-0.17}$ &  1.56$^{+0.32}_{-0.18}$ \\
SDSS J153533.88+025423.3 & 324.3$^{+21.7}_{-21.5}$ & 185.7$^{+16.9}_{-16.3} $ &  138.6$^{+14.2}_{-13.7} $ & 0.75$^{+0.08}_{-0.10}$ & 1.32$^{+0.11}_{-0.08}$\\
SDSS J160528.21+272854.4 & $<61.5$ & 26.8$^{+10.9}_{-10.8}$ & $<28.4$ &  - & - \\
SDSS J161216.75+470253.6 & $23.1_{-6.0}^{+6.6}$ & $19.1_{-5.0}^{+5.7}$ & $<10.1$ & $<0.53$ & $>1.50$ \\
PMN J2134$-$0419& $70.7_{-10.7}^{+11.2}$ & $49.3_{-8.5}^{+9.1}$ & $21.4_{-6.2}^{+7.0}$ & $0.43_{-0.18}^{+0.12}$ & $1.70_{-0.18}^{+0.38}$ \\
SDSS J222032.50+002537.5 & $44.8_{-9.2}^{+9.7}$ & $37.6_{-7.6}^{+8.2}$ & $<15.8$ & $<0.42$ & $>1.76$ \\
\hline
\end{tabular}
    \begin{tablenotes}
    \item[a] {The band ratio here refers to the number of hard-band counts divided by the number of the soft-band counts.}
    \end{tablenotes}
\end{threeparttable}
\end{table*}

\subsection{{\it Swift} data analyses}
Data reduction of the {\it Swift}/X-ray Telescope \citep[XRT;][]{Burrows2005} observations
was performed using standard routines in {\sc ftools} integrated
in HEASoft (v6.21).\footnote{\url{https://heasarc.nasa.gov/lheasoft/}}
Each HRLQ has multiple observations (see Table~\ref{tab:obsLog}).
For each observation, the cleaned event list and exposure map were created
using {\sc xrtpipeline} and {\sc xrtexpomap}, respectively.
We only used XRT data in photon-counting (PC) mode.
The event lists and exposure maps of different observations were
then merged using {\sc xselect} and {\sc ximage}, respectively.
We extracted photons in the three bands
from circular regions centred at the source positions
with radii of $\sim$60$''$ except for SDSS J111323.35+464524.3 and SDSS J161216.75+470253.6,
for which we have adopted a radius of 25$''$ to avoid contamination by nearby sources.
These source-extraction regions enclose $\sim$\mbox{80--90}\% (73\% for SDSS J111323.35+464524.3 and 80\% for SDSS J161216.75+470253.6) of the total energy at 1 keV.
Photons from circular source-free regions of radii that are more than
twice as large as the source region
were extracted to estimate the background level.
We also extracted source and background spectra using {\sc xselect},
and created ARFs using {\sc xrtmkarf},
which simultaneously provides the corresponding RMFs.

\subsection{{\it XMM-Newton} data analyses}
SDSS J102107.57+220921.4 and SDSS J$160528.21+272854.4$ were
serendipitously observed by {\it XMM-Newton} (see Table~\ref{tab:obsLog}).\footnote{These two serendipitously
observed HRLQs are near to the edges of the EPIC-pn CCDs, with off-axis angles of $\approx$15$'$--17$'$.}
Data reduction was performed using SAS (v16.1.0) and the latest Current Calibration Files (as of 2018 March).
We only utilized the data from the pn CCDs of the European Photon Imaging Camera \citep[EPIC-pn;][]{Struder2001} onboard {\it XMM-Newton}.
The data were reprocessed and cleaned using {\sc epproc},
and high-background flaring periods were filtered using {\sc espfilt}.
We created images and exposure maps using {\sc evselect} and {\sc eexpmap}
and then performed source detection using {\sc eboxdetect}.\footnote{\url{https://www.cosmos.esa.int/web/xmm-newton/sas-thread-src-find-stepbystep}}
Both targets were detected in the full and soft bands,
and the offsets between the X-ray positions and optical positions are $\sim$1$''$--2$''$ \citep{Rosen2016}.
We extracted photons from source regions that are defined by a circle with a radius of 40$''$,
centred at the optical position.
Background photons were extracted from source-free circular regions on
the same chips, with radii of 60$''$ and 50$''$
for SDSS J102107.57+220921.4 and SDSS J160528.21+272854.4, respectively.
The encircled-energy fraction is $\approx86$\% for both sources at 1 keV, which
is calculated using the point spread function (PSF) images created by {\sc psfgen}.
We also extracted spectra using {\sc evselect} and created corresponding RMFs and ARFs using {\sc rmfgen} and {\sc arfgen}, respectively.

\subsection{Source detection and photometry}
In the below, analyses of {\it Chandra}/ACIS, {\it Swift}/XRT, and {\it XMM-Newton}/EPIC data were conducted in a unified way.
Using the raw source and background event counts,
we calculated the binomial no-source probability (referred to as $P_{\rm B}$ in this paper; \hbox{\citealt{Weisskopf2007}})\footnote{$P_{\rm B}$
is the chance probability of observing a signal no weaker than the source counts under the null hypothesis
that there is no source in the source-extraction region.  Thus, it is essentially a $p$-value.}
to test the significance of the source signal in each band, and took cases with $P_{\rm B}\le0.01$ as detections.
We calculated net counts from the HRLQs (with aperture corrections) and their $1\sigma$ intervals
using {\sc aprates}\footnote{\url{http://cxc.harvard.edu/ciao/ahelp/aprates.html} and \url{http://cxc.harvard.edu/ciao/threads/aprates/}} within CIAO.
For each band without a detection ($P_{\rm B}>0.01$),
we gave a 90\% confidence upper limit \citep[][]{Kraft1991}.

We then proceeded by calculating the hardness ratio of each source.
The 68\% bounds of hardness ratio were calculated using the Bayesian approach of \cite{Park2006}.
Using the response files and {\sc modelflux} (another CIAO routine),\footnote{See \url{http://cxc.harvard.edu/ciao/ahelp/modelflux.html} and \url{http://cxc.harvard.edu/ciao/why/pimms.html}. Since
we have obtained response files for all the observations,
we also use {\sc modelflux} with {\it XMM-Newton} and {\it Swift} data.}
we calculated the expected HRs of Galactic-absorbed power-law spectra with a range of photon indices,
from which we calculated the effective power-law photon index ($\Gamma_{\rm X}$) of each source.
The results of the photometry are listed in Table~\ref{tab:photometry}.
SDSS J124230.58+542257.3 has a noticeably large effective photon index, and deeper
X-ray observations in the future might help to improve our understanding
of it.\footnote{SDSS J003126.79+150739.5 with $\log R=2.44$ does not strictly satisfy our definition for HRLQs, but it has an even steeper effective photon index.}

\begin{landscape}
\begin{table}
\caption{X-ray, optical/UV, and radio properties.}
\label{tab:GammaWithPhotometry}
\begin{threeparttable}[b]
\begin{tabular}{lcccccccccccccccc}\hline\hline
Object Name&$m_i$&$M_i$&$N_{\rm H}\tnote{a}$&C.R.\tnote{b}&$F_{\rm X}$\tnote{c}&$f_{\rm 2 keV}$\tnote{d}&$\log L_{\rm X}$\tnote{e}&$\Gamma_{\rm X}$\tnote{f}&$f_{2500 \angstrom}$\tnote{g}&$\log L_{2500 \angstrom}$\tnote{h}&$\alpha_{\rm r}$\tnote{i}&$\log L_{\rm r}$\tnote{j}&$\log R$&$\alpha_{\rm ox}$&$\Delta\alpha_{\rm ox, RQQ}$\tnote{k}&$\Delta\alpha_{\rm ox, RLQ}$\tnote{l}\\
(1)&(2)&(3)&(4)&(5)&(6)&(7)&(8)&(9)&(10)&(11)&(12)&(13)&(14)&(15)&(16)&(17) \\\hline
\multicolumn{17}{c}{{\it Chandra} Cycle 17 Objects} \\
\hline
SDSS J003126.79+150739.5&19.99&$-$27.16&$4.43$&$2.33^{+0.74}_{-0.61}$&1.81&21.17&45.61&$2.43^{+1.20}_{-0.33}$ & 0.92 &31.50&0.61&34.07&2.44&$-$1.40&0.31&0.10 \\
B3 0254+434&20.01&$-$27.24&$13.45$&$10.96^{+1.50}_{-1.38}$&7.98&34.94&46.14&$1.44^{+0.21}_{-0.11}$ & 0.56 &31.25&0.06&34.66&3.29&$-$1.23&0.44&0.15 \\
SDSS J030437.21+004653.5&20.15&$-$27.09&$7.26$&$1.60^{+0.60}_{-0.48}$&1.10&7.03&45.35&$>1.79$ & 0.11 &30.57&-&33.85&3.15&$-$1.23&0.35&0.11 \\
SDSS J081333.32+350810.8&19.15&$-$28.30&$4.91$&$2.78^{+0.76}_{-0.64}$&1.72&7.27&45.63&$1.35^{+0.44}_{-0.17}$ & 0.81 &31.54&$-$0.60&34.26&2.61&$-$1.55&0.16&$-$0.07 \\
SDSS J123142.17+381658.9&20.12&$-$26.88&$1.27$&$2.58^{+0.73}_{-0.61}$&1.52&6.22&45.43&$1.37^{+0.45}_{-0.19}$ & 1.13 &31.56&-&33.82&2.14&$-$1.63&0.08&$-$0.10 \\
SDSS J123726.26+651724.4&20.46&$-$26.63&$2.03$&$1.32^{+0.47}_{-0.38}$&0.90&6.59&45.28&$1.92^{+0.90}_{-0.30}$ & 0.61 &31.32&-&33.94&2.50&$-$1.52&0.16&$-$0.05 \\
SDSS J124230.58+542257.3&19.65&$-$27.63&$1.55$&$2.77^{+0.84}_{-0.71}$&1.73&21.96&45.71&$2.38^{+1.13}_{-0.33}$ & 0.42 &31.23&$-$0.56&34.01&2.67&$-$1.24&0.43&0.21 \\
PMN J2314+0201&19.54&$-$27.41&$4.82$&$4.25^{+0.93}_{-0.81}$&2.64&10.36&45.67&$1.33^{+0.33}_{-0.15}$ & 0.72 &31.36&$-$0.27&34.62&3.13&$-$1.47&0.21&$-$0.06 \\
\hline
\multicolumn{17}{c}{Archival Data Objects} \\
\hline
    SDSS J083549.42+182520.0&20.74&$-$26.47&$3.21$&$2.95_{-0.30}^{+0.29}$&6.17&31.74&46.11&$1.56^{+0.15}_{-0.11}$ & 0.27 &30.99&$-$0.20&34.30&3.19&$-$1.12&0.51&0.25 \\
    SDSS J102107.57+220921.4&21.03&$-$26.04&$2.03$&$4.65_{-1.16}^{+1.19}$&2.72&14.32&45.73&$>1.30$ & 0.19 &30.81&$-$0.17&34.69&3.75&$-$1.20&0.41&0.10 \\
    SDSS J111323.35+464524.3&20.58&$-$26.65&$1.31$&$0.63_{-0.12}^{+0.12}$&0.95&4.64&45.30&$1.51^{+0.59}_{-0.21}$ & 0.31 &31.06&$-$0.17&33.99&2.81&$-$1.47&0.17&$-$0.05 \\
    SDSS J134811.25+193523.6&20.21&$-$26.98&$1.93$&$1.13_{-0.21}^{+0.23}$&2.29&11.75&45.68&$1.56^{+0.34}_{-0.17}$ & 0.44 &31.19&$-$0.20&34.28&2.97&$-$1.37&0.29&0.04 \\
    SDSS J153533.88+025423.3&20.08&$-$27.10&$4.44$&$7.03_{-0.62}^{+0.64}$&15.13&59.82&46.48&$1.32^{+0.10}_{-0.09}$ & 0.49 &31.24&$-$0.31&34.50&3.14&$-$1.12&0.55&0.28 \\
    SDSS J160528.21+272854.4&20.85&$-$26.04&$3.94$&$2.44_{-0.98}^{+0.99}$&1.59&8.12&45.44&- & 0.21 &30.82&-&33.56&2.62&$-$1.31&0.30&0.10 \\
    SDSS J161216.75+470253.6&20.29&$-$26.86&$1.33$&$0.39_{-0.10}^{+0.12}$&0.79&4.19&45.21&$>1.50$ & 0.40 &31.14&$-$0.44&34.34&3.07&$-$1.53&0.13&$-$0.13 \\
    PMN J2134$-$0419&19.30&$-$27.84&$3.55$&$1.96_{-0.34}^{+0.36}$&4.19&24.66&45.94&$1.70^{+0.38}_{-0.18}$ & 0.96 &31.53&$-$0.23&35.05&3.40&$-$1.38&0.33&0.02 \\
    SDSS J222032.50+002537.5&19.95&$-$27.20&4.73&$0.86_{-0.17}^{+0.19}$&1.85&11.36&45.56&$>1.76$ & 0.26 &30.93&-&34.32&3.26&$-$1.29&0.34&0.07 \\
\hline
\end{tabular}
\begin{tablenotes}
\item[a] Galactic neutral hydrogen column density in units of 10$^{20}$ cm$^{-2}$.
\item[b] Count rate of the in the observed-frame 0.5--2 keV band, in units of 10$^{-3}$ s$^{-1}$.
\item[c] Galactic absorption-corrected flux in the observed-frame 0.5--2 keV band, in units of 10$^{-14}$ erg cm$^{-2}$ s$^{-1}$.
\item[d] Flux density at $2/(1+z)$ keV (extrapolated from the observed 0.5--8 keV X-ray emission), in units of 10$^{-32}$ erg cm$^{-2}$ s$^{-1}$ Hz$^{-1}$.
\item[e] The logarithm of the X-ray luminosity in the rest-frame 2--10 keV band, in units of erg s$^{-1}$.
\item[f] Effective X-ray power-law photon index.
\item[g] Flux density observed at 2500(1+$z$) \angstrom\ in units of 10$^{-27}$ erg cm$^{-2}$ s$^{-1}$ Hz$^{-1}$.
\item[h] Logarithm of the monochromatic UV luminosity at rest frame 2500 \angstrom\ in units of erg s$^{-1}$ Hz$^{-1}$.
\item[i] Radio spectral index calculated from observed 1.4 GHz and 5 GHz flux, defined as $f_\nu\propto\nu^{\alpha_r}$. 
    If a 5 GHz observation is absent, we take $\alpha_r=0$ in the following calculation.
    The radio spectral index of SDSS J0813+3508 is from \citet{Frey2010}.
\item[j] Logarithm of the monochromatic radio luminosity at rest-frame 5 GHz in units of erg s$^{-1}$ Hz$^{-1}$. 
\item[k] The difference between the measured $\alpha_{\rm ox}$ and the expected $\alpha_{\rm ox}$ for RQQs with similar UV luminosity, defined by Eq.~(3) of \citet{Just2007}.
\item[l] The difference between the measured $\alpha_{\rm ox}$ and the expected $\alpha_{\rm ox}$ for RLQs with similar UV and radio luminosities, defined by the $L_{2 \rm keV}$-$L_{2500 \angstrom}$-$L_{5\rm GHz}$ relation in Table~7 of \cite{Miller2011}.
\end{tablenotes}

\end{threeparttable}
\end{table}
\end{landscape}

\begin{table*}
\centering
\caption{HRLQs at $z>4$ that are analyzed in this paper and from Wu13 (32 objects in total).}
\label{tab:flux-limited}
\begin{threeparttable}[b]
\begin{tabular}{lcccccccccc}
\hline
\hline
    Object Name & $z$ & $m_i$& $M_i$ & $\log R$\tnote{a} & $\alpha_{\rm r}$\tnote{b} & $\alpha_\mathrm{ox}$ & $\Delta\alpha_\mathrm{ox,RQQ}$ & $\Delta\alpha_\mathrm{ox,RLQ}$  & Factor\tnote{c} & FL\tnote{d} \\
\hline
    \multicolumn{11}{c}{From this paper (15 objects)} \\
\hline
    B3 0254+434&4.067&20.01&$-$27.13&3.29&0.06&$-$1.23&0.44&0.15&2.46 & Y\\
    SDSS J030437.21+004653.5&4.266&20.15&$-$27.09&3.15&-&$-$1.23&0.35&0.11&1.93 & Y\\
    SDSS J081333.32+350810.8&4.929&19.15&$-$28.30&2.61&$-$0.60&$-$1.55&0.16&$-$0.07 & 0.66& Y\\
    SDSS J124230.58+542257.3&4.750&19.65&$-$27.63&2.67&$-$0.56&$-$1.24&0.43&0.21&3.53 & Y\\
    SDSS J134811.25+193523.6&4.404&20.20&$-$26.98&2.97&$-$0.20&$-$1.37&0.29&0.04&1.27 &Y\\
    SDSS J153533.88+025423.3&4.388&20.07&$-$27.10&3.14&$-$0.31&$-$1.12&0.55&0.28&5.36 &Y\\
    PMN J2134$-$0419&4.346&19.30&$-$27.82&3.40&$-$0.23&$-$1.38&0.33&0.02&1.13 &Y\\
    SDSS J222032.50+002537.5&4.220&19.95&$-$27.20&3.26&-&$-$1.35&0.31&0.04& 1.27 &Y\\
    PMN J2314+0201&4.110&19.54&$-$27.41&3.13&$-$0.27&$-$1.47&0.21&$-$0.06&0.70 &Y\\
    SDSS J083549.42+182520.0 & 4.412& 20.74 & $-$26.47 & 3.19 & $-$0.20 & $-$1.12 & 0.51 & 0.25 &4.48 &N \\
    SDSS J102107.57+220921.4 & 4.262& 21.03 & $-$26.04 & 3.75 & $-$0.17 & $-$1.20 & 0.41 & 0.10&1.82 & N\\
    SDSS J111323.35+464524.3 & 4.468& 20.58 & $-$26.65 & 2.81 & $-$0.17 & $-$1.53 & 0.11 & $-$0.05 & 0.74& N\\
    SDSS J123726.26+651724.4 & 4.301 & 20.46 & $-$26.63 & 2.50 & - & $-$1.52 & 0.16 & $-$0.05 & 0.74 & N \\
    SDSS J160528.21+272854.4 & 4.024& 20.85 & $-$26.04 & 2.61 & - & $-$1.31 & 0.30 & 0.10& 1.82 & N\\
    SDSS J161216.75+470253.6 & 4.350& 20.29 & $-$26.86 & 3.07 & $-$0.44 & $-$1.53 & 0.13 & $-$0.13&0.46& N\\
\hline
    \multicolumn{11}{c}{From Wu13 (17 objects)} \\
\hline
    PSS 0121+0347&4.130&18.57&$-$28.44&2.57&$-$0.33&$-$1.47&0.28&0.04&1.27& Y\\
    PMN J0324$-$2918&4.630&18.65&$-$28.61&2.95&0.30&$-$1.40&0.35&0.08&1.62 &Y\\
    PMN J0525$-$3343&4.401&18.63&$-$28.52&2.90&0.06&$-$1.17&0.58&0.31&6.42 &Y\\
    Q0906+6930&5.480&19.85&$-$27.76&3.01&0.17&$-$1.31&0.40&0.13&2.18 &Y\\
    SDSS J102623.61+254259.5&5.304&20.03&$-$27.50&3.54&$-$0.38&$-$1.31&0.39&0.07&1.52 &Y\\
    RX J1028.6$-$0844&4.276&19.14&$-$27.95&3.33&$-$0.30&$-$1.09&0.63&0.34&7.69 &Y\\
    PMN J1155$-$3107&4.300&19.28&$-$27.90&2.73&0.53&$-$1.36&0.36&0.12& 2.05&Y\\
    SDSS J123503.03$-$000331.7&4.673&20.10&$-$27.20&3.05&-&$-$1.22&0.39&0.16&2.61 &Y\\
    CLASS J1325+1123&4.415&19.18&$-$28.01&2.72&$-$0.09&$-$1.53&0.19&$-$0.05&0.74 &Y\\
    SDSS J141209.96+062406.9&4.467&19.44&$-$27.74&2.70&-&$-$1.51&0.18&$-$0.06&0.70 &Y\\
    SDSS J142048.01+120545.9&4.027&19.80&$-$27.18&3.05&$-$0.36&$-$1.34&0.33&0.06&1.43 &Y\\
    GB 1428+4217&4.715&19.10&$-$28.18&3.06&0.37&$-$0.93&0.80&0.52&22.6 &Y\\
    GB 1508+5714&4.313&19.92&$-$27.16&3.87&0.13&$-$0.96&0.67&0.34&7.69 &Y\\
    SDSS J165913.23+210115.8&4.784&20.26&$-$27.17&2.56&-&$-$1.39&0.30&0.07&1.52& Y\\
    PMN J1951+0134&4.114&19.69&$-$27.40&3.04&0.24&$-$1.23&0.45&0.20&3.32 &Y\\
    SDSS J091316.55+591921.6\tnote{e} & 5.122 & 20.39 & $-$27.03 & 2.72 & $-$0.67 & $-$1.76 & $-$0.09 & $-$0.32 & 0.15&N\\
    GB 1713+2148 & 4.011 & 21.42 & $-$25.53 & 4.50 & $-$0.30 & $-$1.16 & 0.42 & 0.05 &1.35 &N \\
\hline
\end{tabular}
    \begin{tablenotes}
    \item[a] The objects are sorted in ascending order of RA.
    \item[b] The radio spectral index that is calculated using observed-frame 1.4 GHz and 5 GHz flux densities.
    \item[c] The factor of X-ray enhancement calculated using $10^{\Delta\alpha_{\rm ox,RLQ}/0.3838}$.
    \item[d] This column indicates whether the object belongs to the flux-limited sample or not (see Section~\ref{sec:sample}). The quasars that
        are outside the flux-limited sample are appended below the flux-limited sample, also in ascending order of RA.
    \item[e] We consider SDSS J091316.55+591921.6 as an outlier among the quasar sample.
    \end{tablenotes}
\end{threeparttable}
\end{table*}

\subsection{X-ray, optical/UV, and radio properties}
\label{sec:longTable}
In Table~\ref{tab:GammaWithPhotometry}, we summarize the X-ray, optical/UV, and radio properties of our sample
of HRLQs, utilizing the results of our \mbox{X-ray} data analyses as well as SDSS and FIRST/NVSS surveys.
We explain the content of each column below:

\noindent Column (1): the name of the quasar.

\noindent Column (2): the apparent $i$-band magnitude of the quasar.

\noindent Column (3): the absolute $i$-band magnitude of the quasar.
The values are preferentially taken from SDSS quasar catalogs \citep{Schneider2010, Paris2017}.
For objects that are not in the quasar catalogs, we calculated $M_i$ from $m_i$ by correcting for the
Galactic extinction \citep{Schlafly2011} and using the K-correction in Section~5 of \cite{Richards2006}.

\noindent Column (4): the Galactic neutral hydrogen column density \citep{Dickey1990,Stark1992}.\footnote{\url{http://cxc.harvard.edu/toolkit/colden.jsp}}

\noindent Column (5): the count rate in the observed-frame soft X-ray band for the {\it Chandra} Cycle 17 objects.

\noindent Column (6): the observed X-ray flux in the soft band calculated using {\sc modelflux}, the effective power-law photon
index (see Column 9), and the instrumental response files.
The values have been corrected for Galactic absorption.

\noindent Column (7): following Wu13,
in this column we estimated the observed X-ray flux density at $2/(1+z)$~keV (i.e. rest-frame 2 keV),
corrected for Galactic absorption.
Note that, for objects at $z>4$, the rest-frame 2 keV X-rays are below the lower limit of our observed X-ray bands.
Thus, we have extrapolated their X-ray spectra using the effective power-law photon index (see Column 9) to lower energies.
Note that this is a relatively short extrapolation, generally a factor of $\lesssim1.5$ times below the lowest energy of our observed X-rays.

\noindent Column (8): the logarithm of the rest-frame 2--10 keV luminosity. 

\noindent Column (9): the effective power-law photon index in the X-ray band.
For the sources with only a lower limit or without estimation of $\Gamma_{\rm X}$, we have
adopted a typical value for RLQs \citep[$\Gamma_{\rm X}=1.6$; e.g.][]{Page2005},
and for SDSS J030437.21+004653.5 and SDSS J161216+470253.6 we have used their lower limits ($\Gamma_{\rm X}=1.79$ and 1.76) in the following analysis.
Within a reasonable range \citep[$\Gamma_{\rm X}=$ 1.4--1.9; e.g.][]{Page2005}, the value of $\Gamma_{\rm X}$ does not materially affect the results we presented below.

\noindent Column (10): the observed flux density at $2500(1+z)$~\angstrom\ (i.e. \mbox{rest-frame} 2500~\angstrom).
For objects in the SDSS DR7 quasar catalog, the values were taken from \cite{Shen2011}.
For other objects, the values were calculated from their $i$-band magnitude (Column 3).

\noindent Column (12): the radio spectral index $\alpha_{\rm r}$ ($f_\nu\propto\nu^{\alpha_{\rm r}}$) between
observed-frame 1.4 GHz and 5 GHz. We obtain 1.4 GHz flux densities from the FIRST or NVSS surveys.
The 5 GHz flux densities were mostly from the Green Bank 6-cm survey \citep{Gregory1996}.
We obtain the 5 GHz flux density of PMN J2134$-$0419 from the Parkes-MIT-NRAO survey \citep{Wright1994}.
We took the 5 GHz flux density of SDSS J124230.58+542257.3 from its
VLBI observation \citep{Frey2010}.
The radio counterpart of SDSS J081333.32+350810.8 has a close companion ($\sim$6$''$) in the FIRST catalog,
which cannot be identified in the optical and is likely to be associated with SDSS J081333.32+350810.8 as a jet or lobe.
We thus took the 1.4 GHz flux density of SDSS J081333.32+350810.8 as the sum of the
two radio sources from the FIRST catalog.
Since the radio companion is completely resolved in VLBI imaging,
we take $\alpha_{\rm r}=-0.6$ from \cite{Frey2010}.

\noindent Column (13): the logarithm of the monochromatic luminosity at rest-frame 5 GHz, in units of erg s$^{-1}$ Hz$^{-1}$.
We calculated $\log L_{\rm r}$ using the observed-frame 1.4 GHz flux
and radio spectral index ($\alpha_{\rm r}$; see Column 12).

\noindent Column (14): the logarithm of the radio-loudness parameter, given by
\begin{align}
    \log R &=  \log \bigg(\frac{f_{\rm 5\;GHz}}{f_{4400\rm\angstrom}}\bigg) \\
           &= \log \bigg(\frac{L_{\rm r}}{L_{2500\rm\angstrom}}\bigg) - 0.5\log\bigg(\frac{4400}{2500}\bigg),
\end{align}
where we have calculated the rest-frame 4400 \angstrom\ flux density (monochromatic luminosity) using
the rest-frame 2500 \angstrom\ flux density (monochromatic luminosity)
and an assumed optical spectral index $\alpha=-0.5$ \citep[e.g.][]{Vanden2001}.

\noindent Column (15): the two-point spectral index $\alpha_{\rm ox}$ \citep{Tananbaum1979}, defined by
\begin{align}
    \alpha_{\rm ox} &= \frac{{\rm log}(f_{\rm 2\;keV}/f_{2500\mbox{\rm~\scriptsize\angstrom}})}{{\rm log}(\nu_{\rm 2\;keV} / \nu_{2500\mbox{\rm~\scriptsize\angstrom}})},
\end{align}
which represents the spectral index of an assumed power-law connecting rest-frame 2500 \angstrom\ and 2 keV.

\noindent Column (16): the difference between the measured $\alpha_{\rm ox}$ of our quasar
and the expected $\alpha_{\rm ox}$ for a typical RQQ using the $\alpha_{\rm ox}$-$L_{2500\rm\angstrom}$ relation
in Eq.~3 of \cite{Just2007},
\begin{equation}
    \Delta\alpha_{\rm ox,RQQ} = \alpha_{\rm ox} - \alpha_{\rm ox,RQQ}.
\end{equation}
$\Delta\alpha_{\rm ox,RQQ}$ quantifies the amount of additional X-ray emission from the jet-component of RLQs
compared to the X-ray emission of RQQs.

\noindent Column (17): the difference between the measured $\alpha_{\rm ox}$ of our quasar at $z>4$
and the expected $\alpha_{\rm ox}$ for a typical low-redshift RLQ (mostly at $z=0.3$--2.5 with a median of $z=1.4$)
using the $L_{2\rm\ keV}$-$L_{2500\ \rm\angstrom}$-$L_{5\ \rm GHz}$ relation in Table ~7 of Miller11,
\begin{equation}
    \Delta\alpha_{\rm ox, RLQ} = \alpha_{\rm ox}  - \alpha_{\rm ox, RLQ}.
\end{equation}

\begin{figure*}
\centering
\includegraphics[width=0.6\textwidth, clip]{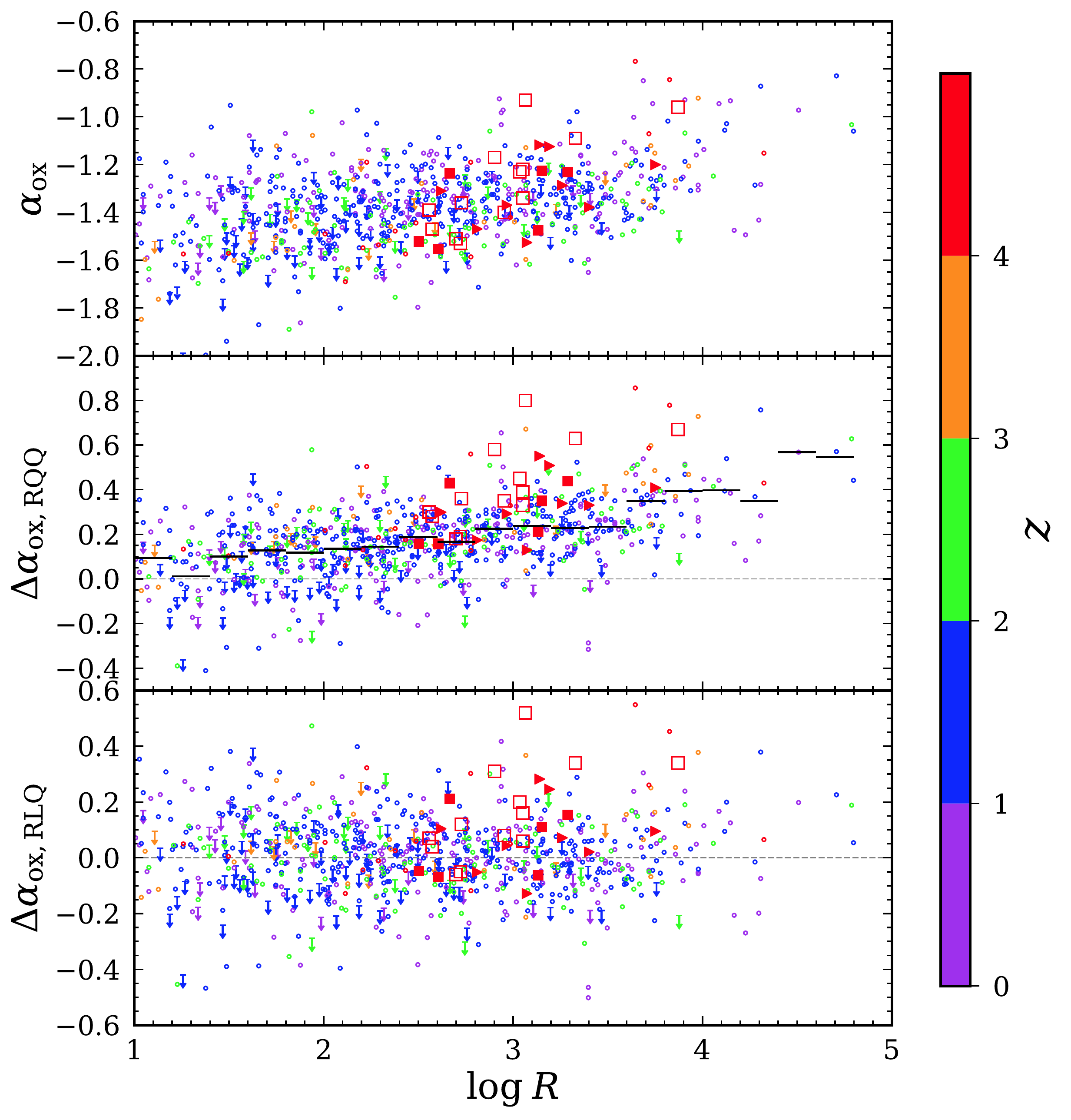}
    \caption{The relations between radio loudness ($\log R$) and $\alpha_{\rm ox}$, $\Delta\alpha_{\rm ox, RQQ}$, and $\Delta\alpha_{\rm ox,RLQ}$, from
top to bottom.
The filled squares and triangles are from the {\it Chandra} Cycle 17 objects and archival data objects, respectively.
The open squares are from Wu13.
    The small open circles represent the radio-loud and radio-intermediate objects in the full sample of
    Miller11
    that are detected
in X-rays, while the downward arrows have only X-ray upper limits.
The dashed lines label the positions of $\Delta\alpha_{\rm ox}=0$.
The thick black lines in the middle panel are the mean $\Delta\alpha_{\rm ox, RQQ}$ 
values for the Miller11 RLQs in $\log R$ bins ($\Delta\log R=0.2$ per bin).
All symbols are color-coded based on their redshifts using the color bar on the right-hand side.}
\label{fig:R-AlphaOX}
\end{figure*}

\begin{figure}
\centering
\includegraphics[width=0.45\textwidth, clip]{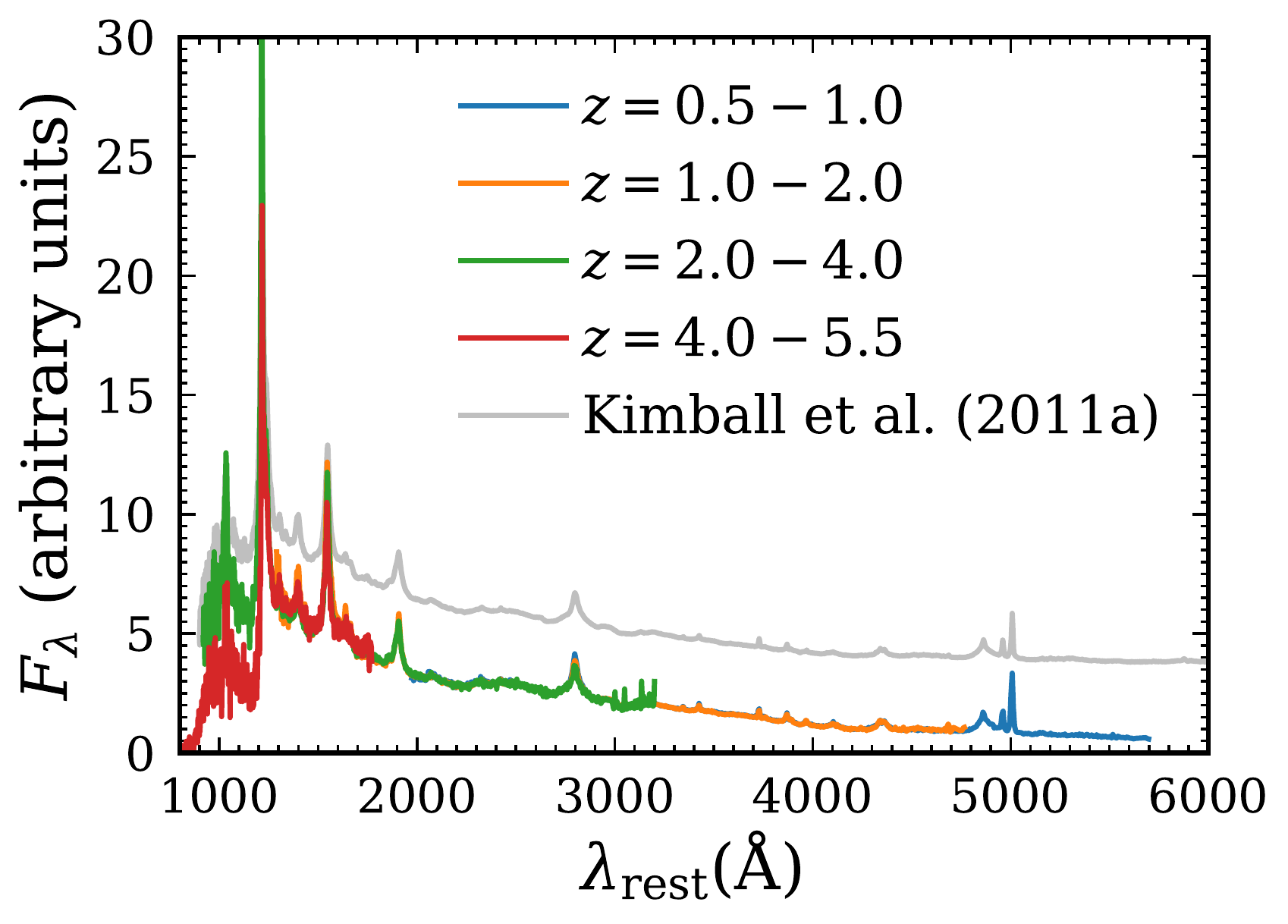}
\caption{The composite SDSS spectra of HRLQs in different redshift bins.
The quasars in bins of $z=$ 0.5--1.0, 1.0--2.0, and 2.0--4.0 are from Miller11, and the quasars in the bin of $z=$ 4.0--5.5 are from this paper.
The grey curve (vertically shifted) is a composite spectrum of SDSS quasars that have radio counterparts with $f_{\rm 1.4\ GHz}\ge2$ mJy \citep{Kimball2011a}.
The composite spectra of HRLQs in different redshift bins match so well that it is hard to notice the overlapping
(except that the $z=$ 4.0--5.5 quasars suffer more severe absorption below the Lyman limit).
The strengths of emission lines do not have an apparent dependence on redshift.}
\label{fig:compoSpec}
\end{figure}

\begin{figure*}
\centering
\includegraphics[width=0.6\textwidth, clip]{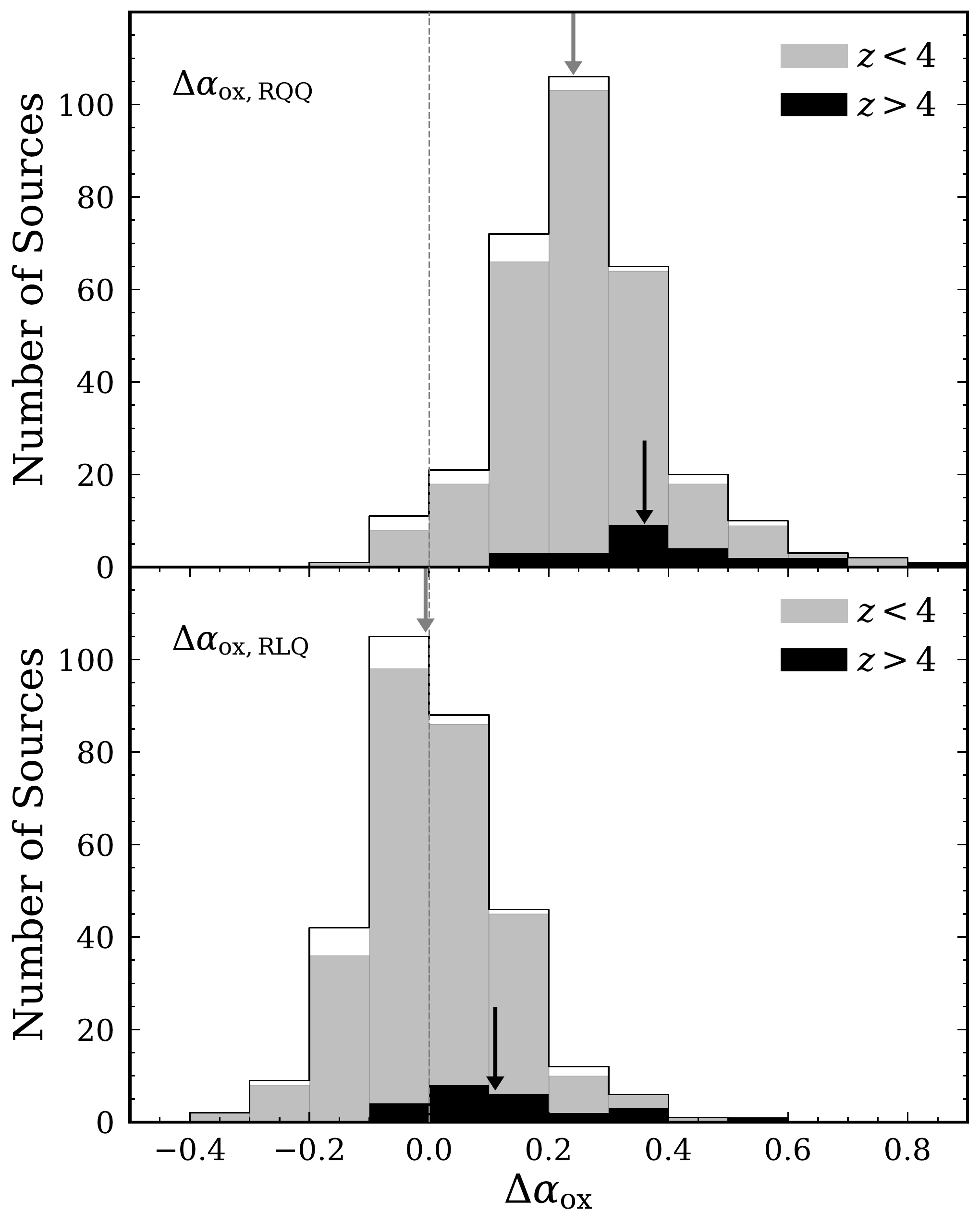}
\caption{The histograms of $\Delta\alpha_{\rm ox,RQQ}$ (top) and $\Delta\alpha_{\rm ox,RLQ}$ (bottom)
for the full-sample objects in Miller11 with $\log R>2.5$, $z<4$, and $m_i\le20.26$.
The gray and open histograms are for X-ray detected and undetected objects in the $z<4$ sample, respectively.
The HRLQs at $z>4$ are plotted in black.
The downward arrows indicate the medians of the $\Delta\alpha_{\rm ox}$ distributions of the two redshift bins.}
\label{fig:DeltaAlphaOXHists}
\end{figure*}

\section{X-ray enhancements of high-redshift HRLQs}

In this section, we perform statistical tests on the
$\Delta\alpha_{\rm ox}$ distributions of HRLQs at $z>4$
against those of their low-redshift counterparts, using the
enlarged and complete sample, compared with Wu13 (see Section~\ref{sec:sample}).
We quantify the typical excess of jet-linked \mbox{X-ray} emission
using the medians of $\Delta\alpha_{\rm ox}$ distributions.
The relevant properties of the flux-limited high-$z$ sample
of HRLQs are compiled in Table~\ref{tab:flux-limited}.

\subsection{Basic comparisons}
We first plot $\alpha_{\rm ox}$, $\Delta\alpha_{\rm ox,RQQ}$, and $\Delta\alpha_{\rm ox,RLQ}$
for all the high-$z$ HRLQs
with sensitive
X-ray coverage
against their $\log R$ in Fig.~\ref{fig:R-AlphaOX}, where the filled squares, filled triangles, and open squares represent
{\it Chandra} Cycle 17 objects, archival data objects, and Wu13 objects, respectively.
The median redshift of the high-$z$ HRLQs is $z=4.3$, and the interquartile (25th percentile to 75th percentile) range is $[4.2, 4.4]$.
For comparison, also plotted in Fig.~\ref{fig:R-AlphaOX} are the radio-loud
and radio-intermediate quasars in the full sample of
Miller11\footnote{Miller11 quantify radio loudness using the ratio of monochromatic luminosities
at rest-frame 5 GHz and 2500 \angstrom.
We have converted their values of radio loudness to the values according to our definition, assuming $\alpha=-0.5$.}
with a median redshift of $z=1.4$ (and an interquartile range of $[1.0, 1.9]$).
The loci of high-$z$ HRLQs are not consistent with those of typical low-$z$ RLQs
in the three panels. Especially in the $\Delta\alpha_{\rm ox, RQQ}$-$\log R$
and $\Delta\alpha_{\rm ox,RLQ}$-$\log R$ planes, HRLQs at $z>4$ have
systematically larger $\Delta\alpha_{\rm ox,RQQ}$ and $\Delta\alpha_{\rm ox, RLQ}$.

We further compared the $\Delta\alpha_{\rm ox}$ distribution
of the \mbox{flux-limited} sample (see Section~\ref{sec:sample}) with that of their low-$z$ counterparts using histograms.
We thus define a flux-limited ($m_i\le20.26$) comparison sample of HRLQs at $z<4$
that is a subset of the full sample of RLQs in Miller11.
The high-$z$ and low-$z$ samples contain 24 and 311 objects,
with median redshifts of 4.4 and 1.3 (and interquartile ranges of $[4.3, 4.7]$ and $[0.9, 1.8]$), respectively.

In addition to the flux-limit and radio-loudness cuts we have applied,
we confirmed that the quasars in the low-$z$ sample show comparably strong emission lines to those in the high-$z$ sample
(i.e. they are largely free from strong boosted non-thermal continuum emission in the optical/UV; see Fig.~\ref{fig:spec}).
We first matched the low-$z$ sample to the DR7 quasar property catalog \citep{Shen2011}
and checked the rest-frame equivalent widths (REWs) of H$\beta$, Mg {\sc ii}, and C {\sc iv}.
187 quasars in the low-$z$ sample are included in the DR7 quasar property catalog; all of them have at least one
emission line (of the three we checked) that has REW$>$5 \angstrom.
We also visually inspected the SDSS spectra of the 214
low-$z$ quasars within the SDSS quasar catalog DR14 \citep{Paris2017}, and
almost all of them show strong emission lines.
To demonstrate the similarly strong emission lines
in the high-$z$ and low-$z$ samples,
we create composite SDSS spectra
for the HRLQs in four redshift bins (quasars at $z<0.5$ are discarded due to their largely different
rest-frame wavelength ranges)
and show them in Fig.~\ref{fig:compoSpec}.
To create these composite spectra, each spectrum was first shifted to its rest frame and normalised to
some continuum window \citep[e.g.][]{Vanden2001}, and then the median flux in each wavelength bin was calculated.
It is apparent from the significant overlapping in Fig.~\ref{fig:compoSpec}
that the composite spectra of HRLQs in the different redshift bins match very well,
especially in the sense that all of them show comparably strong emission lines.
In Fig.~\ref{fig:compoSpec}, we also show the
composite median spectrum of quasars that are detected in the FIRST survey and have $f_{\rm 1.4\ GHz}\ge2$~mJy for comparison \citep{Kimball2011a}.

The histograms of $\Delta\alpha_{\rm ox,RQQ}$ and $\Delta\alpha_{\rm ox, RLQ}$ for the two
samples are shown in Fig.~\ref{fig:DeltaAlphaOXHists},
where the low-$z$ sample contains upper limits as some HRLQs from Miller11 were not detected in X-rays.
The distributions for the high-$z$ and low-$z$ samples are visually different, with not only their peaks differing by $\approx0.1$,
but also their distributions spanning different ranges, in both panels.
From the histograms of $\Delta\alpha_{\rm ox,RLQ}$, the HRLQs at $z<4$ do not show an apparent deviation from
the $L_{2\rm\;keV}$-$L_{2500\rm\;\angstrom}$-$L_{5\rm\;GHz}$ relation that is derived from their parent population of general RLQs.
However, HRLQs at $z>4$ do not follow this relation well
and have an excess of X-ray emission compared with the low-redshift sample.

\subsection{Quantitative statistical tests}
\label{sec:statTest}
Since there are upper limits 
\footnote{Note that Miller11 calculated relatively conservative 95\% confidence upper limits for the X-ray flux when a quasar is not detected in X-rays.
Considering the 19 non-detections in the low-$z$ sample, $\approx18$ are expected to be correct in the sense that 
the flux is actually below the limit value.}
in our comparison sample,
to quantify the statistical significance of the difference in the $\Delta\alpha_{\rm ox}$ distributions of the two samples,
we use the Peto-Prentice test that is implemented in the Astronomy Survival Analysis Package\footnote{Downloaded from \url{http://astrostatistics.psu.edu/statcodes/asurv}. See \cite{Feigelson1985}.}
to perform \mbox{two-sample} tests.
The test shows a 4.56$\sigma$ ($p=2.56\times10^{-6}$)\footnote{The $p$-values in this section
are the probabilities of the data under the null hypothesis and
correspond to the significance level in one-sided tests using a standard normal distribution.}
difference for the
$\Delta\alpha_{\rm ox,RQQ}$ distributions and
a 4.07$\sigma$ ($p=2.35\times10^{-5}$) difference for the $\Delta\alpha_{\rm ox,RLQ}$ distributions,
both of which are better than the corresponding statistical significances measured by Wu13.
Furthermore, while the statistical test results of Wu13 should be considered as suggestive (due to their incomplete X-ray coverage),
our results can be accepted more formally.

The majority of the high-$z$ sample that has multi-frequency
data in the radio band is flat-spectrum ($\alpha_{\rm r}>-0.5$) objects (see Table~\ref{tab:flux-limited}).
We thus performed another statistical test on the subsets of confirmed flat-spectrum quasars
selected from the high-$z$ and low-$z$ samples.
The differences of the high-$z$ subset (17 objects) and low-$z$ subset (100 objects) are 4.18$\sigma$ ($p=1.46\times10^{-5}$)
and 3.18$\sigma$ ($p=7.36\times10^{-4}$)
for the distributions of $\Delta\alpha_{\rm ox,RQQ}$ and $\Delta\alpha_{\rm ox, RLQ}$, respectively.
We performed a Monte Carlo simulation by randomly selecting 17 and 100 objects from the flux-limited high-$z$ (24 objects) and low-$z$ (311 objects) samples
and running statistical tests on these \mbox{sub-samples}.
The statistical significance typically drops by $\approx1\sigma$ compared with that of the flux-limited sample;
we conclude that the smaller statistical significance resulting from the flat-spectrum sub-sample is mainly caused by the smaller sample size,
rather than any strong systematic difference in the degree of high-$z$ enhancement.

Another factor that inevitably affects our statistical tests
and all other such tests in the literature is the
uncertainties of $\alpha_{\rm ox}$ and $\Delta\alpha_{\rm ox}$,
including contributions from flux measurement errors and intrinsic quasar
variability (since the multi-wavelength data have not been simultaneously obtained).
The uncertainties of the X-ray fluxes are in the range of 6\%--30\% (e.g. see Table~\ref{tab:photometry});
the flux errors in the optical ($\sigma_{m_i}\lesssim0.03$ mag)
and radio ($\sigma_{\rm rms}\lesssim0.15$ mJy) are negligible compared with those for the X-rays.
Since the overall variability (in X-rays and optical) dominates the uncertainties,
we adopted a typical value of 20\% for uncertainties due to flux errors in X-rays.
We take a magnitude of 25\% for the variability in the X-ray band \citep[e.g.][]{Gibson2012}
and a magnitude of 25\% for the variability in the optical/UV \citep[considering our sample contains the \mbox{most-luminous} quasars
and RLQs are usually more variable than RQQs, e.g.][]{Vanden2004, MacLeod2010}.
We thus assign a typical value of 0.06 as the uncertainties on $\alpha_{\rm ox}$, which is equivalent to $\approx$43\% uncertainties on the amount of X-ray enhancement.
Both measurement errors and variability are random uncertainties instead of systematic biases,
and they thus broaden the $\Delta\alpha_{\rm ox}$ distributions with the centres unchanged.
The Peto-Prentice tests we performed in the analyses above compare, in fact, the broadened distributions instead of the true distributions.
One of the consequences is that the power of the statistical tests is reduced by the ``smearing'' effect of the uncertainties.
In another words, if we had performed coordinated \mbox{multi-wavelength} observations using
telescopes powerful enough to ignore the measurement errors, the statistical significance would be even higher.
We confirmed the effects of the uncertainties on $\alpha_{\rm ox}$ on two-sample tests using Monte Carlo simulations.
We also confirmed the even higher statistical significance of the ``true'' distribution using Bayesian modelling,
marginalising out the uncertainties of $\alpha_{\rm ox}$. See Appendix~\ref{sec:err} for details on the simulation and modelling process.
We note that the simulation can indicate the direction of the effects of uncertainties while we are not
sure whether a quantitative correction to the statistical significance is well justified or not,
and the modelling analysis depends on assumptions
(e.g. the Gaussian assumption, the magnitudes of the uncertainties, and priors).
We thus conservatively quote the results of the statistical tests using the observed data, following the standard 
practice in the literature.

Among the 7 objects (excluding SDSS J091316.55+591921.6) in Table~\ref{tab:flux-limited} that are outside the flux-limited sample,
three objects have less X-ray emission than predicted from the $L_{2 \rm keV}$-$L_{2500 \angstrom}$-$L_{5\rm GHz}$ relation of low-$z$ RLQs
(negative $\Delta\alpha_{\rm ox,RLQ}$ values).
In comparison, only four out of 24 objects in the flux-limited sample show this deficit of X-ray emission.
Since our multi-wavelength data are not simultaneous,
variability with larger amplitude for fainter objects may generally cause larger scatter of $\alpha_{\rm ox}$.
We cannot conclude from the available data that the X-ray enhancement disappears in the optically faint regime.
We will need complete X-ray coverage of optically fainter HRLQs at $z>4$ to improve understanding of this matter (see Section~\ref{sec:future}).

\subsection{The amount of X-ray enhancement}
\label{sec:statEst}

We quantify the typical amount of X-ray enhancement for the high-$z$ sample relative to the low-$z$ sample
using the difference of the medians of their $\Delta\alpha_{\rm ox}$ distributions.
The medians were calculated using the Kaplan-Meier estimator of the
cumulative distribution function that can cope with upper limits in data.
We describe the relevant methodology in Appendix~\ref{sec:kme}.
The medians of $\Delta\alpha_{\rm ox, RQQ}$ for the high-redshift and low-redshift samples are $0.37\pm0.03$ and $0.24\pm0.01$, respectively.
The medians of $\Delta\alpha_{\rm ox, RLQ}$ for the high-redshift and low-redshift samples are $0.11\pm0.03$ and $0.01\pm0.02$, respectively.
The uncertainties here are estimated using bootstrapping (see the method in Appendix~\ref{sec:kme}).
These medians are also plotted in Fig.~\ref{fig:DeltaAlphaOXHists} as downward arrows.
The difference of the medians of the $\Delta\alpha_{\rm ox,RLQ}$ distributions is $0.11\pm0.04$, and thus the X-ray luminosities of the HRLQs
at $z>4$ are typically $1.9^{+0.5}_{-0.4}$ times those of their low-redshift counterparts.
In addition to the median factor of \mbox{X-ray} enhancement,
we also calculate the interquartile range
from the $\Delta\alpha_{\rm ox,RLQ}$ distribution of the high-$z$ sample,
which is $[0.04, 0.21]$ and corresponds to a factor of X-ray enhancement in the range of $[1.3, 3.5]$.
At the extremes, some objects show no X-ray enhancement while others show an enhancement by a factor of 5--25 (see Table~\ref{tab:flux-limited}).

The factor of $1.9^{+0.5}_{-0.4}$ X-ray enhancement is somewhat smaller than but consistent with
the estimation in Wu13, who found a factor of $\approx 3$ by comparing the means of $\Delta\alpha_{\rm ox}$.
We thus calculate the factor of X-ray enhancement using the median statistic and the 12 HRLQs that were used by Wu13 ($m_i<20$)
resulting in a factor of $2.6_{-1.0}^{+2.0}$.
Therefore, any apparent difference of X-ray enhancement factor is
mainly caused by the large scatter due to the small sample size of Wu13
and partly by the statistic used.

We here point out one potential Malmquist-type bias \citep[e.g.][]{Lauer2007} that
could diminish the X-ray enhancement of the high-redshift sample compared with the low-redshift sample.
A larger fraction of the high-$z$ sample is near the optical flux limit ($m_i=20.26$) than the low-$z$ sample;
if the optical luminosity function is steep, a large fraction of the HRLQs that are near
the flux limit have, intrinsically, a dim X-ray flux.
We thus select another low-$z$ HRLQ sample with $\log L_{2500\rm\angstrom}>30.57$, which is the minimum optical/UV luminosity of the high-$z$ sample.
The resulting low-$z$ sample has a size of 165.
We found that the X-ray enhancement of the high-$z$ sample is a factor of $2.0_{-0.4}^{+0.5}$,
which means this selection effect, if it exists, probably does not significantly affect our results.

\subsection{HRLQs at $3<z<4$}
\label{sec:izbin}
Wu13 performed two-sample tests on HRLQs in different redshift bins below $z=4$ using the RLQs of Miller11 and found an apparent
X-ray enhancement also exists at $z\approx3$. Specifically, the $\Delta\alpha_{\rm ox}$ distributions of HRLQs at $3<z<4$ differ from those of
$z<3$ HRLQs at a $\approx5\sigma$ level. We thus select $3<z<4$ HRLQs from the full RLQ sample
and quantify their typical factor of X-ray enhancement
relative to HRLQs at $z<3$ using the same consistent method described in the previous section.
We have applied an optical flux cut determined by the faintest HRLQ at $3<z<4$ ($m_i=20.38$).
The sample sizes (median redshifts) are 16 ($z=3.4$) and 304 ($z=1.3$) for $3<z<4$ HRLQs and $z<3$ HRLQs, respectively.
We estimated the medians of
$\Delta\alpha_{\rm ox,RLQ}$
for $3<z<4$ and $z<3$ objects are $0.12\pm0.07$ and $-0.01\pm0.01$.
The corresponding factor of X-ray enhancement is $2.0^{+1.1}_{-0.8}$.
The relatively larger error bars of our estimations are largely due to the small sample size
at $3<z<4$.

\subsection{The spectral energy distributions}
\label{sec:SEDs}
We here make another comparison between the X-ray emission of high-$z$ and low-$z$ HRLQs using their SEDs.
We collected photometric data to construct the broadband SEDs of
our objects that cover the radio through X-ray bands from the following sources.

\begin{enumerate}
\item
Radio: the 1.4 GHz flux densities are from the FIRST or NVSS surveys;
the sources for 5 GHz values are the same as those described in Column~(12) in Section~\ref{sec:longTable};
the 150 MHz flux densities are gathered from the GMRT 150 MHz all-sky survey \citep{Intema2017};
the flux densities at other frequencies were retrieved from the NED.

\item
    Mid-infrared: the all-sky catalog of the {\it Wide-field Infrared Survey Explorer} \citep[{\it WISE};][]{Wright2010} provides the \mbox{mid-infrared}
fluxes. All of our objects are detected by {\it WISE} except for SDSS J083549.42+182520.0, SDSS J102107.57+220921.4, and SDSS J160528.21+272854.4.

\item
Near-infrared: we first searched for objects in the Two Micron All Sky Survey \citep[2MASS;][]{Skrutskie2006}.
None of our objects has a 2MASS detection.
We then searched objects in the UKIRT Infrared Deep Sky Survey \citep[UKIDSS;][]{Lawrence2007}.
Five objects (SDSS J030437.21+004653.5, PMN J2314+0201, SDSS J083549.42+182520.0, SDSS J153533.88+025423.3, and SDSS J222032.50+002537.5)
have $Y$, $J$, $H$, and $K$ detections, while SDSS J160528.21+272854.4 has only a $J$-band detection.
We further searched for objects in the VISTA Hemisphere Survey \citep[VHS;][]{McMahon2013}, where we found
$J$-, $H$-, and $K_s$-band detections for SDSS J030437.21+004653.5, PMN J2134$-$0419, and SDSS J222032.50+002537.5,
and an additional $Y$-band detection for PMN J2134$-$0419.

\item
Optical: obtained from SDSS photometry ($u$, $g$, $r$, $i$, and $z$). The bands that are seriously affected by
the Ly$\alpha$ forest are discarded.

\item
X-ray: from this work.

\end{enumerate}

\begin{figure*}
\centering
    \includegraphics[width=0.42\textwidth, clip]{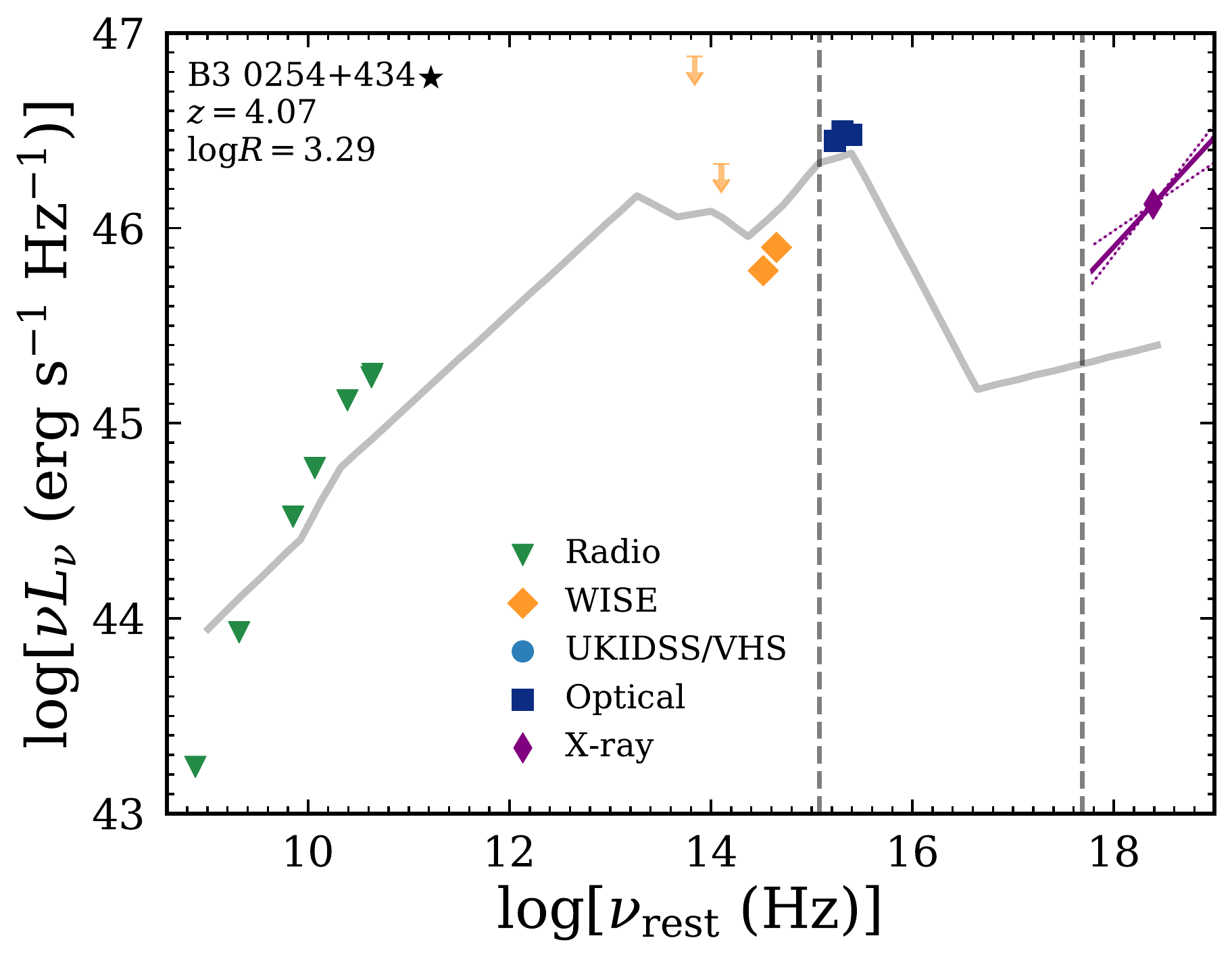}
    \includegraphics[width=0.42\textwidth, clip]{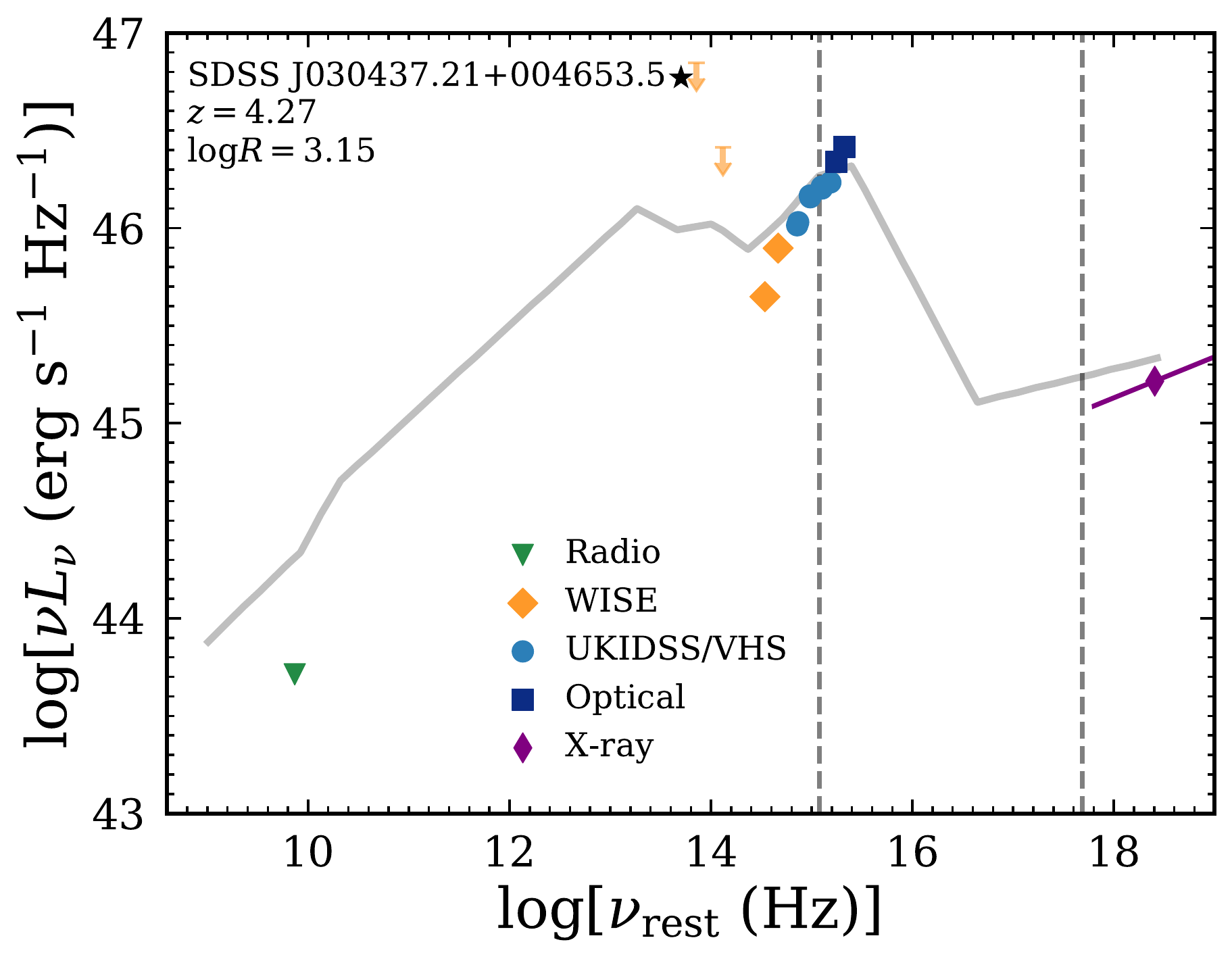}\\
    \includegraphics[width=0.42\textwidth, clip]{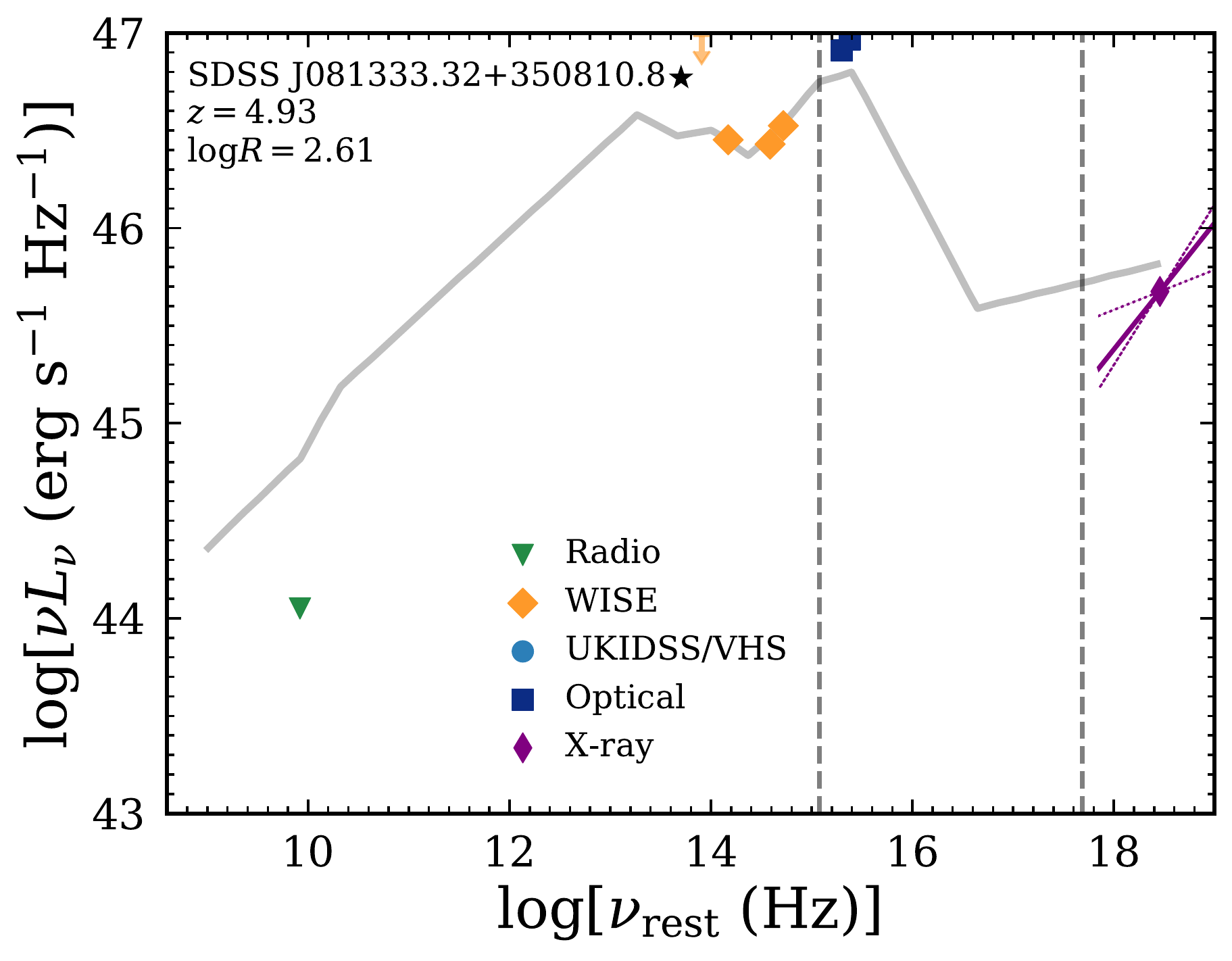}
    \includegraphics[width=0.42\textwidth, clip]{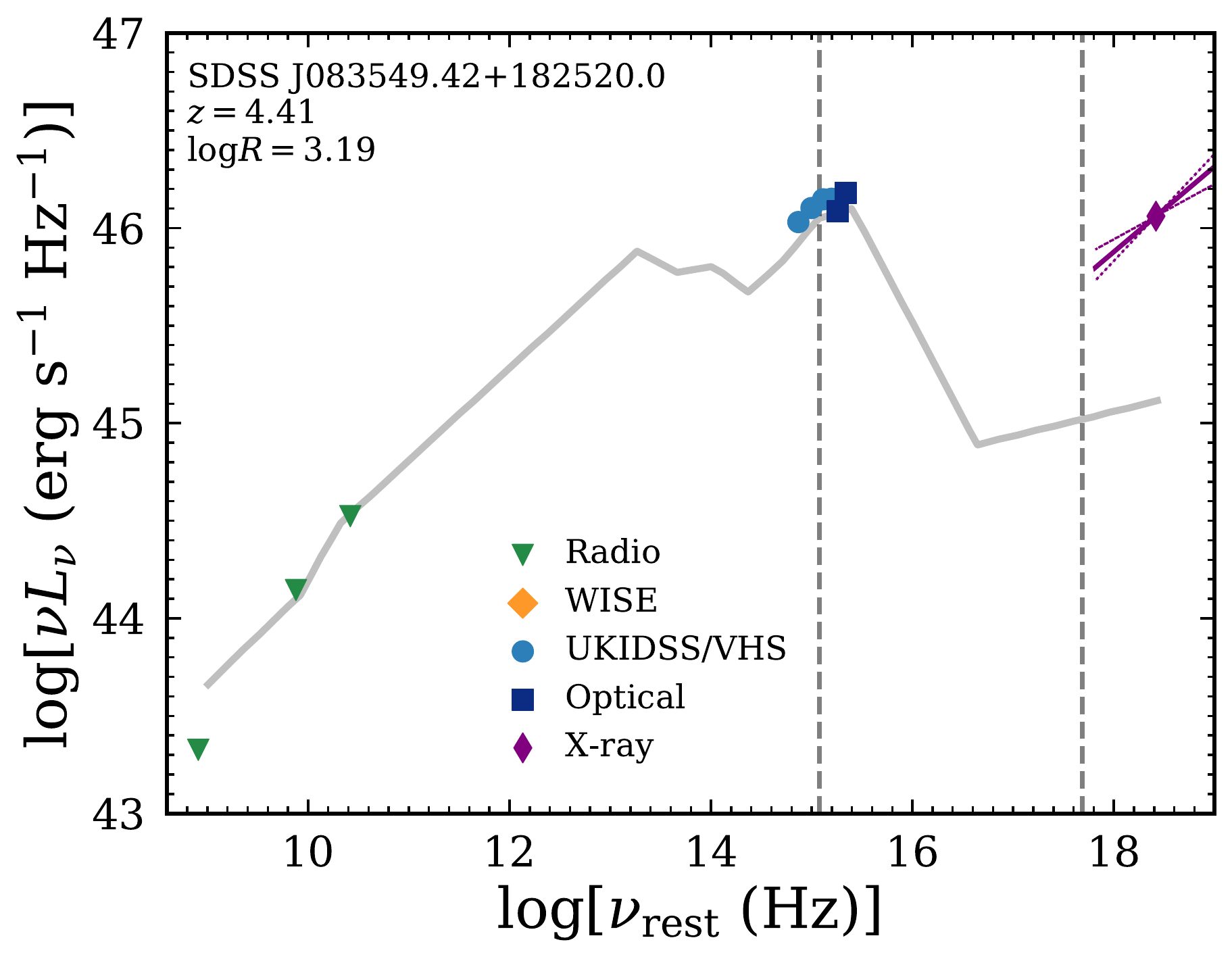}\\
    \includegraphics[width=0.42\textwidth, clip]{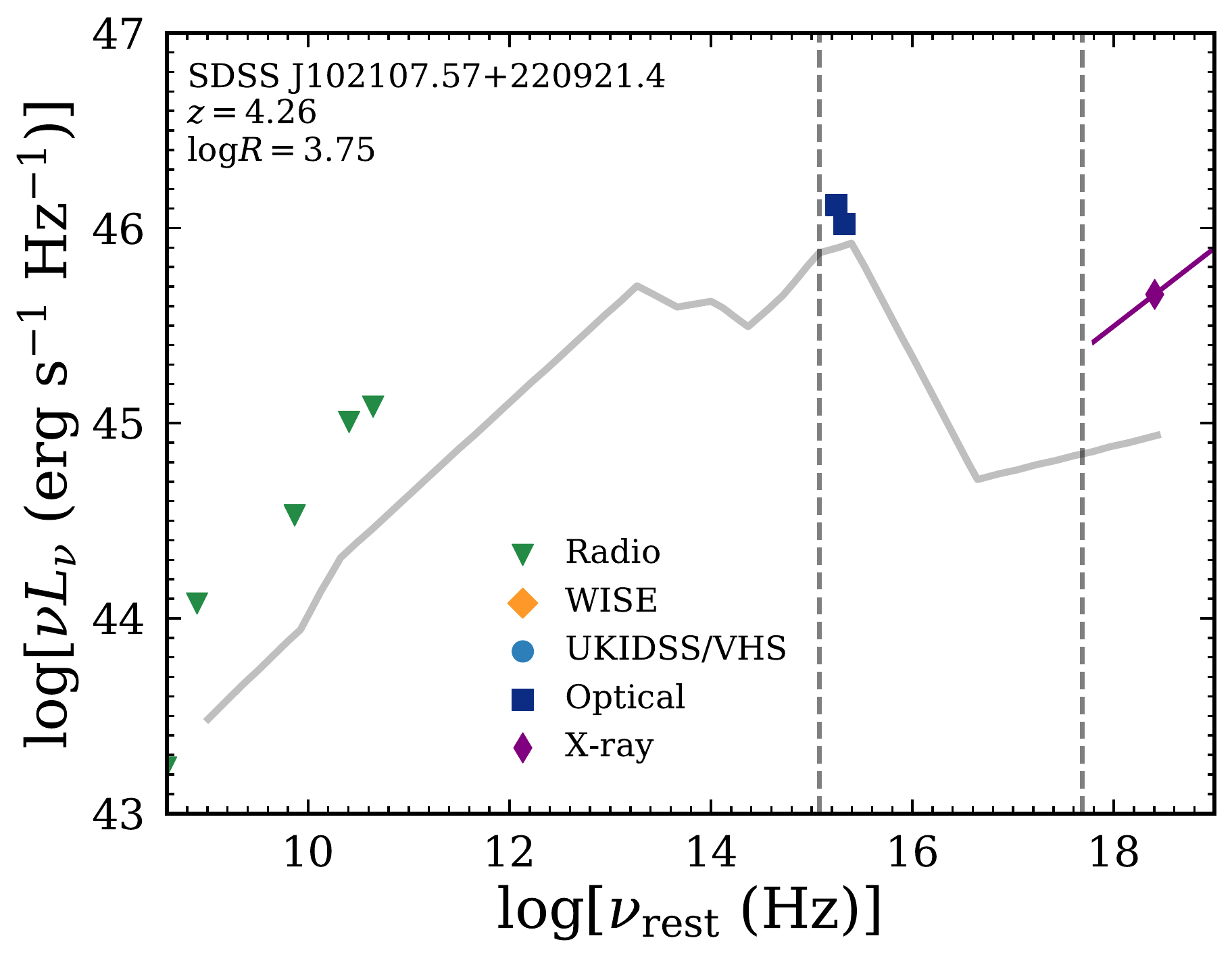}
    \includegraphics[width=0.42\textwidth, clip]{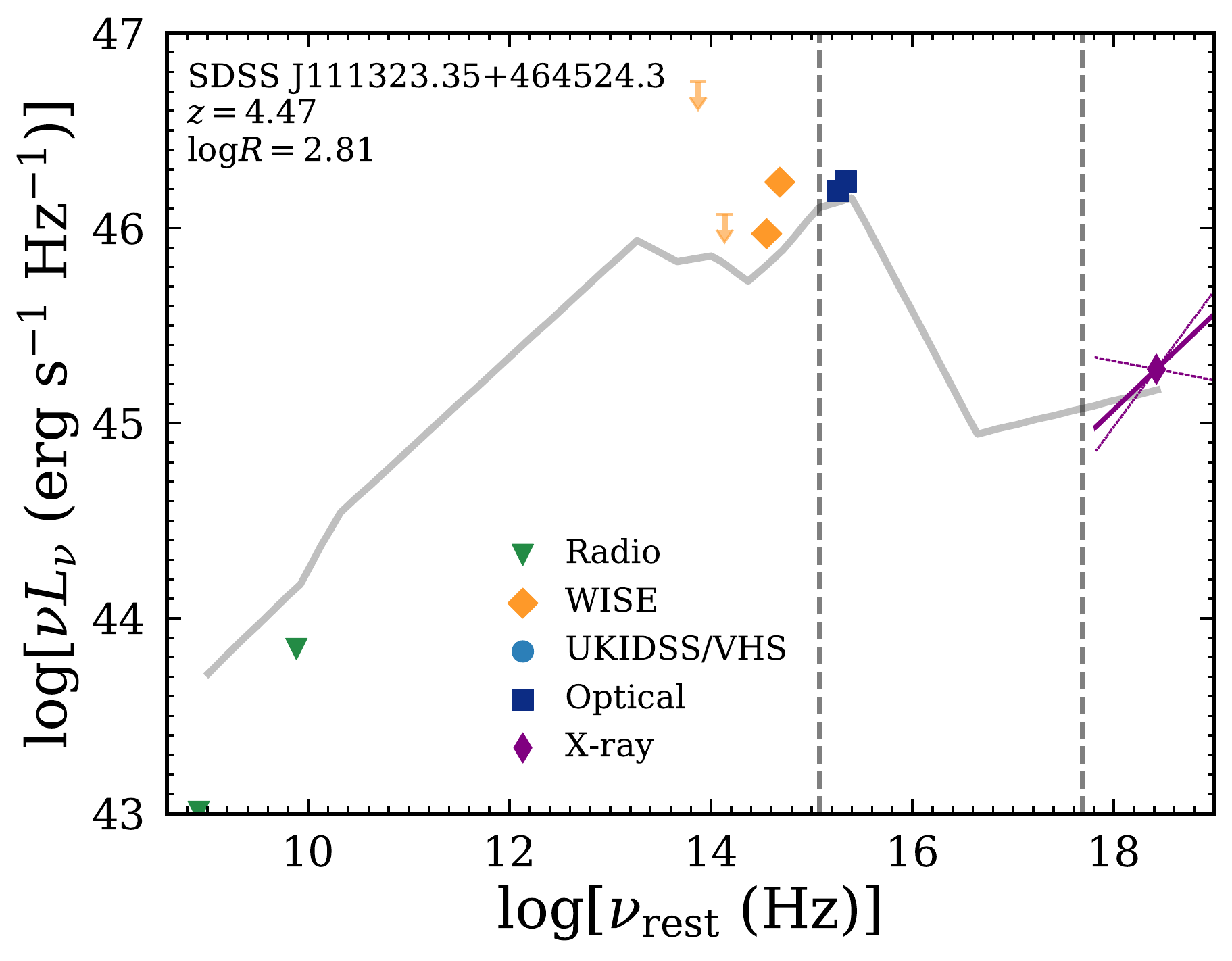}\\
    \caption{The broadband SEDs for the HRLQs in this paper (in ascending order of RA) from radio to X-ray.
    The objects within our flux-limited sample are marked by $\bigstar$ following their names.
    The solid purple lines show the X-ray power-law spectra with their uncertainties as dotted lines (see Column~(9) in Section~\ref{sec:longTable}).
    The purple diamonds represent the observed-frame 2 keV.
    The grey curve is the composite SED for the 10 HRLQs at $z<1.4$ from \citet{Shang2011} with comparable optical luminosity and radio loudness.
    This low-$z$ comparison SED has been normalised to the high-$z$ SEDs at rest-frame 2500 \angstrom.
    The vertical lines indicate rest-frame 2500 \angstrom\ and 2 keV.}
    \label{fig:SED}
\end{figure*}

None of our HRLQs has a counterpart in the {\it Fermi} LAT 4-Year Point Source Catalog \citep{Acero2015}, using a matching radius of 10$'$.
The constructed SEDs of our HRLQs are shown in Fig~\ref{fig:SED}, in ascending order of RA.
Also plotted in Fig~\ref{fig:SED} is a comparison SED (grey curve) that was
constructed by Wu13, using 10 HRLQs at $z<1.4$ from \cite{Shang2011}.
These 10 low-$z$ HRLQs were selected based on their optical/UV
luminosity ($\log \lambda L_\lambda(3000\angstrom)>45.9$), radio-loudness ($2.9<\log R<3.7$), and
useful \mbox{X-ray} data from the literature. Their SEDs were normalised at \mbox{rest-frame} 4215~\angstrom, and 
medians for different waveband bins were calculated \citep[see Section 5.2 of][]{Shang2011}.
Following Wu13, we have normalised the comparison SED to the observed data for our HRLQs at rest-frame 2500 \angstrom.

Five out of the 15 HRLQs have higher X-ray luminosities by a factor of $\approx$3--13 than that of the comparison SED, while
the other 10 HRLQs have comparable X-ray luminosities with that of the comparison SED,
considering the uncertainties of the \mbox{X-ray} luminosities.
Note that the interquartile ranges of the comparison SED are 0.37 dex and 0.56 dex in the radio (5 GHz) and X-ray (2 keV) bands, respectively.
Notably, SDSS J102107.57+220921.4 and SDSS J083549.42+182520.0 show a factor of $\approx10$ enhancement in their X-ray luminosities relative to
the comparison SED; both of them were not included in our flux-limited sample due to their fainter optical fluxes, making future work that
extends our systematic study to the fainter optical regime promising (see Section~\ref{sec:future}).
Including the 17 SEDs of Wu13, half of the high-$z$ HRLQs (16/32) show an apparent excess of X-ray emission by a factor
of $\sim$2.5--20, compared with the
low-$z$ HRLQs with matched optical/UV luminosity and radio-loudness.
Also from their SEDs, the HRLQs do not show weaker optical/UV emission, relative to their infrared emission, which
means the high $\Delta\alpha_{\rm ox}$ values indeed reflect stronger X-ray emission instead of weaker optical/UV emission.
This supports the basic validity of our earlier analyses based on $\Delta\alpha_{\rm ox}$.

\renewcommand{\thefigure}{\arabic{figure} (Continued)}
\addtocounter{figure}{-1}

\begin{figure*}
\centering
    \includegraphics[width=0.42\textwidth, clip]{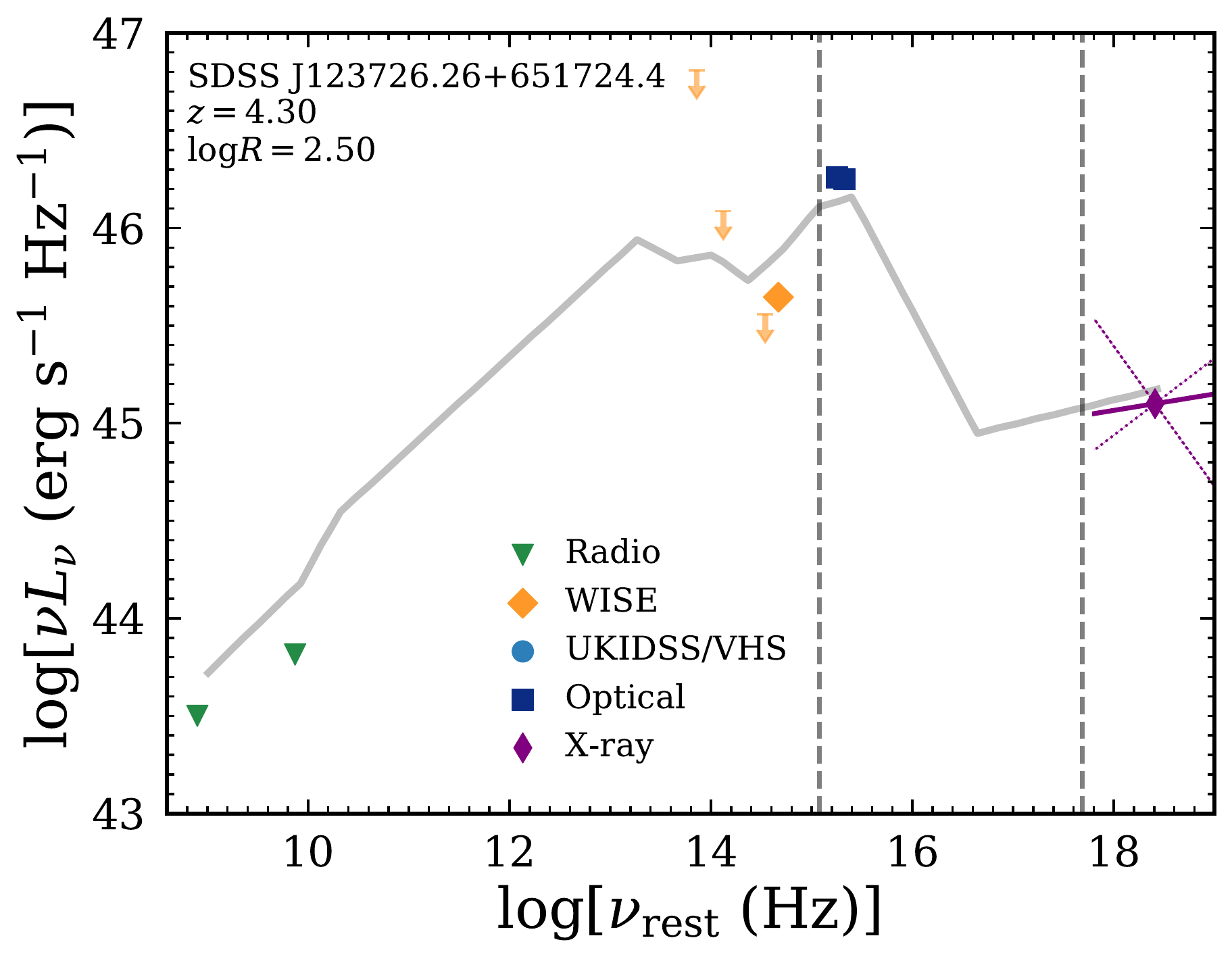}
    \includegraphics[width=0.42\textwidth, clip]{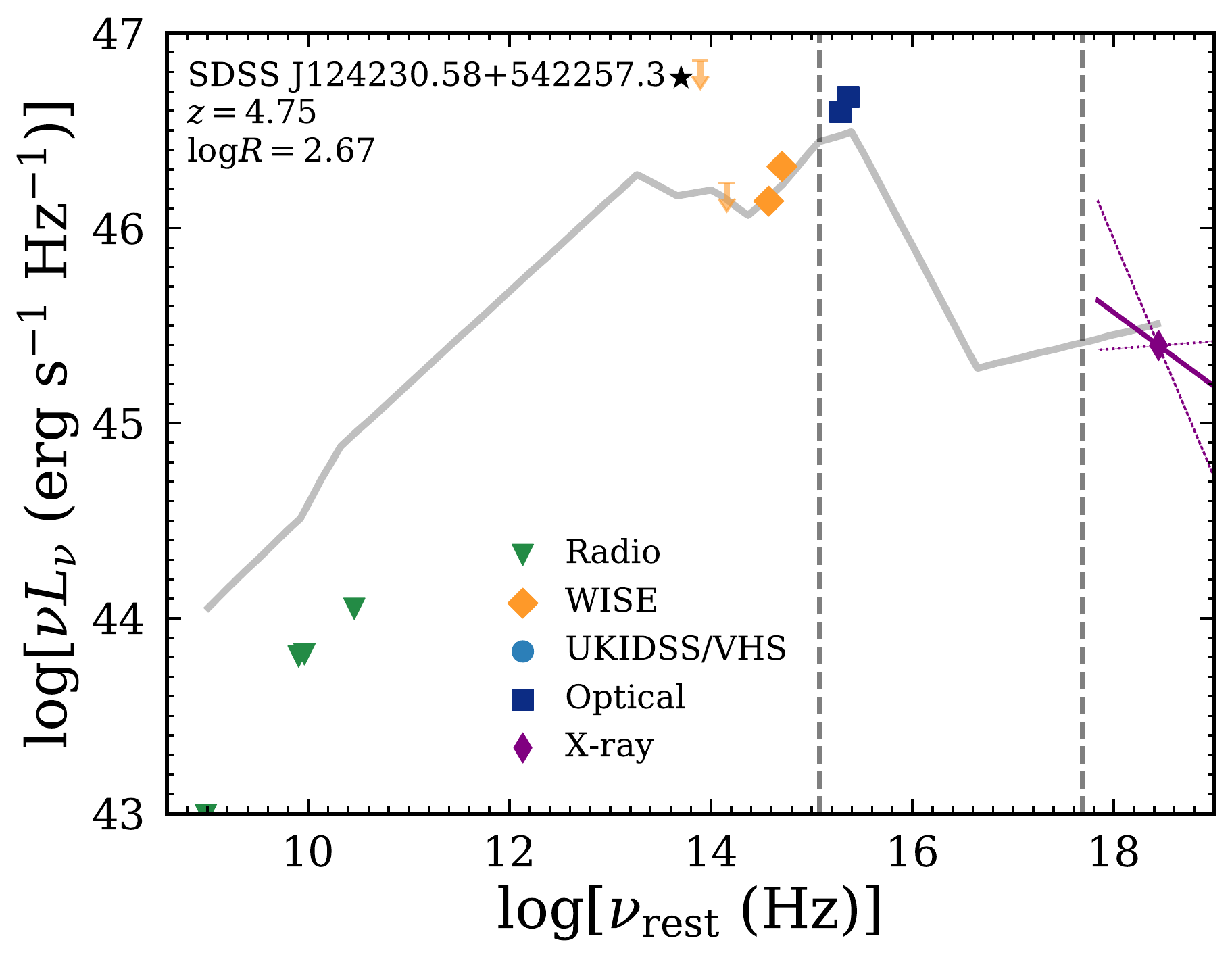}\\
    \includegraphics[width=0.42\textwidth, clip]{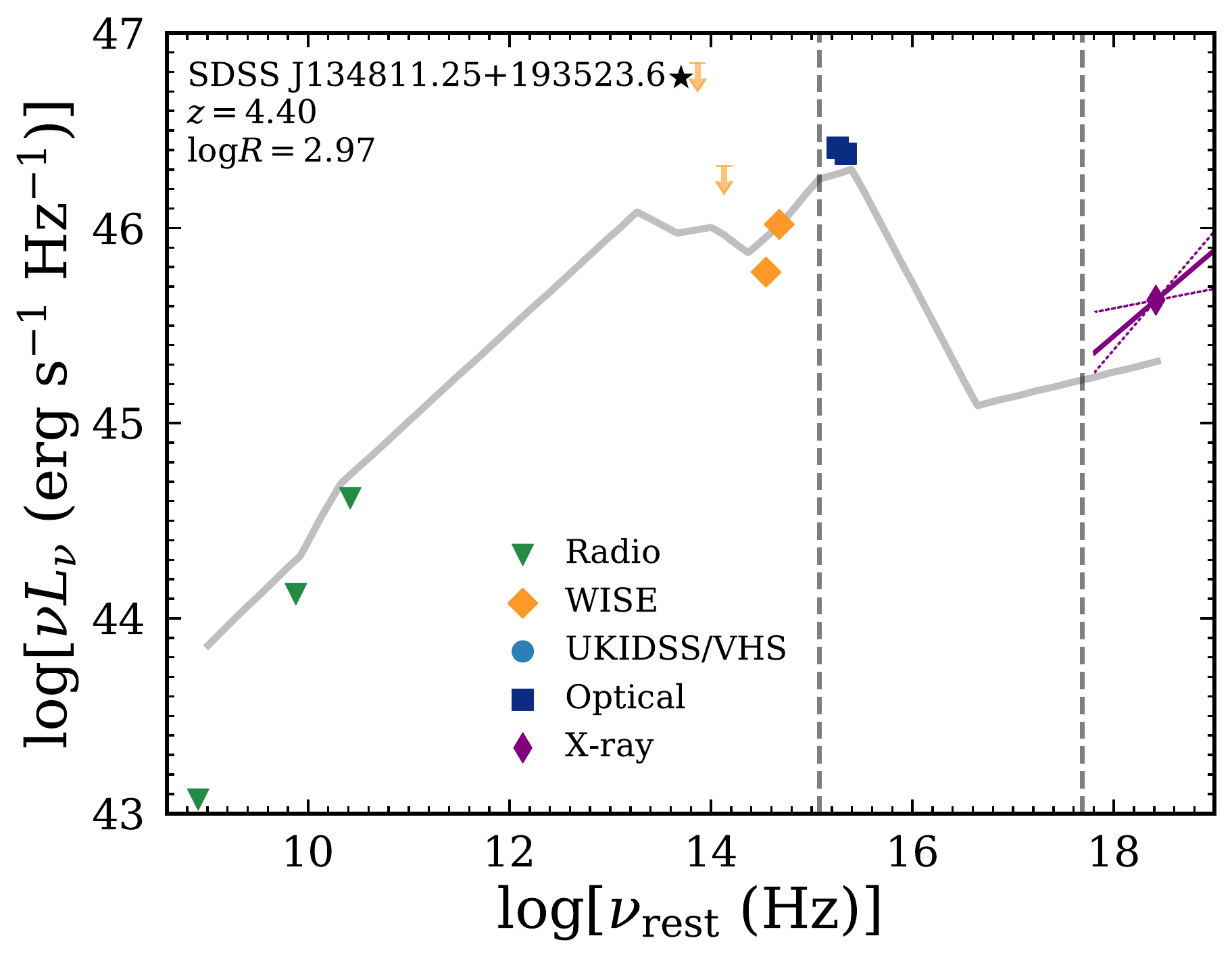}
    \includegraphics[width=0.42\textwidth, clip]{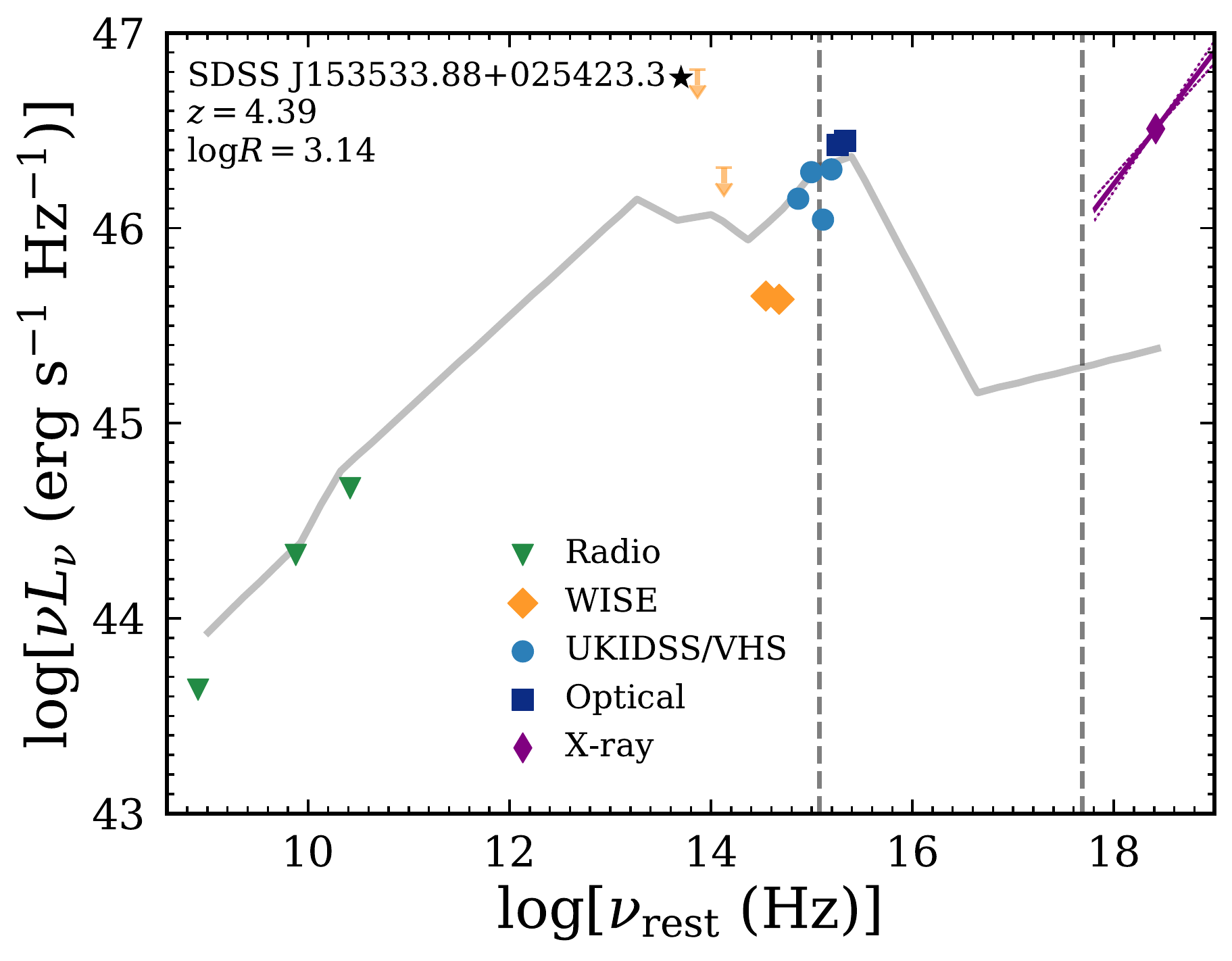}\\
    \includegraphics[width=0.42\textwidth, clip]{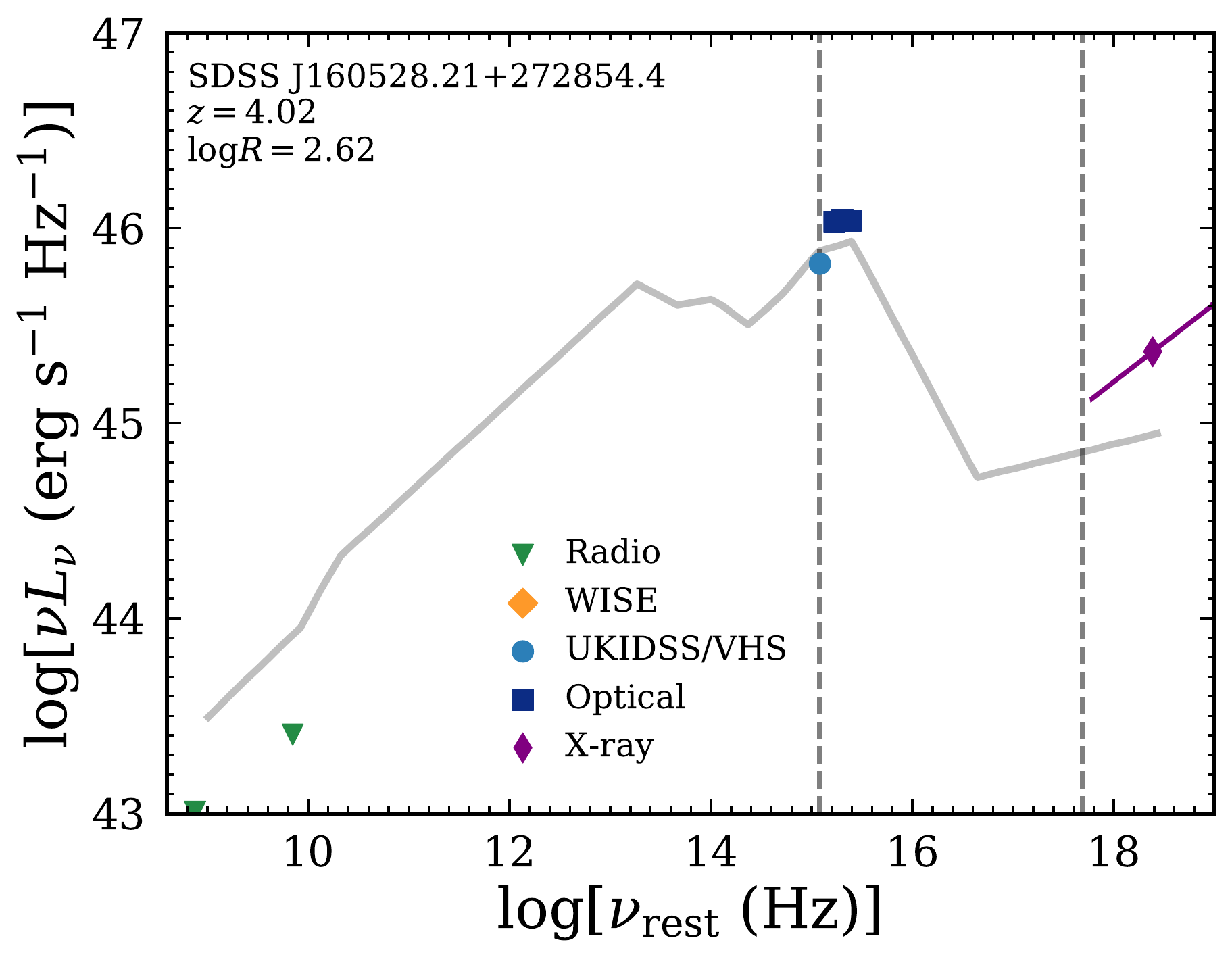}
    \includegraphics[width=0.42\textwidth, clip]{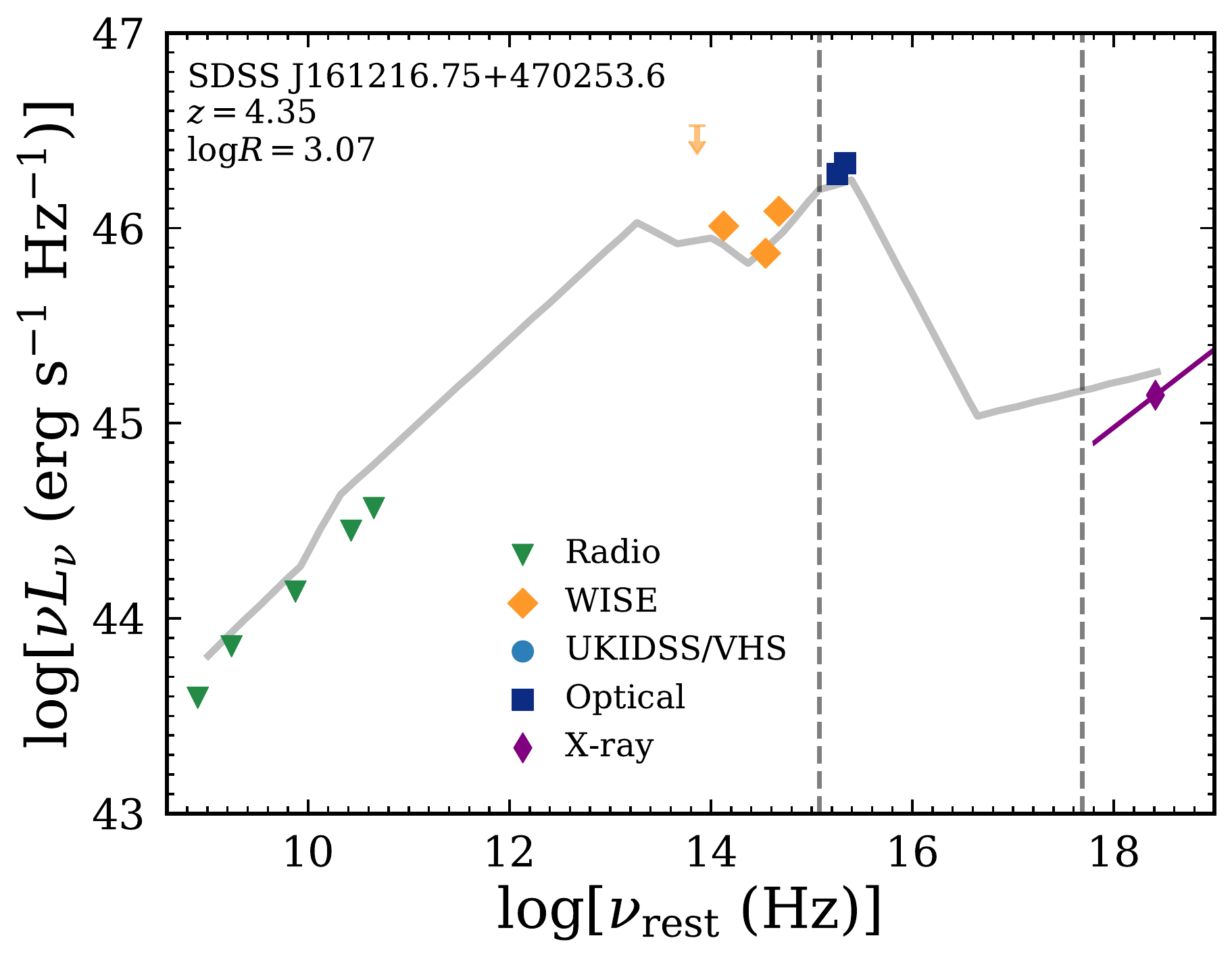}\\
    \includegraphics[width=0.42\textwidth, clip]{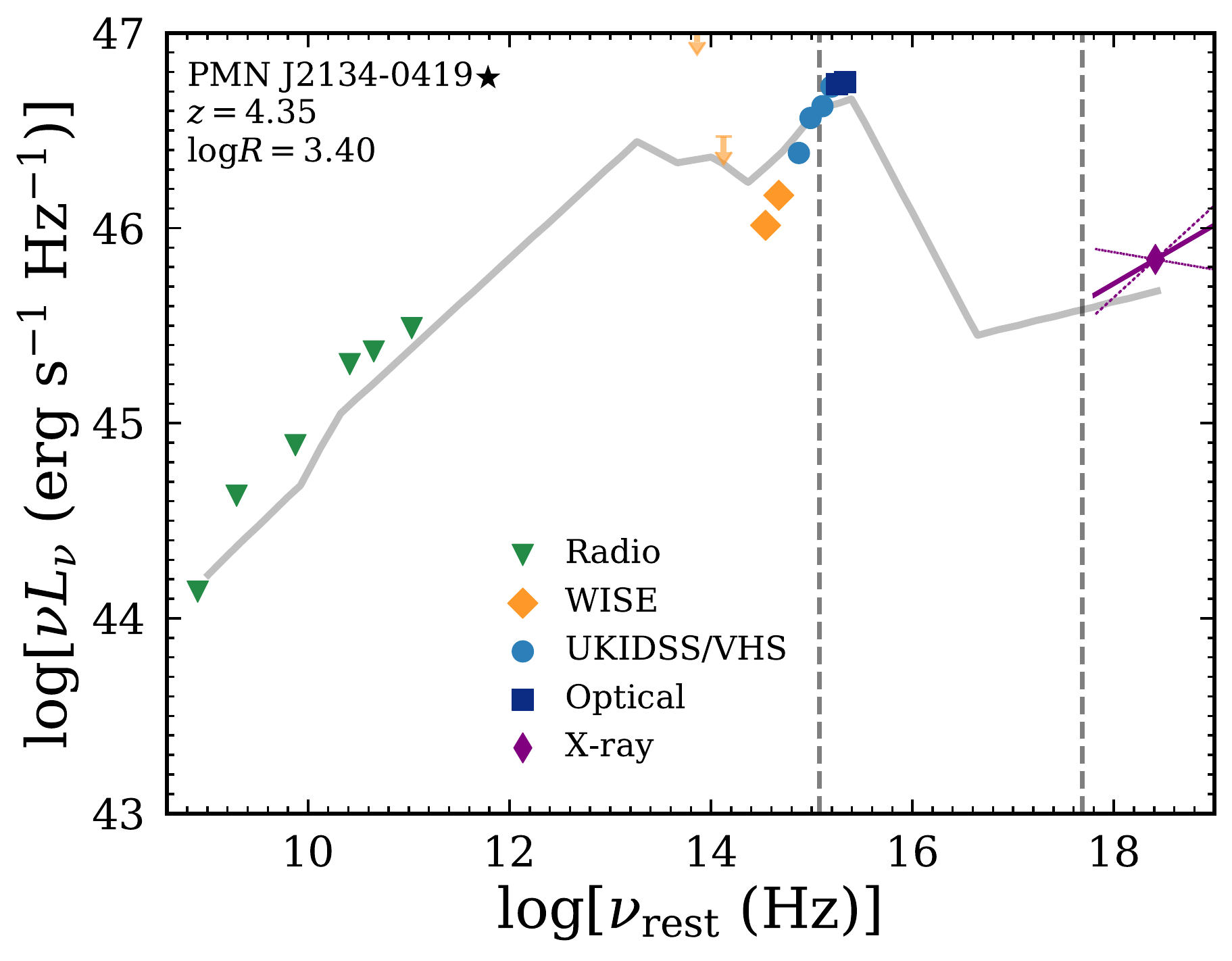}
    \includegraphics[width=0.42\textwidth, clip]{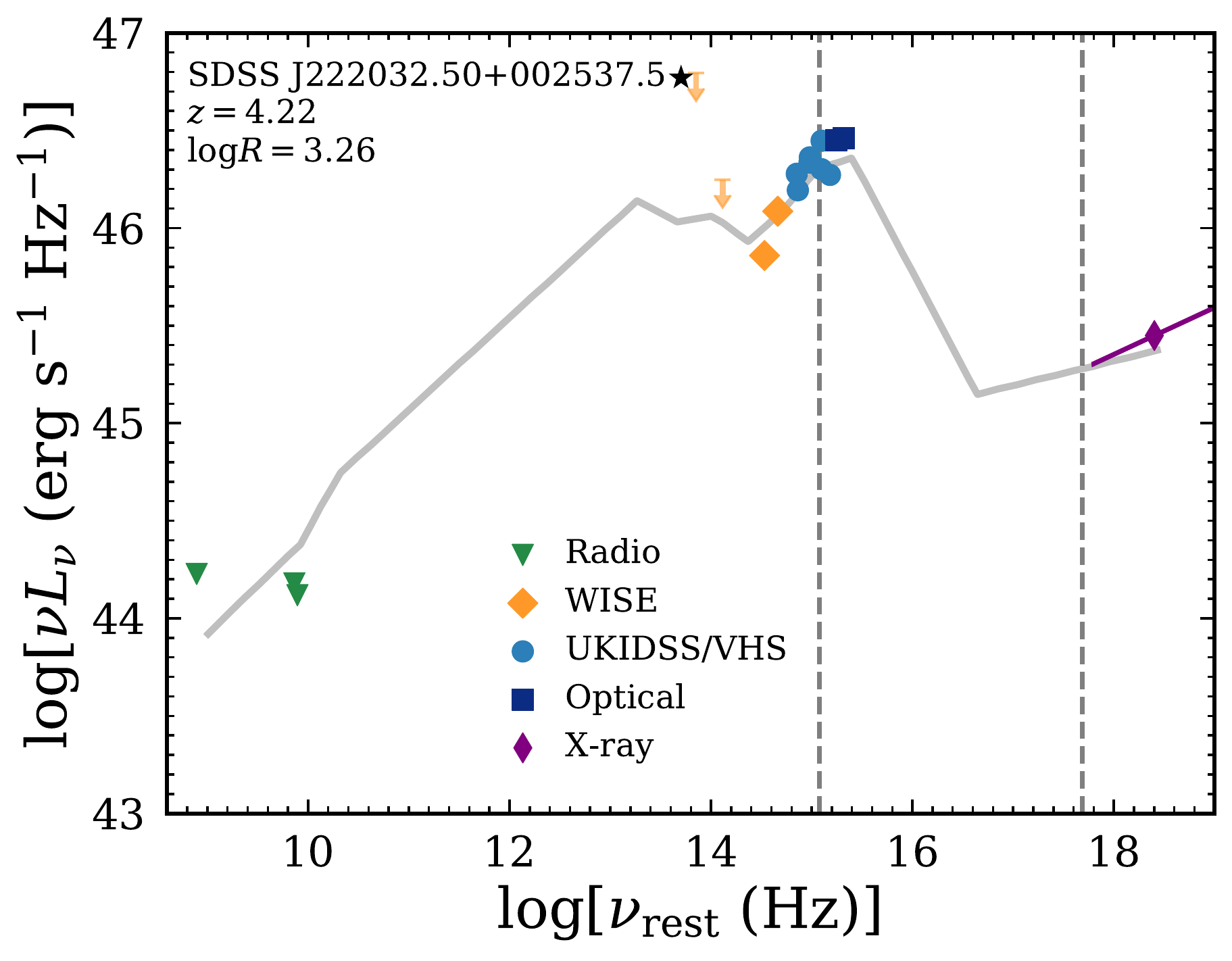}
    \caption{}
\end{figure*}
\renewcommand{\thefigure}{\arabic{figure}}

\renewcommand{\thefigure}{\arabic{figure} (Continued)}
\addtocounter{figure}{-1}

\begin{figure*}
\centering
    \includegraphics[width=0.42\textwidth, clip]{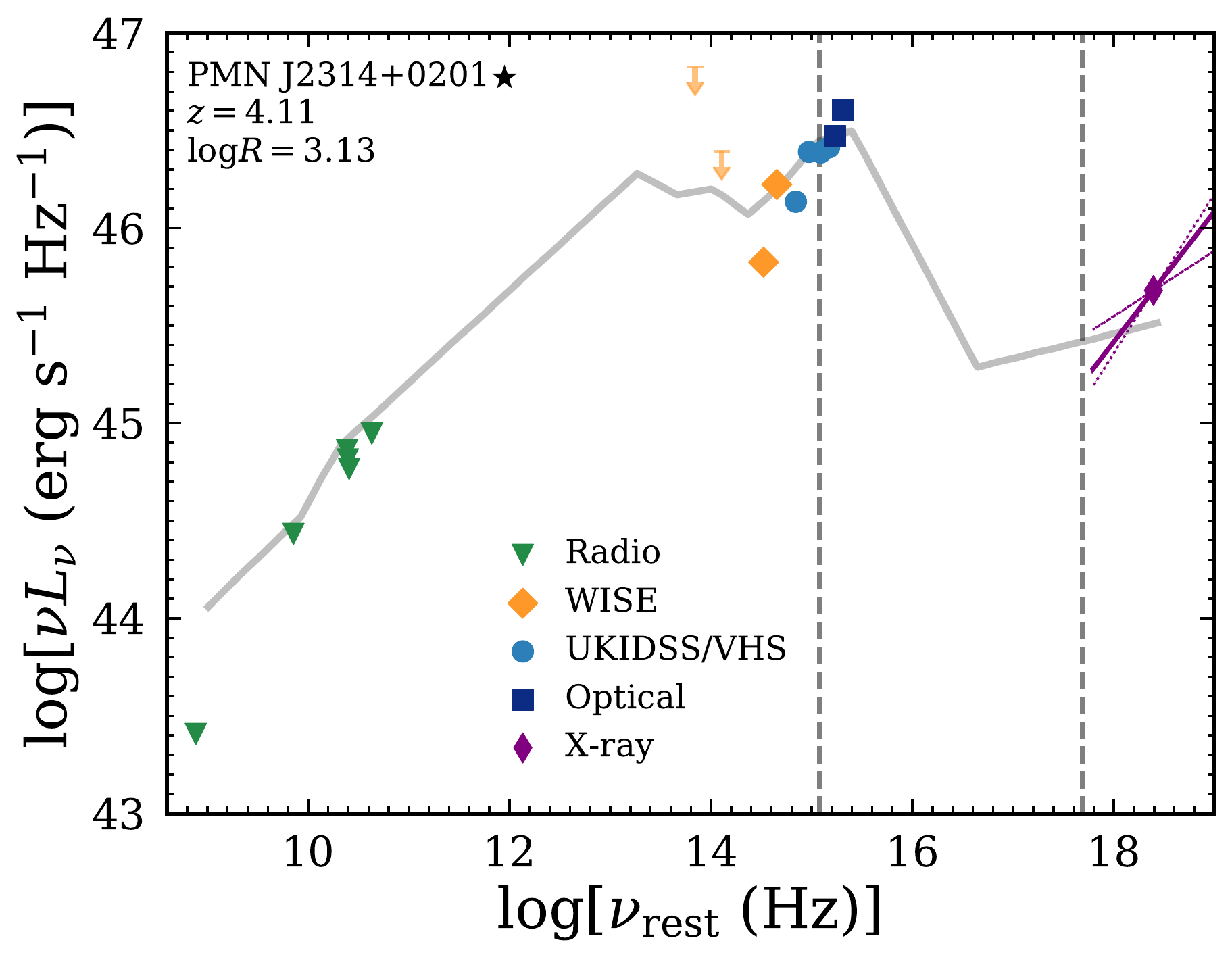}
    \caption{}
\end{figure*}
\renewcommand{\thefigure}{\arabic{figure}}

\section{summary, discussion, and future work}

\subsection{Summary}
In this paper, we have tested and confirmed the X-ray enhancement
of high-redshift ($z>4$) HRLQs ($\log R>2.5$) compared with their low-redshift ($z<4$) counterparts.
We summarize the key points from this work:
\begin{enumerate}
\item
We selected the high-redshift ($z>4$) HRLQs from Wu13 and new objects from sky surveys (SDSS and FIRST) and NED.
We obtained {\it Chandra} observations in Cycle 17 for 6 HRLQs that lacked sensitive X-ray coverage.
We also retrieved archival {\it XMM-Newton} and {\it Swift} X-ray observations that cover another nine high-$z$ HRLQs.
We finally constructed an optically flux-limited sample of 24 HRLQs to $m_i=20.26$ that has complete sensitive X-ray coverage.
See Section~\ref{sec:sample}.
\item
We analyzed the X-ray data and measured HRLQ X-ray photometric properties (see Table~\ref{tab:photometry}).
All the {\it Chandra} Cycle 17 objects were detected in X-rays.
No extended structure was found in the {\it Chandra} images, including for SDSS J$0813+3508$, which is the only {\it Chandra}
Cycle 17 object that has an extended structure in its FIRST image.
See Section~\ref{sec:reduction}.
\item
HRLQs at $z>4$ show an apparent X-ray enhancement compared with matched HRLQs at $z<4$.
The $\Delta\alpha_{\rm ox}$ (including $\Delta\alpha_{\rm ox,RQQ}$ and $\Delta\alpha_{\rm ox,RLQ}$) distributions of the optically flux-limited
high-$z$ sample are significantly different ($\approx$ 4--4.6$\sigma$) from those of the low-$z$ sample.
This result confirms the relevant result of Wu13, in a statistically stronger way and with fewer systematic uncertainties.
See Section~\ref{sec:statTest}.
\item
The typical (median) X-ray enhancement of HRLQs at $z>4$ is a factor of $1.9^{+0.5}_{-0.4}$;
this is smaller than but still consistent with the estimation of Wu13.
See Section~\ref{sec:statEst}.
\item
We constructed the radio--X-ray continuum SEDs for the HRLQs analyzed in this paper, which further illustrate
and support the excess of X-ray emission of high-$z$ HRLQs compared with their low-$z$ counterparts.
See Section~\ref{sec:SEDs}.
\end{enumerate}

\subsection{Discussion}
\label{sec:discuss}

\begin{figure}
\centering
\includegraphics[width=0.45\textwidth, clip]{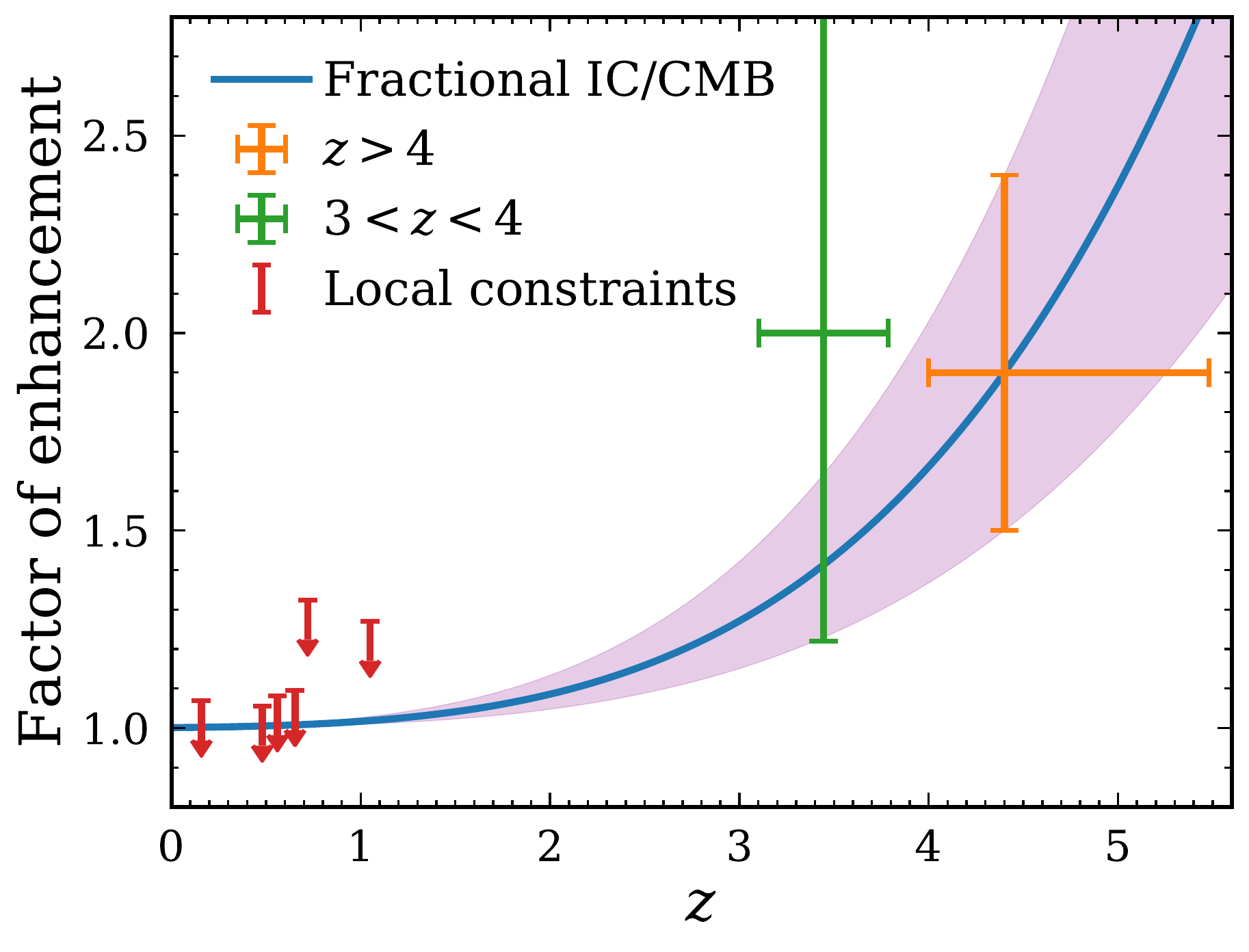}
\caption{The current constraints on the evolution of the X-ray enhancement of HRLQs using the $z>4$ sample from this paper and Wu13 and the $3<z<4$ sample from Miller11.
The blue curve and purple-shaded region
are the prediction from the fractional IC/CMB model and its uncertainties, which are calibrated using the $z>4$ redshift bin.
We constrain the contribution of the IC/CMB mechanism at $z=1.3$ to be $\approx$3\%, with an upper limit of $\approx$5\%.
    The red upper limits are the constraints from non-detections of \mbox{IC/CMB-predicted} $\gamma$-ray emission
from low-$z$ large-scale jets using {\it Fermi} \citep{Meyer2015,Breiding2017,Meyer2017}.
See Section~\ref{sec:discuss} for details.}
\label{fig:iccmb}
\end{figure}

We have confirmed the X-ray enhancement of HRLQs at $z>4$ that was originally proposed by Wu13.
However, we revised the typical amount of X-ray enhancement from a factor of $\approx3$ to $\approx2$.
We plot the factor of X-ray enhancement using the
estimates for HRLQs at $z>4$ in Fig.~\ref{fig:iccmb}.
The fractional IC/CMB model that was suggested by Wu13 can still explain our results if the fraction
of \mbox{X-rays} from the IC/CMB mechanism in HRLQs decreases accordingly.
More specifically, if IC/CMB produces 3\% of the X-ray radiation from HRLQs at $z=1.3$,
a factor of $\approx2$ X-ray enhancement at our median redshift of $z=4.4$ can be
reproduced according to $U_{\rm CMB}\propto(1+z)^4$ evolution, 
assuming that the high-$z$ jets are not physically different
\citep[e.g. severely decelerated on kpc scales;][]{Lopez2006, Volonteri2011, Marshall2018} from the low-$z$ jets.

We have plotted in Fig.~\ref{fig:iccmb} the prediction and
uncertainties of our revised fractional IC/CMB model for the
factor of X-ray enhancement with redshift (blue curve and purple-shaded region).
The curve has the form of $1+A[(1+z)/(1+4.4)]^4$, 
which represents the combination of a non-evolving component and
an evolving IC/CMB-related component.
This curve is calibrated by the analysis of $z>4$ HRLQs.
For example, $A$ needs to be $\approx0.9$ for the blue curve to match the observed enhancement of 1.9 at $z=4.4$.
The boundary of the shaded region is determined accordingly using the error bars of the $z=4.4$ data point.
The curve is consistent with the constraint from $3<z<4$ HRLQs,
which show more substantial uncertainties due to limited sample size.
A larger sample will help to provide tighter constraints on the X-ray enhancement of HRLQs in this redshift bin.
Note that at $z\approx1$ and $z\approx2$ the expected X-ray
enhancements are only factors of $1.02$ and $1.09$, with upper limits of $1.03$ and $1.13$, respectively.

Recall that the quasars with high-resolution {\it Chandra} observations in our sample
do not show any extended structure in their X-ray images,
and the jet-linked \mbox{X-ray} emission is most probably from regions smaller than a few kpc.
The high-$z$ large-scale X-ray jets (if they exist) must lie below the
flux limits of our X-ray observations and are much dimmer than the core region,
in contrast with the prediction of the most-straightforward IC/CMB model under the assumption 
that the radio fluxes of the jets relative to those of the cores do not evolve with redshift
(\citealt{Schwartz2002}; also see \citealt{Bassett2004, Lopez2006}; Miller11).
Either a high-energy synchrotron-emitting electron population or an improved understanding of quasar jets is needed to explain the commonly detected
large-scale X-ray jets of low-$z$ quasars.

The IC/CMB-dominated model for the X-ray emission of
\mbox{large-scale} quasar jets predicts significant radiation in the \mbox{high-energy} $\gamma$-ray band \cite[e.g.][]{Tavecchio2004},
and {\it Fermi} observations have thus been suggested to be used to test the IC/CMB model \cite[e.g.][]{Dermer2004}.
The first such test was performed by \cite{Meyer2014} on the large-scale jet of 3C 273 and their results disfavour the IC/CMB-dominated model.
We here compare the constraints on the IC/CMB X-ray emission from our high-$z$ quasars with the constraints from $z\approx \mbox{0.1--1}$ \mbox{large-scale} jets, which
are shown as upper limits in Fig.~\ref{fig:iccmb}.
The six upper limits are the following: 1.068 \citep[3C 273, $z=0.160$;][]{Meyer2015}, 1.095 \citep[PKS 0637$-$752, $z=0.650$;][]{Meyer2017},
1.055 (PKS 2209+080, $z=0.480$),
1.082 (PKS 1136$-$135, $z=0.560$),
1.324 (PKS 1354+195, $z=0.720$),
and 1.269 \citep[PKS 1229$-$021, $z=1.05$;][]{Breiding2017}.
Following \cite{Meyer2015}, we have assumed the ``angle-averaged''
\mbox{jet-linked} X-ray emission from those low-$z$ RLQs is dominated by large-scale jets,
after correcting for the beaming effect of the radiation from the cores.
We calculated the ratio between the upper limit from the {\it Fermi} data and the predicted IC/CMB $\gamma$-rays,
which is equivalent to the upper limit on the fraction of IC/CMB X-ray emission to total jet-linked X-rays.
Since the literature has divided the {\it Fermi} bandpass into multiple sub-bands, we have chosen the band 
giving the most-stringent constraint.

Note that early X-ray studies using representative samples of
moderately radio-loud to highly radio-loud ($1\lesssim\log R\lesssim 4$)
quasars at $z>4$ (e.g. \citealt{Bassett2004, Lopez2006}; Wu13)
have argued against the scenario where the IC/CMB mechanism plays
a dominant role in the jet-linked X-ray emission from these
high-redshift objects \citep[e.g.][]{Schwartz2002}.
In addition, Miller11 found no evidence supporting
an apparent redshift dependence of X-ray properties
in their large-sample (607 objects) study of RLQs that
spans $0<z<5$ and $1<\log R<5$.

Indeed, considering the X-ray jets of
PKS 0637$-$752 \citep[$z=0.650$, $L_{2-10\ \rm keV}\approx4\times10^{44}$ erg s$^{-1}$;][]{Schwartz2000}
and B3 0727+409 \citep[$z=2.5$, $L_{2-10\ \rm keV}\approx6\times10^{44}$ erg s$^{-1}$;][]{Simionescu2016},
if the IC/CMB model were responsible for their X-ray emission,
their analogs at $z\approx4.4$ would have $L_{2-10\ \rm keV}\approx5\times10^{46}$ erg s$^{-1}$
and $L_{2-10\ \rm keV}\approx 3\times10^{45}$ erg s$^{-1}$, both of which would outshine their cores in X-rays.
However, X-ray observations of RLQs at $z>4$ do
not support this prediction (e.g. \citealt{Bassett2004, Lopez2006, Saez2011, Miller2011}; Wu13).
\footnote{B3 0727+409 has a large-scale jet that is bright at X-ray and faint at radio, which
is thought to be consistent with the prediction of IC/CMB model \citep{Simionescu2016}.
However, B3 0727+409 has a core that is extremely radio-loud ($\log R\approx 6$), and
thus its analog at $z>4$ will not be missed by our selection criterion.}
Note that the X-ray luminosities of the few resolved $z>4$ kpc-scale
jets are only a few percent that of the quasar cores \citep[e.g.][]{Yuan2003, Cheung2012},
consistent with the results for low-$z$ jets.

\cite{McKeough2016} investigated the redshift dependence of the X-ray-to-radio flux ratios ($\alpha_{\rm xr}$) of
11 quasars and found that the $z>3$ quasars have marginally stronger X-ray emission
relative to the $z<3$ quasars in their sample.
\cite{Marshall2018} studied the $\alpha_{\rm rx}$
distribution of 56 quasar jets at $z\lesssim2$ and found weak redshift 
dependence with their $0.95<z<2.05$ sub-sample showing marginally larger X-ray flux densities relative to that of radio than their $0.55<z<0.95$ sub-sample.
Their results disfavour the scenario where the IC/CMB mechanism dominates the
jets' X-ray emission (without changing the properties of high-$z$ jets) and are consistent with our previous result from $z>4$ RLQs.

Wu13 also discussed another possible cause of the X-ray enhancements in which the photon field
of the host galaxy inverse-Compton scatters off
the relativistic electrons in the jets.
This mechanism requires the host galaxies at high redshifts to have enhanced
star-formation activity that produces dense
infrared photon fields \citep[e.g.][]{Wang2011, Mor2012, Netzer2014}.
Our results can still be explained by this scenario.
While the cosmological evolution of the CMB energy density can be easily predicted,
the evolution of the star-forming activity of the hosts of quasars with powerful relativistic jets at different redshifts
has not been established \citep[e.g.][]{Archibald2001}.
However, if future X-ray studies of HRLQs that extend to $z\approx$ 0.5--4 detect any
deviation from the prediction of the blue curve in Fig.~\ref{fig:iccmb} and
disfavour the fractional IC/CMB model, alternative models like this will gain more credit.

\cite{Ajello2009} found that the number density of
\mbox{flat-spectrum} radio quasars (FSRQs) selected by the {\it Swift}/Burst Alert Telescope (BAT; in hard X-rays) has a peak at 
a notably high redshift of $z\approx$ 3--4.
The interpretation of such a number-density peak can be affected by the X-ray luminosity enhancement of HRLQs at $z>4$ we confirmed here.
Qualitatively, this X-ray enhancement might cause high-$z$ HRLQs to be more easily
picked up by {\it Swift}/BAT, and their apparent peak in number density will correspondingly be biased toward a higher redshift.
A quantitative discussion of this issue is beyond the scope of this paper.

\subsection{Future work}
\label{sec:future}
There are several ways the results in this work might be productively
extended. First, the sample statistics of the $z>4$ HRLQs could be
improved by future X-ray observations of the additional objects
listed in Table~\ref{tab:suplist}. Several objects in
Table~\ref{tab:suplist} have already been scheduled for {\it Chandra}
observations, and a {\it Chandra} snapshot survey of the remaining
objects would extend complete X-ray coverage to an optically flux-limited
sample reaching $m_i=21$ with a size of 37. With this larger sample size,
the level of X-ray enhancement of $z>4$ HRLQs could be better constrained.

Furthermore, one could now substantially enlarge the sample of HRLQs at
$z<4$ with sensitive X-ray coverage via systematic archival data mining.
The Miller11 sample used for our $z<4$ comparisons here was largely based
on SDSS Data Release~5 (DR5) from 2007 \citep[e.g.][]{Schneider2007},
and it utilized X-ray coverage from {\it Chandra}, {\it XMM-Newton},
and {\it ROSAT}. Over the past decade, more than 450,000 new quasars
have been spectroscopically identified by the SDSS \citep[e.g.][]{Paris2017}, including many new HRLQs at
\hbox{$z\approx 0.5$--4}. Furthermore, the sizes of the {\it Chandra}
and {\it XMM-Newton} archives have grown substantially since the
work of Miller11, and more sensitive radio data have been gathered
in the SDSS footprint (e.g. via the ongoing VLA Sky
Survey\footnote{\url{https://science.nrao.edu/science/surveys/vlass}}). Systematic
archival X-ray analyses of these new \hbox{$z\approx 0.5$--4} HRLQs
should allow more precise measurements of the factor of X-ray enhancement
vs.\ redshift (see Figure~\ref{fig:iccmb}), thereby testing and quantifying the
fractional IC/CMB model.

Finally, alternative explanations of the observed X-ray enhancement
should also be explored. For example, ALMA measurements of star-formation
rates for $z>4$ HRLQ hosts could test if their star formation
is sufficiently elevated to drive the X-ray enhancement via a stronger
host seed photon field (see Section~\ref{sec:intro}).

\section*{Acknowledgements}
We thank M. Ajello, D. P. Schneider, and W. M. Yi for helpful discussions.
We thank the referee for helpful comments.
SFZ and WNB acknowledge support from the Penn State ACIS
Instrument Team Contract SV4-74018 (issued by the {\it Chandra}
X-ray Center, which is operated by the Smithsonian Astrophysical
Observatory for and on behalf of NASA under contract
NAS8-03060) and CXC grant AR8-19011X.
The Chandra Guaranteed Time Observations (GTO) for some of the quasars
studied herein were selected by the ACIS Instrument Principal
Investigator, Gordon P. Garmire, currently of the Huntingdon Institute
for X-ray Astronomy, LLC, which is under contract to the Smithsonian
Astrophysical Observatory via Contract SV2-82024.

\bibliographystyle{mnras}
\bibliography{mn} 

\begin{thebibliography}{}
\bibitem[Acero et al.(2015)]{Acero2015} Acero, F., Ackermann, M., Ajello, M., et al.\ 2015, \apjs, 218, 23
\bibitem[Ajello et al.(2009)]{Ajello2009} Ajello, M., Costamante, L., Sambruna, R.~M., et al.\ 2009, \apj, 699, 603
\bibitem[Amirkhanyan \& Mikhailov(2006)]{Amirkhanyan2006} Amirkhanyan, V.~R. \& Mikhailov, V.~P.\ 2006, Astrophysics, 49, 184
\bibitem[Archibald et al.(2001)]{Archibald2001} Archibald, E.~N., Dunlop, J.~S., Hughes, D.~H., et al.\ 2001, \mnras, 323, 417
\bibitem[Atoyan \& Dermer(2004)]{Atoyan2004} Atoyan, A. \& Dermer, C.~D.\ 2004, \apj, 613, 151
\bibitem[Avni et al.(1980)]{Avni1980} Avni, Y., Soltan, A., Tananbaum, H., \& Zamorani, G.\ 1980, \apj, 238, 800
\bibitem[Bassett et al.(2004)]{Bassett2004} Bassett, L.~C., Brandt, W.~N., Schneider, D.~P., et al.\ 2004, \aj, 128, 523
\bibitem[Becker et al.(1995)]{Becker1995} Becker, R.~H., White, R.~L., \& Helfand, D.~J.\ 1995, \apj, 450, 559
\bibitem[Begelman et al.(1984)]{Begelman1984} Begelman, M.~C., Blandford, R.~D. \& Rees, M.~J.\ 1984, Reviews of Modern Physics, 56, 255
\bibitem[Brandt \& Alexander(2015)]{Brandt2015} Brandt, W.~N., \& Alexander, D.~M.\ 2015, \aapr, 23, 1
\bibitem[Brandt et al.(2002)]{Brandt2002} Brandt, W.~N., Schneider, D.~P., Fan, X., et al.\ 2002, \apjl, 569, L5
\bibitem[Breiding et al.(2017)]{Breiding2017} Breiding, P., Meyer, E.~T., Georganopoulos, M., et al.\ 2017, \apj, 849, 95
\bibitem[Brunetti et al.(1999)]{Brunetti1999} Brunetti, G., Comastri, A., Setti, G., et al.\ 1999, \aap, 342, 57
\bibitem[Burrows et al.(2005)]{Burrows2005} Burrows, D.~N., Hill, J.~E., Nousek, J.~A., et al.\ 2005, \ssr, 120, 165
\bibitem[Cao et al.(2017)]{Cao2017} Cao, H.-M., Frey, S., Gab{\'a}nyi, K.~{\'E}., et al.\ 2017, \mnras, 467, 950
\bibitem[Celotti et al.(2001)]{Celotti2001} Celotti, A., Ghisellini, G., \& Chiaberge, M.\ 2001, \mnras, 321, L1
\bibitem[Chartas et al.(2000)]{Chartas2000} Chartas, G., Worrall, D.~M., Birkinshaw, M., et al.\ 2000, \apj, 542, 655
\bibitem[Cheung et al.(2012)]{Cheung2012} Cheung, C.~C., Stawarz, {\L}., Siemiginowska, A., et al.\ 2012, \apj, 756, L20
\bibitem[Conrad(2015)]{Conrad2015} Conrad, J.\ 2015, Astroparticle Physics, 62, 165
\bibitem[Condon et al.(1998)]{Condon1998} Condon, J.~J., Cotton, W.~D., Greisen, E.~W., et al.\ 1998, \aj, 115, 1693
\bibitem[Dermer \& Atoyan(2004)]{Dermer2004} Dermer, C.~D. \& Atoyan, A.\ 2004, \apj, 611, L9
\bibitem[Dickey \& Lockman(1990)]{Dickey1990} Dickey, J.~M., \& Lockman, F.~J.\ 1990, \araa, 28, 215
\bibitem[Fan(2012)]{Fan2012} Fan, X.\ 2012, Research in Astronomy and Astrophysics, 12, 865
\bibitem[Feigelson et al.(1995)]{Feigelson1995} Feigelson, E.~D., Laurent-Muehleisen, S.~A., Kollgaard, R.~I., et al.\ 1995, \apj, 449, L149
\bibitem[Feigelson \& Nelson(1985)]{Feigelson1985} Feigelson, E.~D., \& Nelson, P.~I.\ 1985, \apj, 293, 192
\bibitem[Felten \& Morrison(1966)]{Felten1966} Felten, J.~E. \& Morrison, P.\ 1966, \apj, 146, 686
\bibitem[Freeman et al.(2002)]{Freeman2002} Freeman, P.~E., Kashyap, V., Rosner, R., \& Lamb, D.~Q.\ 2002, \apjs, 138, 185
\bibitem[Frey et al.(2010)]{Frey2010} Frey, S., Paragi, Z., Gurvits, L.~I., Cseh, D., \& Gab{\'a}nyi, K.~{\'E}.\ 2010, \aap, 524, A83
\bibitem[Hook et al.(2002)]{Hook2002} Hook, I.~M., McMahon, R.~G., Shaver, P.~A., et al.\ 2002, \aap, 391, 509
\bibitem[Garmire et al.(2003)]{Garmire2003} Garmire, G.~P., Bautz, M.~W., Ford, P.~G., Nousek, J.~A., \& Ricker, G.~R., Jr.\ 2003, \procspie, 4851, 28
\bibitem[Ghisellini \& Tavecchio(2009)]{Ghisellini2009} Ghisellini, G., \& Tavecchio, F.\ 2009, \mnras, 397, 985
\bibitem[Gibson \& Brandt(2012)]{Gibson2012} Gibson, R.~R. \& Brandt, W.~N.\ 2012, \apj, 746, 54
\bibitem[Gibson et al.(2008)]{Gibson2008} Gibson, R.~R., Brandt, W.~N., \& Schneider, D.~P.\ 2008, \apj, 685, 773
\bibitem[Gregory et al.(1996)]{Gregory1996} Gregory, P.~C., Scott, W.~K., Douglas, K., \& Condon, J.~J.\ 1996, \apjs, 103, 427
\bibitem[Hardcastle \& Croston(2011)]{Hardcastle2011} Hardcastle, M.~J., \& Croston, J.~H.\ 2011, \mnras, 415, 133
\bibitem[Hardcastle(2006)]{Hardcastle2006} Hardcastle, M.~J.\ 2006, \mnras, 366, 1465
\bibitem[Harris \& Grindlay(1979)]{Harris1979} Harris, D.~E. \& Grindlay, J.~E.\ 1979, \mnras, 188, 25
\bibitem[Harris \& Krawczynski(2002)]{Harris2002} Harris, D.~E., \& Krawczynski, H.\textbackslash 2002, \apj, 565, 244
\bibitem[Harris \& Krawczynski(2006)]{Harris2006} Harris, D.~E., \& Krawczynski, H.\ 2006, \araa, 44, 463
\bibitem[Hewett \& Wild(2010)]{Hewett2010} Hewett, P.~C., \& Wild, V.\ 2010, \mnras, 405, 2302
\bibitem[Hodge et al.(2011)]{Hodge2011} Hodge, J.~A., Becker, R.~H., White, R.~L., et al.\ 2011, \aj, 142, 3
\bibitem[Hogan et al.(2011)]{Hogan2011} Hogan, B.~S., Lister, M.~L., Kharb, P., et al.\ 2011, \apj, 730, 92
\bibitem[Intema et al.(2017)]{Intema2017} Intema, H.~T., Jagannathan, P., Mooley, K.~P., et al.\ 2017, \aap, 598, A78
\bibitem[Ivezi{\'c} et al.(2002)]{Ivezic2002} Ivezi{\'c}, {\v Z}., Menou, K., Knapp, G.~R., et al.\ 2002, \aj, 124, 2364
\bibitem[Ivezi{\'c} et al.(2004)]{Ivezic2004} Ivezi{\'c}, Z., Richards, G., Hall, P., et al.\ 2004, AGN Physics with the Sloan Digital Sky Survey, 311, 347
\bibitem[Jiang et al.(2006)]{Jiang2006} Jiang, L., Fan, X., Hines, D.~C., et al.\ 2006, \aj, 132, 2127
\bibitem[Just et al.(2007)]{Just2007} Just, D.~W., Brandt, W.~N., Shemmer, O., et al.\ 2007, \apj, 665, 1004
\bibitem[Kaplan \& Meier(1958)]{Kaplan1958} Kaplan, E.~L., \& Meier, P.\ 1958, J. Amer. Statist. Assn., 53, 457
\bibitem[Kataoka \& Stawarz(2005)]{Kataoka2005} Kataoka, J. \& Stawarz, {\L}.\ 2005, \apj, 622, 797
\bibitem[Kellermann et al.(1989)]{Kellermann1989} Kellermann, K.~I., Sramek, R., Schmidt, M., Shaffer, D.~B., \& Green, R.\ 1989, \aj, 98, 1195
\bibitem[Kelly(2007)]{Kelly2007} Kelly, B.~C.\ 2007, \apj, 665, 1489
\bibitem[Kimball et al.(2011a)]{Kimball2011a} Kimball, A.~E., Ivezi{\'c}, {\v{Z}}., Wiita, P.~J., et al.\ 2011a, \aj, 141, 182.
\bibitem[Kimball et al.(2011b)]{Kimball2011b} Kimball, A.~E., Kellermann, K.~I., Condon, J.~J., Ivezi{\'c}, {\v Z}., \& Perley, R.~A.\ 2011b, \apjl, 739, L29
\bibitem[Kraft et al.(1991)]{Kraft1991} Kraft, R.~P., Burrows, D.~N., \& Nousek, J.~A.\ 1991, \apj, 374, 344
\bibitem[Lauer et al.(2007)]{Lauer2007} Lauer, T.~R., Tremaine, S., Richstone, D., \& Faber, S.~M.\ 2007, \apj, 670, 249
\bibitem[Lawrence et al.(2007)]{Lawrence2007} Lawrence, A., Warren, S.~J., Almaini, O., et al.\ 2007, \mnras, 379, 1599
\bibitem[Lopez et al.(2006)]{Lopez2006} Lopez, L.~A., Brandt, W.~N., Vignali, C., et al.\ 2006, \aj, 131, 1914
\bibitem[Lu et al.(2007)]{Lu2007} Lu, Y., Wang, T., Zhou, H., et al.\ 2007, \aj, 133, 1615
\bibitem[Lynden-Bell(1971)]{Lynden-Bell1971} Lynden-Bell, D.\ 1971, \mnras, 155, 95
\bibitem[MacLeod et al.(2010)]{MacLeod2010} MacLeod, C.~L., Ivezi{\'c}, {\v{Z}}., Kochanek, C.~S., et al.\ 2010, \apj, 721, 1014
\bibitem[Marshall et al.(2005)]{Marshall2005} Marshall, H.~L., Schwartz, D.~A., Lovell, J.~E.~J., et al.\ 2005, \apjs, 156, 13
\bibitem[Marshall et al.(2018)]{Marshall2018} Marshall, H.~L., Gelbord, J.~M., Worrall, D.~M., et al.\ 2018, \apj, 856, 66
\bibitem[McGreer et al.(2013)]{McGreer2013} McGreer, I.~D., Jiang, L., Fan, X., et al.\ 2013, \apj, 768, 105
\bibitem[McKeough et al.(2016)]{McKeough2016} McKeough, K., Siemiginowska, A., Cheung, C.~C., et al.\ 2016, \apj, 833, 123
\bibitem[McMahon et al.(2013)]{McMahon2013} McMahon, R.~G., Banerji, M., Gonzalez, E., et al.\ 2013, The Messenger, 154, 35
\bibitem[Meyer \& Georganopoulos(2014)]{Meyer2014} Meyer, E.~T. \& Georganopoulos, M.\ 2014, \apj, 780, L27
\bibitem[Meyer et al.(2015)]{Meyer2015} Meyer, E.~T., Georganopoulos, M., Sparks, W.~B., et al.\ 2015, \apj, 805, 154
\bibitem[Meyer et al.(2016)]{Meyer2016} Meyer, E.~T., Sparks, W.~B., Georganopoulos, M., et al.\ 2016, \apj, 818, 195
\bibitem[Meyer et al.(2017)]{Meyer2017} Meyer, E.~T., Breiding, P., Georganopoulos, M., et al.\ 2017, \apjl, 835, L35
\bibitem[Miller et al.(2011)]{Miller2011} Miller, B.~P., Brandt, W.~N., Schneider, D.~P., et al.\ 2011, \apj, 726, 20 (Miller11)
\bibitem[Mor et al.(2012)]{Mor2012} Mor, R., Netzer, H., Trakhtenbrot, B., Shemmer, O., \& Lira, P.\ 2012, \apjl, 749, L25
\bibitem[Mullin \& Hardcastle(2009)]{Mullin2009} Mullin, L.~M., \& Hardcastle, M.~J.\ 2009, \mnras, 398, 1989
\bibitem[Nanni et al.(2017)]{Nanni2017} Nanni, R., Vignali, C., Gilli, R., Moretti, A., \& Brandt, W.~N.\ 2017, \aap, 603, A128
\bibitem[Netzer et al.(2014)]{Netzer2014} Netzer, H., Mor, R., Trakhtenbrot, B., Shemmer, O., \& Lira, P.\ 2014, \apj, 791, 34
\bibitem[Padovani(2017)]{Padovani2017} Padovani, P.\ 2017, Nature Astronomy, 1, 194
\bibitem[Page et al.(2005)]{Page2005} Page, K.~L., Reeves, J.~N., O'Brien, P.~T., \& Turner, M.~J.~L.\ 2005, \mnras, 364, 195
\bibitem[Planck Collaboration et al.(2016)]{Planck2016} Planck Collaboration, Ade, P.~A.~R., Aghanim, N., et al. 2016, \aap, 594, A13
\bibitem[P{\^a}ris et al.(2018)]{Paris2017} P{\^a}ris, I., Petitjean, P., Aubourg, {\'E}., et al.\ 2018, \aap, 613, A51
\bibitem[Park et al.(2006)]{Park2006} Park, T., Kashyap, V.~L., Siemiginowska, A., et al.\ 2006, \apj, 652, 610
\bibitem[Richards et al.(2006)]{Richards2006} Richards, G.~T., Strauss, M.~A., Fan, X., et al.\ 2006, \aj, 131, 2766
\bibitem[Rosen et al.(2016)]{Rosen2016} Rosen, S.~R., Webb, N.~A., Watson, M.~G., et al.\ 2016, \aap, 590, A1
\bibitem[Saez et al.(2011)]{Saez2011} Saez, C., Brandt, W.~N., Shemmer, O., et al.\ 2011, \apj, 738, 53
\bibitem[Sbarrato et al.(2015)]{Sbarrato2015} Sbarrato, T., Ghisellini, G., Tagliaferri, G., et al.\ 2015, \mnras, 446, 2483
\bibitem[Shang et al.(2011)]{Shang2011} Shang, Z., Brotherton, M.~S., Wills, B.~J., et al.\ 2011, \apjs, 196, 2
\bibitem[Siemiginowska et al.(2003)]{Siemiginowska2003} Siemiginowska, A., Smith, R.~K., Aldcroft, T.~L., et al.\ 2003, \apjl, 598, L15
\bibitem[Sikora et al.(2007)]{Sikora2007} Sikora, M., Stawarz, {\L}., \& Lasota, J.-P.\ 2007, \apj, 658, 815
\bibitem[Simionescu et al.(2016)]{Simionescu2016} Simionescu, A., Stawarz, {\L}., Ichinohe, Y., et al.\ 2016, \apjl, 816, L15
\bibitem[Singal et al.(2013)]{Singal2013} Singal, J., Petrosian, V., Stawarz, {\L}., \& Lawrence, A.\ 2013, \apj, 764, 43
\bibitem[Schmidt(1968)]{Schmidt1968} Schmidt, M.\ 1968, \apj, 151, 393
\bibitem[Schmitt(1985)]{Schmitt1985} Schmitt, J.~H.~M.~M.\ 1985, \apj, 293, 178
\bibitem[Schneider et al.(2007)]{Schneider2007} Schneider, D.~P., Hall, P.~B., Richards, G.~T., et al.\ 2007, \aj, 134, 102
\bibitem[Schneider et al.(2010)]{Schneider2010} Schneider, D.~P., Richards, G.~T., Hall, P.~B., et al.\ 2010, \aj, 139, 2360
\bibitem[Schwartz et al.(2000)]{Schwartz2000} Schwartz, D.~A., Marshall, H.~L., Lovell, J.~E.~J., et al.\ 2000, \apjl, 540, 69
\bibitem[Schwartz(2002)]{Schwartz2002} Schwartz, D.~A.\ 2002, \apjl, 569, L23
\bibitem[Schlafly \& Finkbeiner(2011)]{Schlafly2011} Schlafly, E.~F., \& Finkbeiner, D.~P.\ 2011, \apj, 737, 103
\bibitem[Shemmer et al.(2017)]{Shemmer2017} Shemmer, O., Brandt, W.~N., Paolillo, M., et al.\ 2017, \apj, 848, 46
\bibitem[Shemmer et al.(2006)]{Shemmer2006} Shemmer, O., Brandt, W.~N., Schneider, D.~P., et al.\ 2006, \apj, 644, 86
\bibitem[Shen et al.(2011)]{Shen2011} Shen, Y., Richards, G.~T., Strauss, M.~A., et al.\ 2011, \apjs, 194, 45
\bibitem[Skrutskie et al.(2006)]{Skrutskie2006} Skrutskie, M.~F., Cutri, R.~M., Stiening, R., et al.\ 2006, \aj, 131, 1163
\bibitem[Spergel et al.(2003)]{Spergel2003} Spergel, D.~N., Verde, L., Peiris, H.~V., et al.\ 2003, \apjs, 148, 175
\bibitem[Stark et al.(1992)]{Stark1992} Stark, A.~A., Gammie, C.~F., Wilson, R.~W., et al.\ 1992, \apjs, 79, 77
\bibitem[Str{\"u}der et al.(2001)]{Struder2001} Str{\"u}der, L., Briel, U., Dennerl, K., et al.\ 2001, \aap, 365, L18
\bibitem[Sambruna et al.(2004)]{Sambruna2004} Sambruna, R.~M., Gambill, J.~K., Maraschi, L., et al.\ 2004, \apj, 608, 698
\bibitem[Tananbaum et al.(1979)]{Tananbaum1979} Tananbaum, H., Avni, Y., Branduardi, G., et al.\ 1979, \apj, 234, L9
\bibitem[Tavecchio et al.(2000)]{Tavecchio2000} Tavecchio, F., Maraschi, L., Sambruna, R.~M., \& Urry, C.~M.\ 2000, \apjl, 544, L23
\bibitem[Tavecchio et al.(2004)]{Tavecchio2004} Tavecchio, F., Maraschi, L., Sambruna, R.~M., et al.\ 2004, \apj, 614, 64
\bibitem[Uchiyama et al.(2006)]{Uchiyama2006} Uchiyama, Y., Urry, C.~M., Cheung, C.~C., et al.\ 2006, \apj, 648, 910
\bibitem[Vanden Berk et al.(2001)]{Vanden2001} Vanden Berk, D.~E., Richards, G.~T., Bauer, A., et al.\ 2001, \aj, 122, 549
\bibitem[Vanden Berk et al.(2004)]{Vanden2004} Vanden Berk, D.~E., Wilhite, B.~C., Kron, R.~G., et al.\ 2004, \apj, 601, 692
\bibitem[Vignali et al.(2003)]{Vignali2003a} Vignali, C., Brandt, W.~N., Schneider, D.~P., Garmire, G.~P., \& Kaspi, S.\ 2003, \aj, 125, 418
\bibitem[Volonteri et al.(2011)]{Volonteri2011} Volonteri, M., Haardt, F., Ghisellini, G., \& Della Ceca, R.\ 2011, \mnras, 416, 216
\bibitem[Wall \& Jenkins(2012)]{Wall2012} Wall, J.~V., \& Jenkins, C.~R.\ 2012, Practical Statistics for Astronomers, Cambridge, UK: Cambridge University Press
\bibitem[Wang et al.(2011)]{Wang2011} Wang, R., Wagg, J., Carilli, C.~L., et al.\ 2011, \aj, 142, 101
\bibitem[Weisskopf et al.(2007)]{Weisskopf2007} Weisskopf, M.~C., Wu, K., Trimble, V., et al.\ 2007, \apj, 657, 1026
\bibitem[Wilkes \& Elvis(1987)]{Wilkes1987} Wilkes, B.~J., \& Elvis, M.\ 1987, \apj, 323, 243
\bibitem[Worrall et al.(1987)]{Worrall1987} Worrall, D.~M., Tananbaum, H., Giommi, P., \& Zamorani, G.\ 1987, \apj, 313, 596
\bibitem[Wright et al.(2010)]{Wright2010} Wright, E.~L., Eisenhardt, P.~R.~M., Mainzer, A.~K., et al.\ 2010, \aj, 140, 1868
\bibitem[Wright et al.(1994)]{Wright1994} Wright, A.~E., Griffith, M.~R., Burke, B.~F., \& Ekers, R.~D.\ 1994, \apjs, 91, 111
\bibitem[Wu et al.(2013)]{Wu2013} Wu, J., Brandt, W.~N., Miller, B.~P., et al.\ 2013, \apj, 763, 109 (Wu13)
\bibitem[Wu et al.(2017)]{Wu2017} Wu, J., Ghisellini, G., Hodges-Kluck, E., et al.\ 2017, \mnras, 468, 109
\bibitem[Yang et al.(2016)]{Yang2016a} Yang, J., Wang, F., Wu, X.-B., et al.\ 2016, \apj, 829, 33
\bibitem[Yuan et al.(2003)]{Yuan2003} Yuan, W., Fabian, A.~C., Celotti, A., \& Jonker, P.~G.\ 2003, \mnras, 346, L7
\bibitem[Zamfir et al.(2008)]{Zamfir2008} Zamfir, S., Sulentic, J.~W., \& Marziani, P.\ 2008, \mnras, 387, 856
\end{thebibliography}

\appendix
\section{Statistical testing for extended structure}
\label{sec:imgTest}

The {\it Chandra} Cycle 17 objects mostly have limited photon counts.
It is not feasible to directly compare their images with the PSF images to check for extended X-ray jets \citep[e.g.][]{Wu2017}.
We first produced the PSF image for each observation using ray-tracing\footnote{\url{http://cxc.harvard.edu/cal/Hrma/SAOTrace.html} and \url{http://space.mit.edu/CXC/MARX/}} with
a large number of simulated events. The statistical method we used is as follows.

We only consider $D\times D$ patches of pixels
centred at the source position on the X-ray image and PSF image.
The net source counts ($N_{\rm src}$) and background counts ($N_{\rm bkg}$) of the X-ray image are estimated using photometry (see Section~\ref{sec:reduction}).
We add background events to the PSF image with a total number of

\begin{equation}
    N_{\rm bkg,psf} = \frac{N_{\rm bkg}}{N_{\rm src}}N_{\rm src,psf},
\end{equation}
where $N_{\rm src,psf}$ is the total number of simulated events in the PSF image.
We then ``flatten'' the 2-dimensional images to sequences of pixels of length $L=D\times D$.
The distribution of events in the PSF image is multinomial, with the estimated probability of $i$-th pixel

\begin{equation}
    p_i=\frac{n_{i\textrm{,psf}} + N_{\rm bkg,psf}/L}{N_{\rm src,psf}+N_{\rm bkg,psf}},
\end{equation}
where $n_{i\textrm{,psf}}$ is the number of events in the $i$th pixel of the PSF image with $N_{\rm src,psf}=\sum_in_{i,\rm psf}$.

\begin{figure*}
\centering
\includegraphics[width=0.28\textwidth, clip]{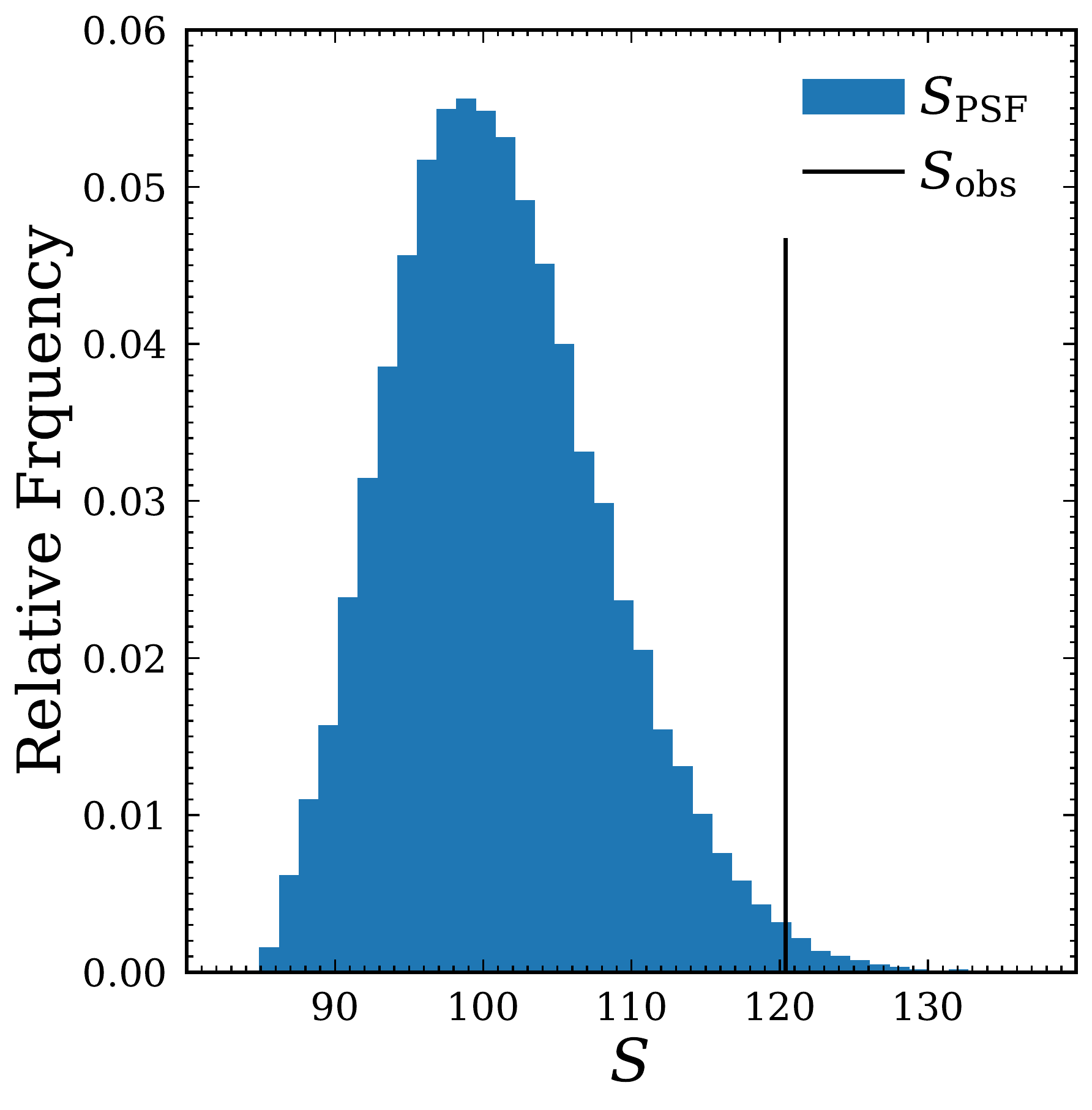}\quad\quad
\includegraphics[width=0.6\textwidth, clip]{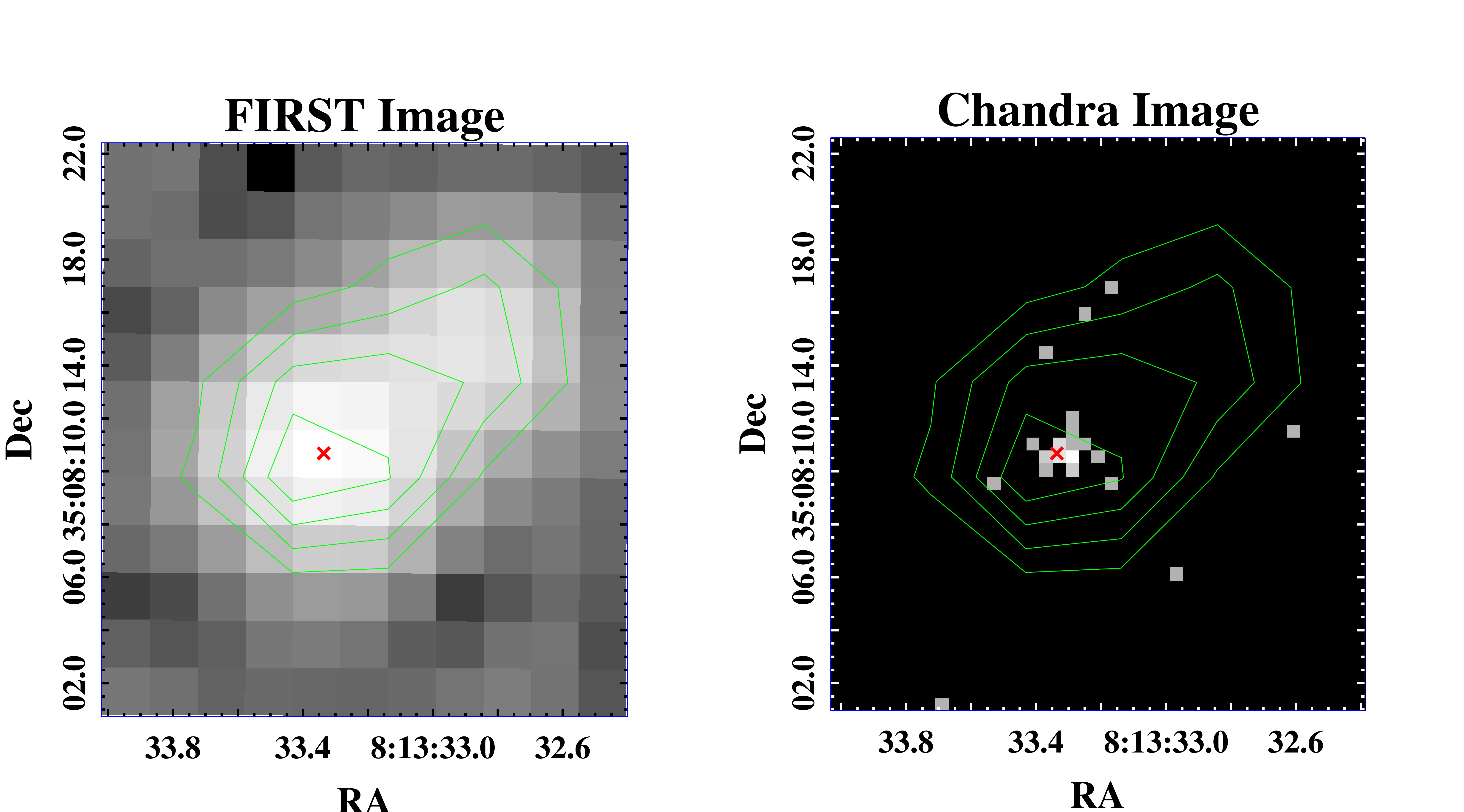}
\caption{Left: The empirical distribution of $S$ (blue histogram) calculated using simulated images of the {\it Chandra} PSF,
together with the $S$ value (vertical black line) calculated using the observed 0.5--8 keV {\it Chandra} image of SDSS J$0813+3508$.
Centre: Radio image of SDSS J$0813+3508$ from FIRST at 1.4 GHz. Right: X-ray image of SDSS J$0813+3508$ in the 0.5--8 keV band.
The red ``$\times$'' symbols in both the centre and right panels indicate the optical position of the quasar.
Contours of the radio image are plotted in both the centre and right panels, 
where the radio surface brightness increases from 2.9~mJy~beam$^{-1}$ (outermost) to 12.2~mJy~beam$^{-1}$ (innermost) with a step size of 3.3 mJy beam$^{-1}$.
Even though $S_{\rm obs}$ appears marginally inconsistent with the distribution predicted by the {\it Chandra} PSF (left, $p\approx0.011$),
the X-ray image (right) does not show an extended structure that corresponds to the structure in the radio image (centre).}
\label{fig:radioXimg}
\end{figure*}

The likelihood of observing the X-ray image of a point source given the PSF's multinomial distribution is
\begin{equation}
    P({\rm X\textnormal{-}ray\;image}|{\rm PSF}) = \frac{N_{\rm src}!}{n_1!n_2! \dots n_L!}\prod_{i=0}^Lp_i^{n_i},
\end{equation}
where $n_i$ is the number of events in the $i$-th pixel of the observed X-ray image with $N_{\rm src}=\sum_in_i$.
We define a simpler statistic
\begin{equation}
    S=-\sum_{i=1}^L[n_i\ln p_i - \ln (n_i!)],
\end{equation}
which is the negative log-likelihood, omitting constants. The calculation is reduced by realizing
that $n_i$ is mostly 0 or 1 in the case of low counts.
By drawing a large set of samples from the multinomial distribution $(p_1,\dots,p_L)$, we obtain the empirical
distribution of $S$, from which the $p$-value can be calculated. See Fig.~\ref{fig:radioXimg} (left) for the example of SDSS J$0813+3508$.

In our calculation, we used $D=20$ (i.e. the size of the patch is $10''\times10''$).
We found a detection for the 0.5--8 keV image of SDSS J$0813+3508$ with $p=0.0108$,
which becomes less significant (the $p$-value increases to $p=0.0832$)
after taking into account the number
of tests we have performed \citep[e.g.][]{Conrad2015}.
We then compared the radio and X-ray images of SDSS J$0813+3508$,
and did not find extended X-ray structure that corresponds to the radio jet,
as shown in Fig.~\ref{fig:radioXimg} (centre and right).
Note that the extended radio component that is $\approx7''$ away from SDSS J$0813+3508$
has a peak radio flux density (11.9 mJy) that is about half that of the core (20.0 mJy).
We estimated the (observed-frame \mbox{0.5--2}~keV) surface brightness coincident with
the extended radio component to be $<1.43\times10^{-16}$ erg cm$^{-2}$ s$^{-1}$ arcsec$^{-2}$,
which means that the extended jet (if it exists) must be more than 10 times dimmer than the core in X-rays.
Other sources are consistent with point sources.
We conclude that no statistically convincing extended structure is found in the X-ray images.

We investigated further the constraints coming from the \mbox{non-detections}
on the relative brightness of X-ray jets.
In the X-ray images, we put an artificial point source $2''$
(4 pixels; the typical size of the resolved X-ray jets for quasars at $z>4$) away from the core and increase its
intensity until it is detected with the statistical tests above.
Typically 3--4 photons are needed for the artificial jets to be detected. Therefore, any X-ray jets have to be
$\gtrsim$ \mbox{3--25} times fainter than the cores for the {\it Chandra} Cycle 17 quasars,
consistent with previous X-ray upper limits for high-$z$ RLQs \citep[e.g.][]{Bassett2004, Lopez2006}.

\section{The effects of non-simultaneous data and measurement errors}
\label{sec:err}
Since both types of uncertainties under consideration, measurement errors and variability, are stochastic in nature, it is impossible
to apply corrections to the observed data to obtain underlying ``true'' values.
The purpose of the Monte Carlo simulation is to add more fluctuations to the data
and observe the consequence of the enlarged uncertainties.
We created degraded samples by adding random numbers
drawn from $N(0, 0.06^2)$ (a Gaussian distribution with zero mean and standard deviation $\sigma=0.06$)
to the observed $\Delta\alpha_{\rm ox}$ (only detections).
The statistical significance typically drops by $\sim1\sigma$ using these worsened values, which is
expected because more noise will tend to wash out the differences between the two distributions.

To demonstrate further the effects of uncertainties, we performed Bayesian fitting using a Gaussian structural model.
What we provide below is a largely simplified version of the model of \cite{Kelly2007}.
We assume the uncertainties on $\Delta\alpha_{\rm ox}$ due to measurement errors
and variability are Gaussian distributed with standard deviation $\sigma_{\rm e}=0.06$.
We assume $\Delta\alpha_{\rm ox}$ follows a Gaussian distribution with mean $\mu$ (identical to the median) and standard deviation $\sigma_{\rm i}$.
Here, $\sigma_{\rm i}$ represents the part of the scatter of the $\alpha_{\rm ox}$-$L_{2500\rm\angstrom}$
and $L_{2\rm\ keV}$-$L_{2500\ \rm\angstrom}$-$L_{5\ \rm GHz}$ relations
that cannot be explained by measurement error and variability \citep[e.g.][]{Gibson2008}.
The structural model is formulated as
\begin{align}
    \mu&\sim {\rm Uniform}(-1,1),\\
    \sigma_{\rm i}&\sim {\rm Uniform}(0,1),\\
    \Delta\alpha_{\rm ox}^{\rm true}&\sim N(\mu,\sigma_{\rm i}^2), \\
    \Delta\alpha_{\rm ox}^{\rm obs}&\sim N(\mu,\sigma_{\rm i}^2+\sigma_{\rm e}^2).
\end{align}
Following Section 5.2 of \cite{Kelly2007} and considering cases with both detections and non-detections, the likelihood of the model parameters is
\begin{equation}
    \begin{split}
        \ln\mathcal{L}&=-\sum_{j=1}^{N_{\rm det}}\frac{(\mu-\Delta\alpha_{\rm ox,j})^2}{2(\sigma_{\rm i}^2+\sigma_{\rm e}^2)}-\frac{1}{2}N_{\rm det}\ln[2\pi(\sigma_{\rm i}^2+\sigma_{\rm e}^2)]\\
        &+\sum_{k=1}^{N_{\rm non\textnormal{-}det}}\ln\Big[\int_{-\infty}^{\Delta\alpha_{\rm ox,k}}N(\mu, \sigma_{\rm i}^2+\sigma_{\rm e}^2)dx\Big],
    \end{split}
\end{equation}
where $N_{\rm det}$ and $N_{\rm non\textnormal{-}det}$ are the numbers of detections and \mbox{non-detections}, respectively.
We have used $\Delta\alpha_{\rm ox,j}$ and $\Delta\alpha_{\rm ox,k}$ to denote the values of detections and assigned upper limits for non-detections, respectively.
The above likelihood has taken the uncertainties into account by marginalising them out.

\begin{figure*}
\centering
\includegraphics[width=0.80\textwidth, clip]{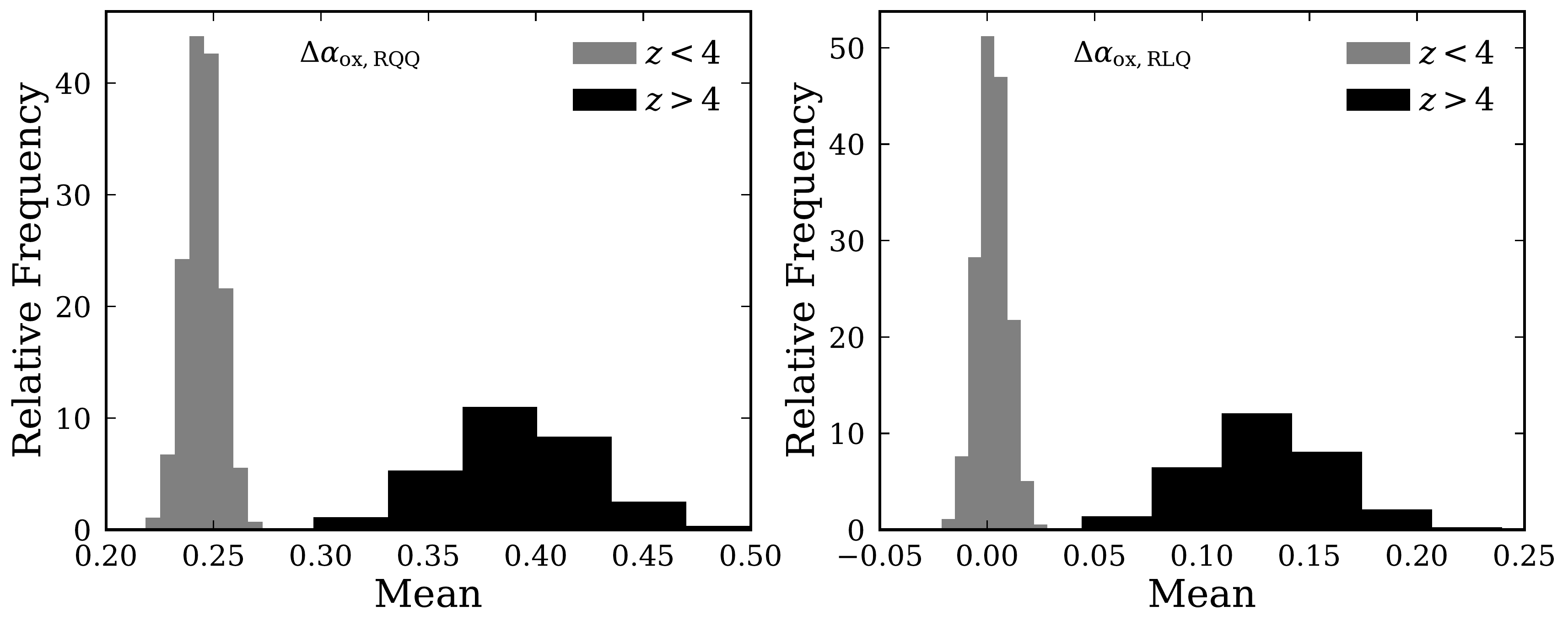}
\caption{The posterior distribution of the means (also medians) of $\Delta\alpha_{\rm ox,RQQ}$ (left) and $\Delta\alpha_{\rm ox,RLQ}$ (right),
where grey denotes the low-$z$ sample and black denotes the high-$z$ sample.}
\label{fig:errTest}
\end{figure*}

We have drawn samples from the posterior distributions of $\mu$ and $\sigma_{\rm i}$ for $\Delta\alpha_{\rm ox,RQQ}$ and $\Delta\alpha_{\rm ox,RLQ}$ for both redshift bins.
We plot in Fig.~\ref{fig:errTest} the comparison of $\mu$ for different redshift bins.
The impact of the sample size reflects itself in the concentration of the distribution:
the mean of the low-$z$ sample is better constrained than that of the high-$z$ sample.
Student's $t$-tests return, practically, $p=0.0$, i.e.
it is almost impossible for the centres of the $\Delta\alpha_{\rm ox}$ distributions for the different redshift bins to be consistent with each other,
after considering the smearing effect of uncertainties.

In addition to hypothesis testing, the modelling process above can also be used to calculate the amount of X-ray enhancement (Section~\ref{sec:statEst}).
Fig.~\ref{fig:errTest} (right) indicates that the $\Delta\alpha_{\rm ox, RLQ}$ of HRLQs at $z>4$ are larger that that of HRLQs at $z<4$ by $0.13\pm0.03$ on average,
which is consistent with the result of the Kaplan-Meier estimator.

\section{Kaplan-Meier estimator and bootstrapping}
\begin{figure*}
\centering
\includegraphics[width=0.8\textwidth, clip]{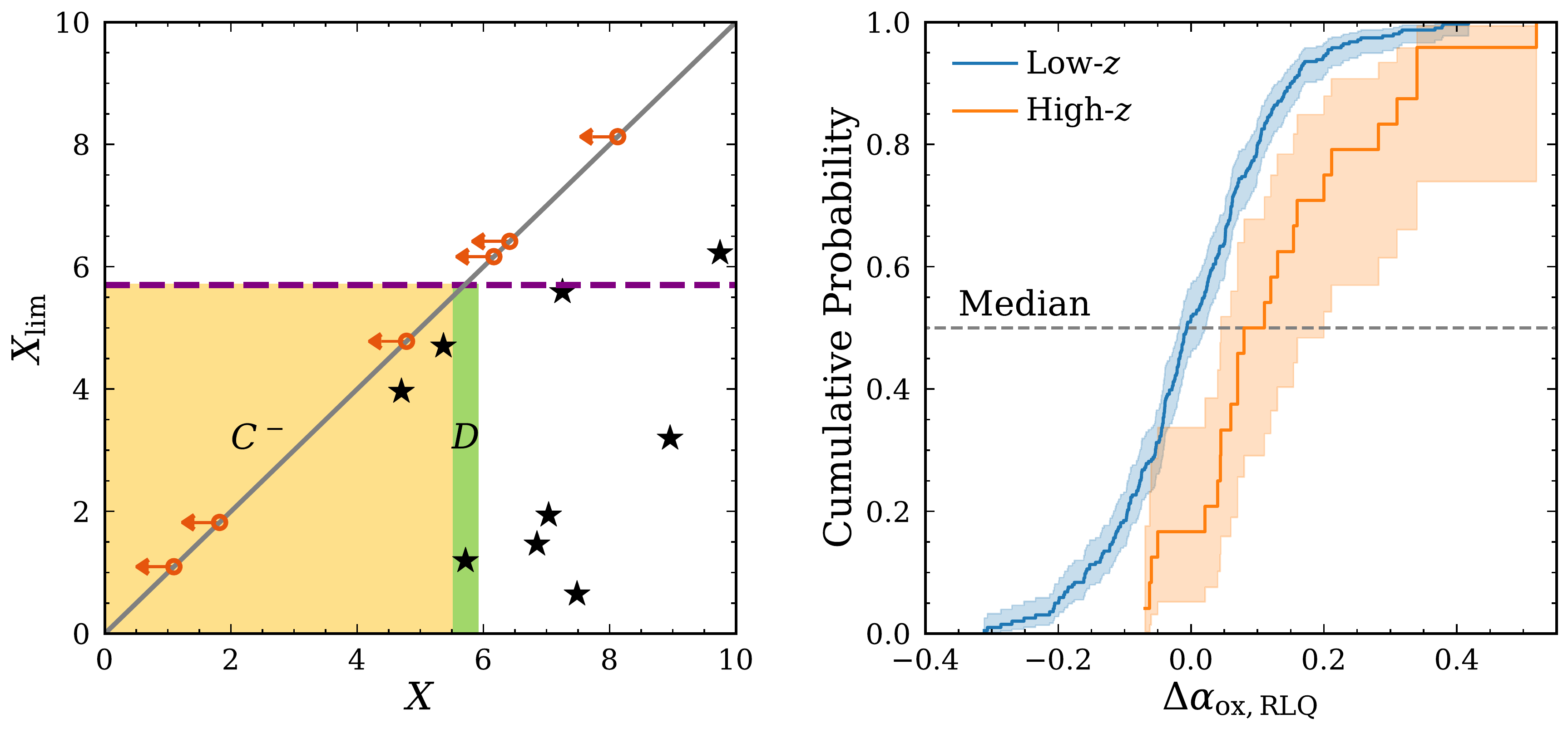}
\caption{Left: The $x$-axis is the value of detection,
and the $y$-axis is the limit of each observation.
Upper limits are shown as leftward arrows,
in contrast to detections that are shown as stars.
The number of detections in the narrow green area is noted as $D$;
the number of observations (including detections and upper limits)
in the non-overlapping yellow-shaded area is noted as $C^-$.
A sub-sample is defined by the dashed purple line,
under which the data points can represent the distribution of that of the complete sample on $X$ direction
if $X$ and $X_{\rm lim}$ are independent of each other.
Right: The cumulative probability distribution of $\Delta\alpha_{\rm ox,RLQ}$ of
the flux-limited samples of HRLQs at high redshift (orange) and low redshift (blue).
The crossing points of the horizontal dashed lines with the curves are medians.}
\label{fig:sa}
\end{figure*}

\label{sec:kme}
Some HRLQs in the Miller11 sample have only upper limits for their X-ray fluxes, which leads to the
corresponding measurements of $\Delta\alpha_{\rm ox}$ also being upper limits.
Therefore, we measure the median of $\Delta\alpha_{\rm ox}$ using the Kaplan-Meier curve \citep{Kaplan1958},
which is a maximum-likelihood estimator
of the survival function (or, equivalently, the cumulative distribution function).
To justify its application to our problem,
we briefly provide in the below a heuristic derivation of the Kaplan-Meier estimator
based on the assumption that whether an HRLQ is detected or not in X-rays is
independent of the true value of its $\Delta\alpha_{\rm ox}$,
which is essential for the application of survival analysis \citep[e.g.][]{Avni1980, Wall2012}.
The upper limits of $\Delta\alpha_{\rm ox, RQQ}$ and $\Delta\alpha_{\rm ox, RLQ}$
from Miller11 spread across a wide dynamic range (see Fig.~\ref{fig:DeltaAlphaOXHists}),
and thus we think this assumption is reasonable.
The derivation deals with left-censored data (data with upper limits) that are
common in the astronomical context, while most of the statistical literature deals
with right-censored data (data with lower limits).\footnote{\cite{Feigelson1985} circumvent this by providing a prescription that converts
left-censored data to right-censored data.}

We treat the result of each observation as a pair of the underlying value and the observational limit $(X, X_{\rm lim})$;
for the observations in the region of $X<X_{\rm lim}$, only the upper limit is recorded.
We plot a mock experiment in Fig.~\ref{fig:sa}~(left),
where detections are shown as stars in the bottom-right triangle on the plane, and upper limits are shown as leftward arrows on the diagonal.
We define the cumulative distribution function $\Psi(x)\equiv P(X\le x)$.
The key observation of the left panel of Fig.~\ref{fig:sa} is that,
under the assumption of the independence of $X$ and $X_{\rm lim}$,
\begin{equation}
    \frac{d\Psi}{\Psi(x)}=\frac{dD}{C(x)}
\end{equation}
follows, where $D(x)$ denotes the number of detections in a narrow strip at $x$, below the line of $X=X_{\rm lim}$ (green-shaded region), and
$C(x)$ denotes the number of observations (including detections and upper limits) in the rectangle left of $x$ (yellow-shaded region).
This is because the data points below the dashed purple line in Fig.~\ref{fig:sa}~(left)
form a sub-sample that has the same $\Psi(x)$ as that of the complete sample.
The solution of $\Psi(x)$ is
\begin{align}
    \Psi(x)= A\exp{\Big[\int\frac{dD}{C}\Big]} & = A\exp{\Big[\int_{-\infty}^x \frac{dx'}{C(x')}\frac{dD(x')}{dx'}\Big]} \\
    & = A\prod_{i:x_i\le x} \frac{C^-(x_i)+D(x_i)}{C^-(x_i)}
\end{align}
where we explicitly differentiate the regions of $C$ excluding $D$ and including $D$, using $C^-$ and $C^+$, respectively.
Note that $i$ is only the index for detections.
Therefore, the Kaplan-Meier estimator of the cumulative
distribution function here is a series of increasing step functions,
and the jumps only happen at the values of detections.
The factor $A$ can be chosen as
\begin{equation}
    A=\prod_i\frac{C^-(x_i)}{C^-(x_i)+D(x_i)}
\end{equation}
so that $\Psi(x)=1$ at $x\ge x_{\max}$.
Therefore, the Kaplan-Meier estimator for the cumulative distribution function is
\begin{equation}
    \Psi(x) = \prod_{i:x_i>x}\frac{C^-(x_i)}{C^-(x_i)+D(x_i)}= \prod_{i:x_i>x}\Big[1-\frac{D(x_i)}{C^+(x_i)}\Big]
\end{equation}
For most cases, each measured $x_i$ is a discrete quantity, i.e. $D(x_i)=1$.
However, in the bootstrapping we performed below, we will also have $D(x_i)>1$.
We have used a Python package\footnote{\url{http://lifelines.readthedocs.io/en/latest/}} to calculate the Kaplan-Meier estimator.
We show an example of the cumulative distribution function in Fig.~\ref{fig:sa} (right), where the crossing points
of the horizontal dashed line with the S-shaped curves are the median estimates.
We calculate the difference of the medians to quantify the X-ray enhancement of HRLQs at $z>4$ relative to HRLQs at low redshifts.
We bootstrapped the samples of $\Delta\alpha_{\rm ox}$ (1000 times) to estimate the dispersion of medians,
as well as the scattering of the difference of medians.

This derivation is motivated by that of the $C^-$ method in \cite{Lynden-Bell1971},
based on the similarity between their mathematical forms.
The estimator of the survival function for \mbox{right-censored} data can be easily derived from a plot that is similar to Fig.~\ref{fig:sa}~(left), where
the detections are in the top-left triangle of the $X_{\rm lim}$-$X$ plane and lower limits are rightward arrows on the diagonal.
See \cite{Feigelson1985} and \cite{Schmitt1985} for a complete discussion on astronomical applications of
the Kaplan-Meier estimator
and \cite{Avni1980} for a different
algorithm that works with singly censored astronomical data.

\bsp	
\label{lastpage}
\end{document}